\documentclass[12pt]{article}
\usepackage{jheppub}
\pdfoutput=1
\usepackage{epstopdf}

\usepackage{jheppub}
\usepackage{mathrsfs}
\usepackage{psfrag}

\usepackage[table]{xcolor}

\usepackage{amsmath,bbm,array,amsfonts,graphicx,wrapfig,lscape,float,slashbox,multirow,longtable,rotating,subfigure}
\usepackage{url}

\newcommand{\be}{\begin{equation}}
\newcommand{\ee}{\end{equation}}
\newcommand{\beq}{\begin{equation}}
\newcommand{\beql}[1]{\begin{equation}\label{#1}}
\newcommand{\eeq}{\end{equation}}
\newcommand{\ba}{\begin{array}}
\newcommand{\ea}{\end{array}}
\newcommand{\bea}{\begin{eqnarray}}
\newcommand{\beal}[1]{\begin{eqnarray}\label{#1}}
\newcommand{\eea}{\end{eqnarray}}
\newcommand{\ben}{\begin{enumerate}}
\newcommand{\een}{\end{enumerate}}
\newcommand{\bean}{\begin{eqnarray*}}
\newcommand{\eean}{\end{eqnarray*}}
\newcommand{\eref}[1]{(\ref{#1})}
\newcommand{\sref}[1]{\S\ref{#1}}
\newcommand{\tref}[1]{Table~\ref{#1}}

\newcommand{\fref}[1]{Figure \ref{#1}}
\newcommand{\btab}[1]{\begin{tabular}{#1}}
\newcommand{\etab}{\end{tabular}}

\newcommand{\comment}[1]{}

\newcommand{\qed}{\nobreak \ifvmode \relax \else
      \ifdim\lastskip<1.5em \hskip-\lastskip
      \hskip1.5em plus0em minus0.5em \fi \nobreak
      \vrule height0.75em width0.5em depth0.25em\fi}

\newcolumntype{C}[1]{>{\centering\arraybackslash}m{#1}}

\newcommand{\Gr}{$Gr_{k,n}$}
\newcommand{\pl}{Pl\"ucker }

\newcommand{\rotxc}[1]{\begin{sideways}#1\end{sideways}}
\newcommand{\invert}[1]{\rotxc{\rotxc{#1}}}
\def\Le{\hbox{\invert{$\Gamma$}}}

\title{The Geometry of On-Shell Diagrams}

\author{Sebasti\'an Franco,}
\author{Daniele Galloni,}
\author{Alberto Mariotti}

\affiliation{
Institute for Particle Physics Phenomenology, Department of Physics\\
Durham University, Durham DH1 3LE, United Kingdom
}

\emailAdd{sebastian.franco@durham.ac.uk, daniele.galloni@durham.ac.uk,alberto.mariotti@durham.ac.uk}

\abstract{The fundamental role of on-shell diagrams in quantum field theory has been recently recognized. On-shell diagrams, or equivalently bipartite graphs, provide a natural bridge connecting gauge theory to powerful mathematical structures such as the Grassmannian. We perform a detailed investigation of the combinatorial and geometric objects associated to these graphs. We mainly focus on their relation to polytopes and toric geometry, the Grassmannian and its stratification. Our work extends the current understanding of these connections along several important fronts, most notably eliminating restrictions imposed by planarity, positivity, reducibility and edge removability. We illustrate our ideas with several explicit examples and introduce concrete methods that considerably simplify computations. We consider it highly likely that the structures unveiled in this article will arise in the on-shell study of scattering amplitudes beyond the planar limit. Our results can be conversely regarded as an expansion in the understanding of the Grassmannian in terms of bipartite graphs.
}

\preprint{
\begin{flushright}IPPP/13/89\end{flushright} \vspace{-0.9cm}
\begin{flushright}DCPT/13/178\end{flushright}
}

\begin{document}

\maketitle


\section{Introduction}

We are in the midst of what might become a profound reformulation of quantum field theory, one which privileges hidden infinite dimensional symmetries over manifest locality and unitarity \cite{ArkaniHamed:2009dn,Mason:2009qx,ArkaniHamed:2009vw,ArkaniHamed:2010kv,Bourjaily:2010kw}.  The main laboratory for the new ideas is planar $\mathcal{N}=4$ SYM. This approach has led to a focus on on-shell diagrams, equivalently bipartite graphs, which determine well-defined physical quantities exhibiting all the symmetries of the quantum field theory \cite{ArkaniHamed:2012nw}. On-shell diagrams can be used as building blocks of scattering amplitudes. In addition, they reveal a profound and only recently explored role in physics of mathematical concepts such as cluster algebras, the Grassmannian and matroids. Most probably this is just the tip of an iceberg, with useful insights flowing between the physics and mathematics worlds in both directions. The latest addition to this story is the amplituhedron, a new type of geometric object whose volume gives the scattering amplitudes of the quantum field theory \cite{ref_amplituhedron}.

The main goal of this article is to investigate the geometric structures associated to on-shell diagrams. In particular, our work constitutes a concrete step in reducing several of the assumptions often made in the interplay between bipartite graphs and the Grassmannian: dropping the conditions of reducibility and removable edges often invoked when discussing stratification, planarity of graphs and positivity. This paper also admits an alternative, more formal, reading. It can be regarded as an investigation of the description of the Grassmannian in terms of bipartite graphs, extending it beyond the well-studied planar case. 

The on-shell diagram approach to quantum field theories is part of an ambitious program which starts from planar $\mathcal{N}=4$ SYM and might eventually lead to a new understanding of gravity and even string theory. We expect that the new structures we develop in this article should naturally appear when moving forward to the next stage, into non-planar theories.

This paper is organized as follows. In \sref{introBFT}, we review bipartite graphs and related concepts, including bipartite field theories.  The Grassmannian and its different possible decompositions are discussed in \sref{section_Gr_decompositions}. In \sref{sec:planarboundmeas} we begin the discussion of the parametrization of the Grassmannian in terms of edge weights of bipartite graphs, by means of the boundary measurement. The relevance of the stratification of the Grassmannian for the singularity structure of on-shell diagrams is briefly reviewed in \sref{singularities}. In \sref{section_graphs_polytopes_geometry}, we present various complementary perspectives and methods for determining the matching and matroid polytopes, equivalently toric geometries, associated to general bipartite graphs. \sref{section_graph_equivalence_and_reduction} is dedicated to the notions of graph equivalence and reduction. A useful criterion for quantifying the degree of reducibility of a graph is presented in \sref{sec:QuantGraphRed}. In \sref{Strat_NewRegionNewMethod}, we introduce a new decomposition of the Grassmannian in terms of bipartite graphs. For planar graphs, this is a new way of obtaining its positroid decomposition. Including non-planar graphs allows us to cover new regions of the Grassmannian, providing what can be regarded as a partial matroid stratification. In \sref{section_boundary_measurement_non-planar} we extend the boundary measurement to graphs with an arbitrary number of boundaries. Our tools are applied to explicit non-planar examples in \sref{sec:NonplDecomp}. \sref{section_matroid_graphs} presents some thoughts on the possibility of constructing the matroid stratification by considering multiple graphs, planar and non-planar. We conclude in \sref{section_conclusions}. Three appendices collect auxiliary material.

\bigskip

\section{Overview of Bipartite Graphs and Related Objects}

\label{introBFT}

In this section we review basic aspects of bipartite graphs and their
combinatorial properties. We describe the notion of perfect matchings, perfect orientations, flows,
and define an edge parameterization that will be used in the rest of the paper.
We also introduce two relevant physics 
applications of such graphs:
on-shell diagrams for scattering amplitudes in $\mathcal{N}=4$ SYM,
and an infinite class of $\mathcal{N}=1$ gauge theories.

\bigskip

\paragraph{Bipartite Graphs.}

A graph is a collection
of nodes and of edges connecting them. 
The graphs we consider have two types of nodes, distinguished by a white or black color. 
If white nodes are only connected to black nodes and vice-versa, the graph is \textit{bipartite}. We denote the number of edges connected to a given node as its {\it valence}. The framework introduced in this paper deals with general bipartite graphs containing nodes of arbitrary valence.
  
In many applications, it can often be useful to consider embeddings of the graphs onto Riemann surfaces with boundaries. 
We shall call \textit{planar} a graph which can be embedded on the disk without crossing.
Instead, those graphs whose embedding involves edge crossings or multiple boundaries are referred to as \textit{non-planar}.

We divide nodes into two distinct categories: \emph{external}
nodes are defined as those nodes which must lie on a boundary in any embedding of the graph,
the remaining nodes are \emph{internal}.
We shall only consider monovalent external nodes.

Once an embedding of the graph on a Riemann surface is specified, one can define \textit{faces} 
as those regions on the surface surrounded by edges and/or by boundaries. Faces are also divided in two categories: 
\emph{internal faces} are those which are only surrounded by edges, and \emph{external faces} are those whose perimeter includes at least one boundary.

\bigskip

\begin{figure}[h]
\begin{center}
\includegraphics[scale=0.7]{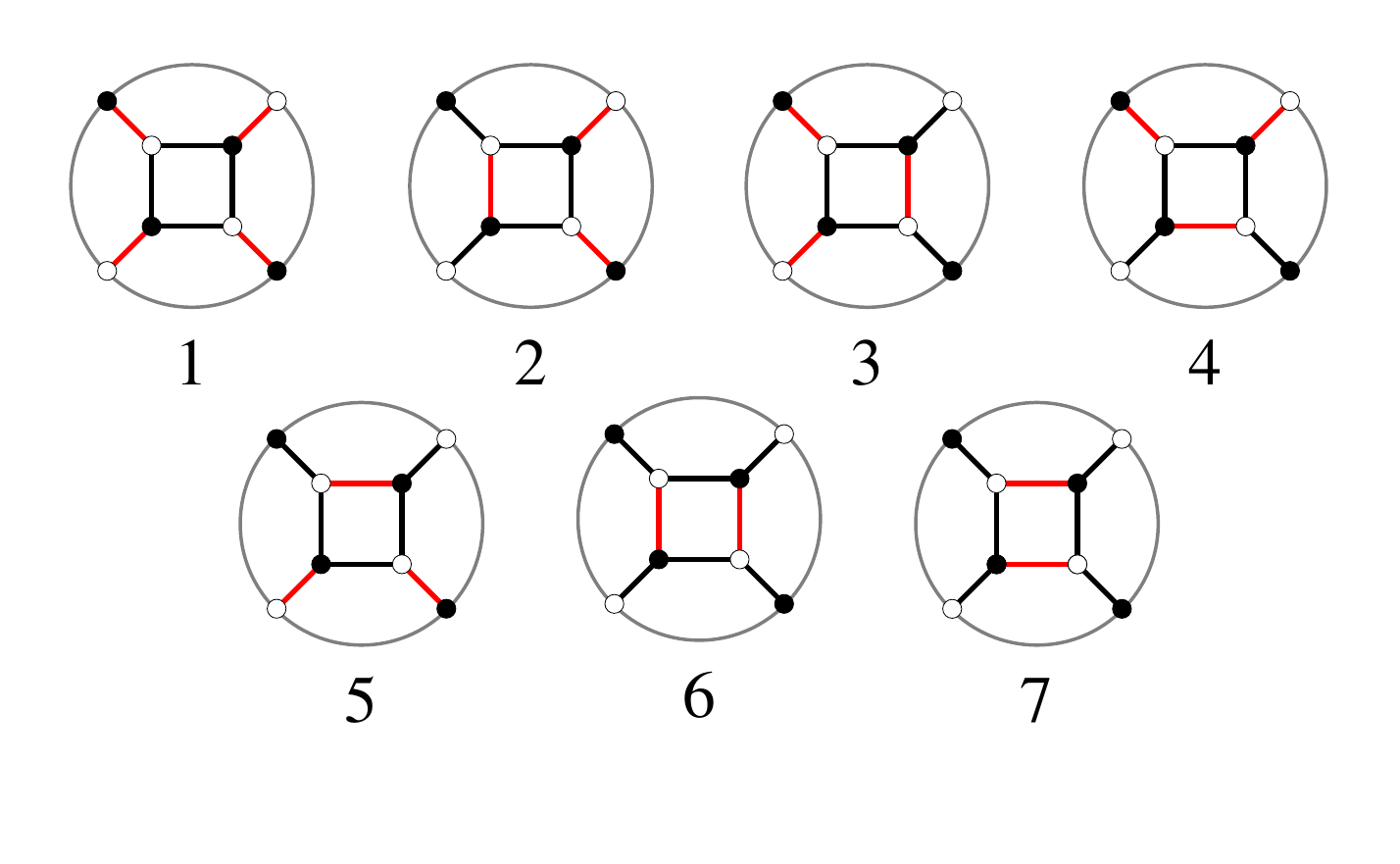}
\vspace{-1cm}\caption{All seven perfect matchings for a bipartite graph with four external nodes. Edges in the perfect matchings are shown in red. The graph is embedded in a disk, the boundary is shown in gray.}
\label{fig:sqbpms}
\end{center}
\end{figure}

\paragraph{Perfect Matchings.}
Perfect matchings are key combinatorial objects of bipartite graphs. A perfect matching is a sub-collection of edges such that every internal node is the endpoint of only one edge, 
while external nodes may or may not contained in the perfect matching.\footnote{Strictly speaking, this is known as an {\it almost perfect matching}. For brevity, we simply refer to these objects as perfect matchings in what follows.}
Usually, there are several ways to select sub-collections of edges with this property, and each of
these is a different perfect matching.
An example of a bipartite graph and its perfect matchings is provided in \fref{fig:sqbpms}.

In \sref{Kasteleyn} we will show how to find all perfect matchings for a given bipartite graph in a systematic way.

\bigskip

\paragraph{Perfect Orientations.}
A bipartite graph can equally be characterized by its perfect orientations. A perfect orientation is a way of assigning arrows to the edges of a graph in such a way that for 3-valent nodes we have:

\begin{itemize}
\item {\bf White node:} 1 incoming and $2$ outgoing arrows.
\item {\bf Black node:} 1 outgoing and $2$ incoming arrows.
\end{itemize}
In addition, 2-valent nodes have one incoming and one outgoing arrow.

General bipartite graphs with nodes of arbitrary valence $v$ can be constructed in terms of graphs containing only $v=2$ and $3$ nodes. We refer the reader to \cite{Franco:2012mm} for a detailed discussion of how this is achieved. The rules controlling perfect orientations for arbitrary $v$ can thus be derived from those for $v=2,3$. For general graphs, a perfect orientation is such that, for a node with valence $v\geq 3$, we have:

\begin{itemize}
\item {\bf White node:} 1 incoming and $v-1$ outgoing arrows.
\item {\bf Black node:} 1 outgoing and $v-1$ incoming arrows.
\end{itemize}

\begin{figure}[h]
\begin{center}
\includegraphics[width=8cm]{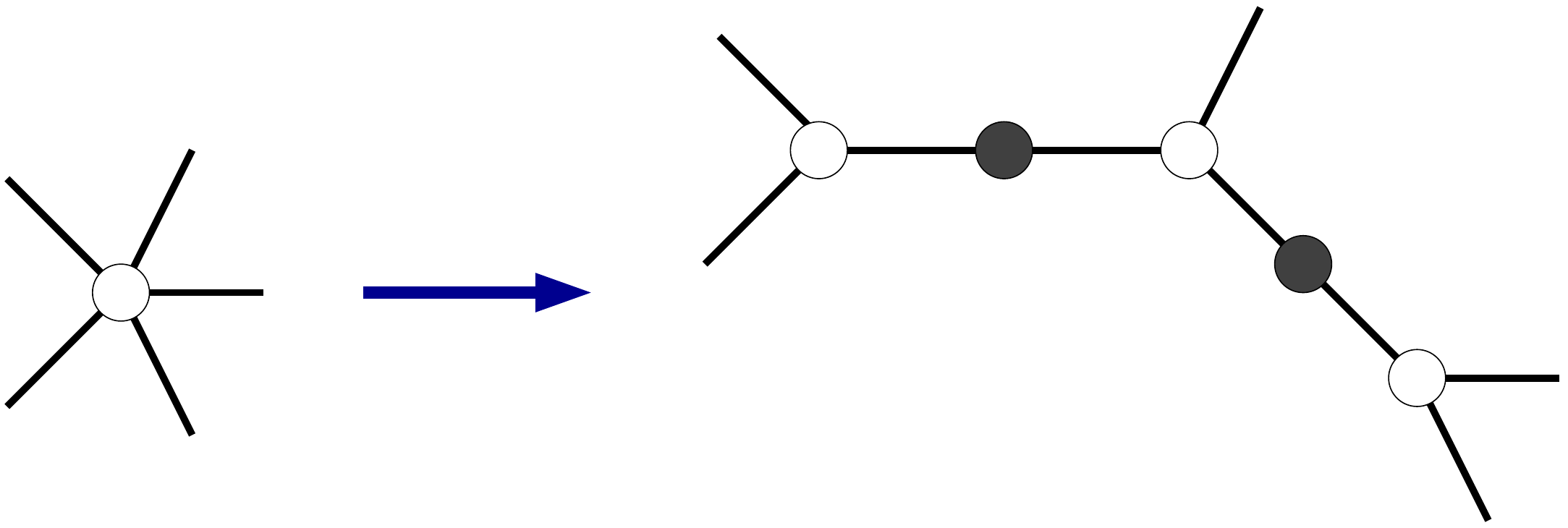}
\caption{Decomposition of a $v$-valent node into $(v-2)$ 3-valent and $(v-3)$ black 2-valent ones.}
\label{node_decomposition}
\end{center}
\end{figure}

\begin{figure}[h]
\begin{center}
\includegraphics[width=8cm]{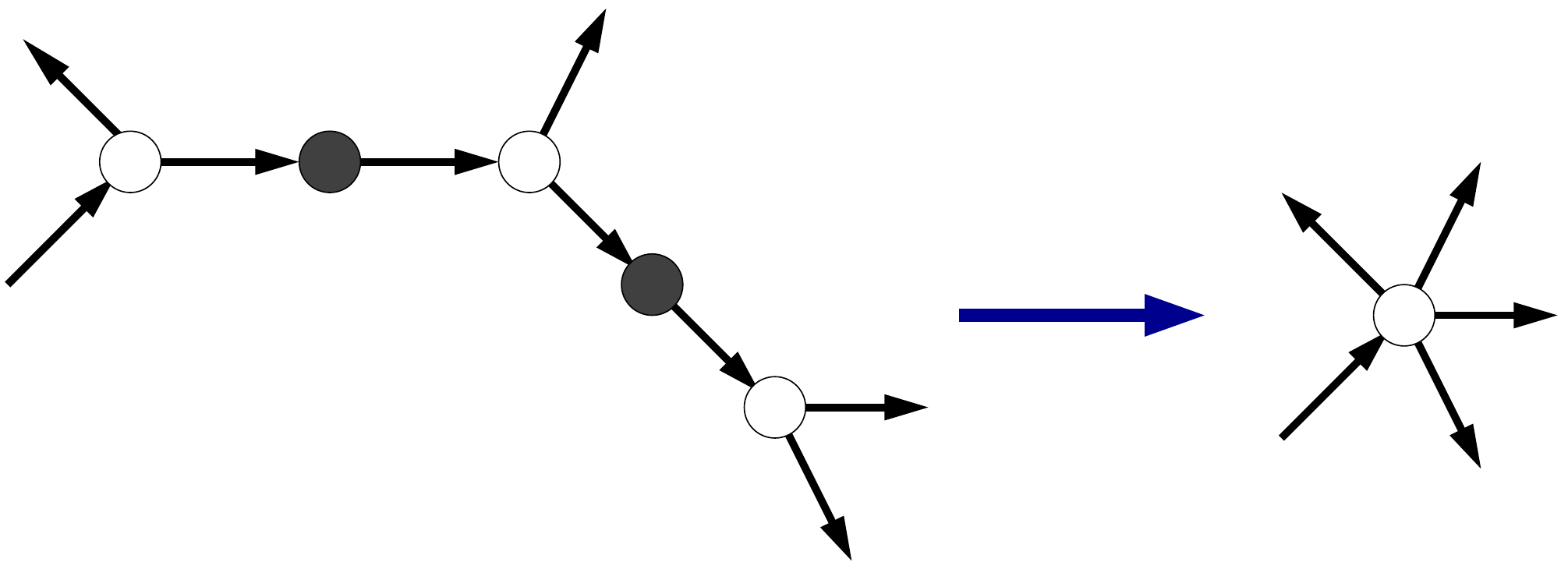}
\caption{Perfect orientation resulting from integrating out 2-valent nodes.}
\label{node_decomposition_perfect_orientation}
\end{center}
\end{figure}

\noindent It is straightforward to prove this based on the behavior of 2 and 3-valent nodes. For concreteness, consider a $v$-valent white node. As explained in \cite{Franco:2012mm}, it can be decomposed into $(v-2)$ white 3-valent nodes and $(v-3)$ black 2-valent nodes, as shown in \fref{node_decomposition}. 
%
%
The fact that this decomposition is in general not unique does not affect our conclusions. The white nodes give $2(v-2)$ outgoing and $(v-2)$ ingoing arrows. Out of these, $(v-3)$ in-out pairs are contracted along the black nodes, giving the result shown in \fref{node_decomposition_perfect_orientation}. The reasoning for black nodes is identical up to inversion of arrows. 

\begin{figure}[h]
\begin{center}
\includegraphics[scale=1.2]{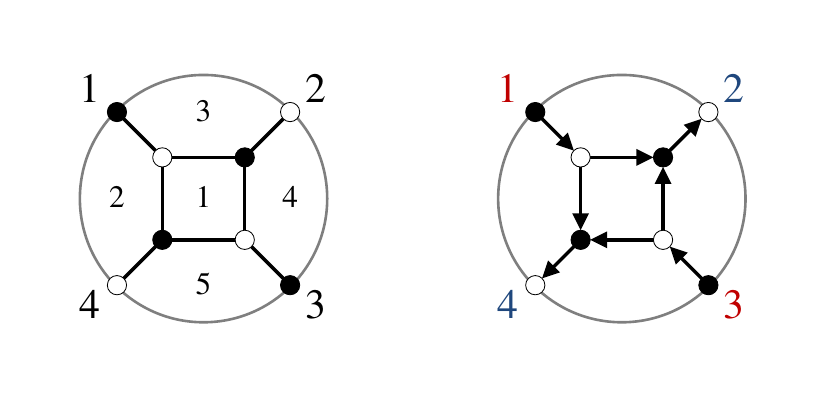}
\caption{A bipartite graph and a possible perfect orientation. Sources are marked in red and sinks in blue.}
\label{fig:sqbpo}
\end{center}
\end{figure}

\fref{fig:sqbpo} provides an example of a perfect orientation for a bipartite graph on a disk,
with 3-valent nodes.

Given a perfect orientation, external nodes can be naturally divided into {\it sources} and {\it sinks}. The number of elements in each of these two sets does not depend on the choice of the perfect orientation and is a characteristic of the graph itself.

\bigskip

\paragraph{Flows.}
Given a graph and a perfect orientation, it is possible to specify the latter by listing all oriented non self-intersecting paths in it. We refer to such paths as {\it flows} and denote them as $\mathfrak{p}_{\mu}$.  Flows may involve more than one disjoint component. These components can connect external nodes or correspond to closed loops. The trivial flow, i.e.\ the one which does not involve any edge of the graph, is also included.

\bigskip

\subsection{Relation Between Perfect Orientations, Flows and Perfect Matchings}
\label{RelationFlowPerfOrientMatch}

Perfect orientations are in bijection with perfect matchings. 
Given a perfect matching, the way to obtain the corresponding perfect orientation is to assign arrows to the edges as follows:

\begin{itemize}
\item Edges belonging to the perfect matching point from the black node to the white node.
\item All other edges point out of white nodes and into black nodes.
\end{itemize}
Since a perfect matching only touches each internal node once, the above definition automatically satisfies the rules for arrows in a perfect orientation. Conversely, it is possible to obtain the perfect matching from a perfect orientation by selecting the incoming arrow for white nodes and the outgoing arrow for black nodes. 

There is also a bijection between flows and perfect matchings. 
In order to find it, we begin by choosing a perfect matching $p_{\text{ref}}$, called the \textit{reference perfect matching} (or just reference matching for short), and assigning to all of its edges an orientation that points from white nodes to black nodes. We orient the edges of all other perfect matchings in a similar way. Subtracting $p_{\text{ref}}$ from all perfect matchings, i.e.\ reversing the arrows in $p_{\text{ref}}$ before combining them, creates a set of oriented paths. These paths necessarily live in the perfect orientation associated to $p_{\text{ref}}$, because all arrows point out of white nodes and into black nodes except for the ones belonging to $p_{\text{ref}}$, which have opposite orientation. These paths are thus precisely the flows in the perfect orientation defined by $p_{\text{ref}}$, i.e.\ we can think about them as $\mathfrak{p}_{\mu}=p_\mu - p_{\text{ref}}$.

In summary, for each perfect matching there is an associated perfect orientation. The number of flows in each perfect orientation is equal to the number of perfect matchings, and they are found by subtracting the reference perfect matching from the corresponding perfect matchings.

\bigskip

\subsection{Oriented Edge Weights}
\label{orientedge}

We will often be interested in relating edge weights, which strictly speaking have no associated orientation, to oriented paths. It is thus useful to devise a formalism that consistently deals with such a connection. We will refer to edge weights as $X_{i}$, where the index $i=1,\ldots, E$ runs over all edges of the graph.

\begin{figure}[h]
\begin{center}
\includegraphics[scale=0.5]{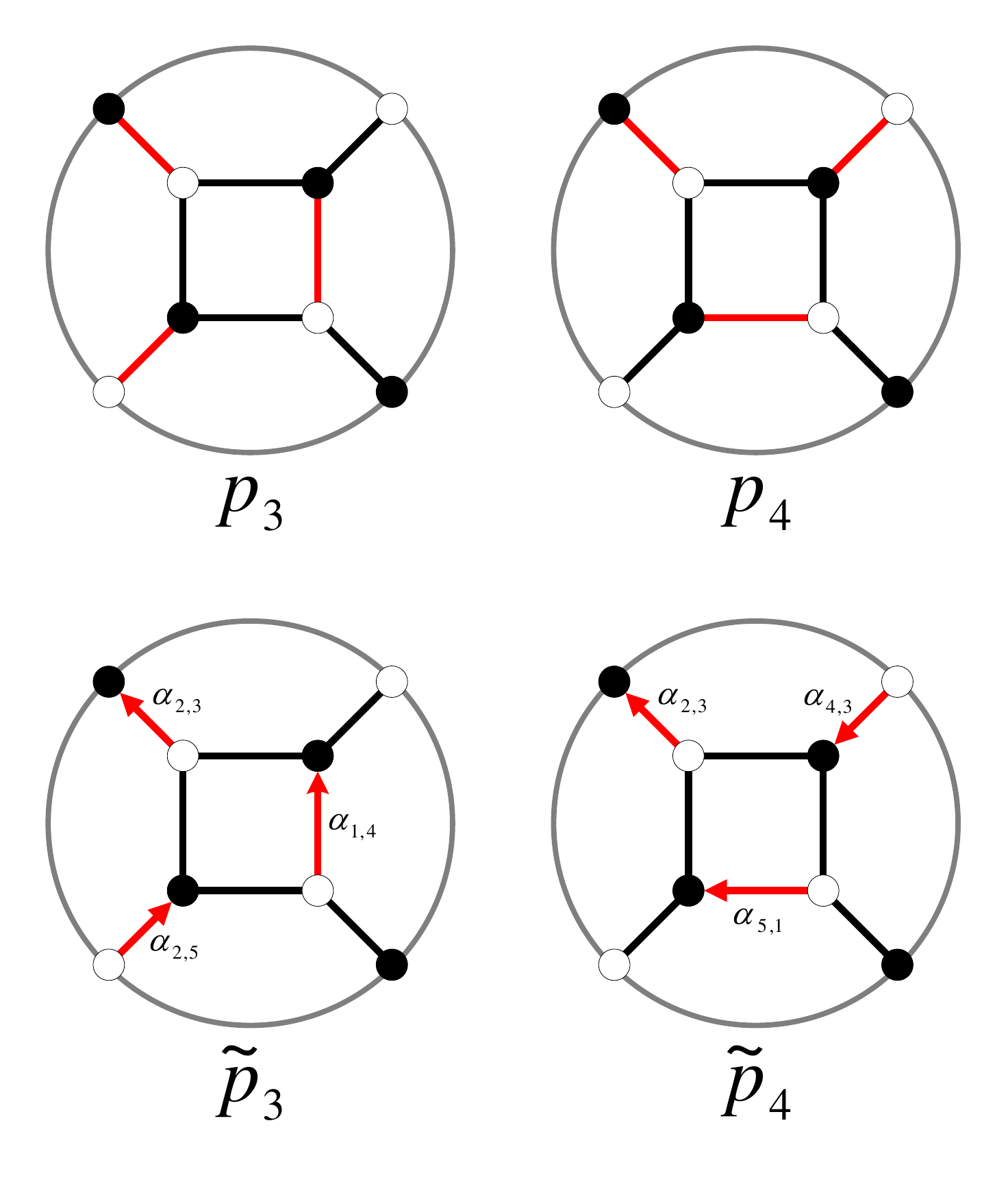}
\caption{Example of ordinary perfect matchings $p_i$ and oriented perfect matchings $\tilde{p}_i$. Edges $\alpha_{i,j}$ are oriented from white nodes to the black nodes.}
\label{AlphaPmsbis}
\end{center}
\end{figure}

With the goal of describing oriented paths, we introduce new variables $\alpha_i$, which are edge weights endowed with an orientation. In our convention the orientation goes from white to black nodes. We can thus associate an {\it oriented perfect matching} $\tilde{p}_\mu$ to every perfect matching $p_\mu$. The oriented perfect matching is given by the product of the $\alpha_{i}$ variables over all edges in the corresponding perfect matching. For example, for \fref{fig:sqbpms} the oriented perfect matchings are\footnote{
Here we have switched to a convenient bifundamenal notation for the $\alpha$'s, i.e.\ $\alpha_{i,j}$ corresponds to an edge separating faces $i$ and $j$. We hope the reader is not confused by this choice.}
\beq
\begin{array}{cclcccl}
\tilde{p}_1 & = & \alpha _{2,3} \alpha _{2,5} \alpha _{4,3} \alpha _{4,5} & \ \ \ \ \ \ \ & \tilde{p}_5 & = & \alpha _{2,5} \alpha _{3,1} \alpha _{4,5} \\
\tilde{p}_2 & = & \alpha _{1,2} \alpha _{4,3} \alpha _{4,5} & & \tilde{p}_6 & = & \alpha _{1,2} \alpha _{1,4}  \\
\tilde{p}_3 & = & \alpha _{1,4} \alpha _{2,3} \alpha _{2,5} & & \tilde{p}_7 & = & \alpha _{3,1} \alpha _{5,1} \\
\tilde{p}_4 & = & \alpha _{2,3} \alpha _{4,3} \alpha _{5,1} & & &
\end{array} .
\eeq
\fref{AlphaPmsbis} shows two perfect matchings $p_3, p_4$ and their corresponding oriented perfect matchings $\tilde{p}_3, \tilde{p}_4$.

\begin{figure}[h]
\begin{center}
\includegraphics[scale=0.6]{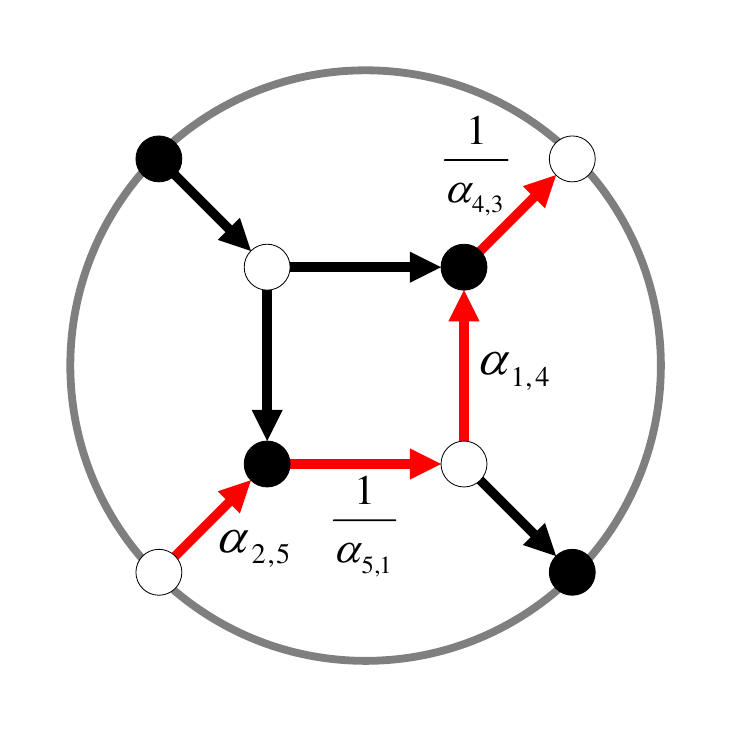}
\caption{A flow in the perfect orientation corresponding to the reference perfect matching $p_4$. The flow shown is $\mathfrak{p}_3 =\frac{\tilde{p}_3}{\tilde{p}_4} = \frac{\alpha_{2,5}\alpha_{1,4}}{\alpha_{5,1}\alpha_{4,3}}$.}
\label{fig:alphapath}
\end{center}
\end{figure}

We can in fact write any oriented path on the graph as a product or ratio of these new variables: if a segment of the path goes from a white node to a black node, the relevant $\alpha_{i,j}$ contributes to the expression of the path in the numerator; if the segment goes from a black node to a white node, its $\alpha_{i,j}$ contributes to the denominator. In particular, flows can be written in terms of these variables; an example is provided in \fref{fig:alphapath}, where the perfect orientation corresponds to the perfect matching $p_4$. Here the flow is expressed as
\begin{equation}
\frac{\alpha_{2,5}\alpha_{1,4}}{\alpha_{5,1}\alpha_{4,3}} \quad .
\end{equation}

Moreover, in this parameterization all flows can be expressed as ratios $\mathfrak{p}_i = \tilde{p}_i / \tilde{p}_{\text{ref}}$, where $\tilde{p}_{\text{ref}}$ is the reference matching defining the underlying perfect orientation. In the example in the figure, the flow is $\mathfrak{p}_3 = \frac{\tilde{p}_3}{\tilde{p}_4} = \frac{\alpha_{2,5}\alpha_{1,4}}{\alpha_{5,1}\alpha_{4,3}}$.
Note that the trivial flow is $\mathfrak{p}_{\text{ref}}=1$.

This parameterization is very convenient for the study of the connection between bipartite graphs and the Grassmannian, and will be extensively used in the rest of the paper.

\bigskip

\subsection{Finding Perfect Matchings}
\label{Kasteleyn}

Flows, perfect orientations and perfect matchings contain equivalent combinatorial information about the bipartite graph. Among the three, perfect matchings are those which are obtained most efficiently. This is done using a generalization of Kasteleyn matrix techniques, which will be briefly outlined here. The reader is referred to \cite{Franco:2012mm} for a detailed discussion of these techniques. 

The starting point for finding the perfect matchings is the construction of a weighted adjacency matrix, known as the {\it master Kasteleyn matrix} $K_0$. When there are multiple edges between two nodes their contributions are added. Denoting internal white and black nodes $W_i$ and $B_i$, respectively, and external white and black nodes $W_e$ and $B_e$, $K_0$ takes the form:

\begin{equation}
K_0 = \left(\begin{array}{c|c|c} & \ \ B_i \ \ & \ \ B_e \ \ \\ \hline
W_i & * & * \\ \hline
W_e & * & 0 \\
\end{array}
\right) \; .
\end{equation}

\noindent The zero in the bottom-right corner arises because external nodes are only paired with internal nodes. $K_0$ is not necessarily square.

For any subsets $W_{e,del} \subseteq W_e$ and $B_{e,del} \subseteq B_e$ of the external nodes, we define the {\it reduced Kasteleyn matrix} $K_{\left( W_{e,del},B_{e,del} \right)}$ as the matrix resulting from starting from $K_0$ and deleting the rows in $W_{e,del}$ and the columns in $B_{e,del}$.

All perfect matchings in the graph are given by the polynomial:
\beq
\mathcal{P}=\sum_{W_{e,del},B_{e,del}} \text{perm } K_{\left( W_{e,del},B_{e,del} \right)},
\label{PM_Kastel}
\eeq
where the sum runs over all possible subsets $W_{e,del}$ and $B_{e,del}$ of the external nodes such that the resulting reduced Kasteleyn matrices are square.\footnote{The permanent of a matrix is the determinant where all signs in the final expression are positive.} Every term in this polynomial corresponds to the product of edges in a perfect matching.

\bigskip

\subsection{On-Shell Diagrams}
\label{onshellDiag}

Recently, a remarkable new formalism based on {\it on-shell diagrams} has been developed for $\mathcal{N}=4$ SYM \cite{ArkaniHamed:2012nw}. This approach naturally relates scattering amplitudes to the Grassmannian. The connection between gauge theory and the Grassmannian has been exhaustively investigated in earlier works, such as \cite{ArkaniHamed:2009dn,Mason:2009qx,ArkaniHamed:2009vw,ArkaniHamed:2010kv,Bourjaily:2010kw}.

On-shell diagrams are constructed by gluing 3-particle MHV (maximally helicity violating) and $\overline{\rm{MHV}}$ amplitudes. They are characterized by $k$, the number of external particles with negative helicity, and $n$, the total number of external particles. In these diagrams, all lines represent particles whose momentum is on-shell. Integrating over the on-shell phase space of internal particles, with helicity and momentum-conserving delta functions at each vertex, they produce a function of the external kinematical data. 

Being constructed in terms of two types of building blocks, on-shell diagram are naturally bi-colored graphs. Indeed, as explained in \cite{Franco:2012mm}, it is straightforward to relate general on-shell diagrams to bipartite graphs. For this reason, we will simply regard the two classes of objects as synonyms in what follows. Given an on-shell diagram, the possible assignations of helicity flows consistent with the rules for MHV and $\overline{\rm{MHV}}$ vertices correspond to perfect orientations.

Bipartite graphs are mapped to elements of the Grassmannian via a map known as the {\it boundary measurement}, which we will study in \sref{sec:planarboundmeas} and \sref{section_boundary_measurement_non-planar}. Hence we have a connection among:
\begin{equation}
\text{Bipartite Graphs/On-Shell Diagrams}~~ \Leftrightarrow~~ \text{Elements in the Grassmannian} \nonumber
\end{equation}
Much of this article is devoted to investigating these relations.

\bigskip

\subsection{Bipartite Field Theories}

Bipartite Field Theories (BFTs) are a class of 4d, $\mathcal{N}=1$ gauge theories whose Lagrangians are defined by bipartite graphs on (bordered) Riemann surfaces \cite{Franco:2012mm,Franco:2012wv}.\footnote{As we explain below, a certain sub-class of BFTs is independent of the underlying Riemann surface which, nevertheless, is a helpful intermediate object for defining the theory.} BFTs provide an alternative, and sometimes very powerful, perspective on bipartite graphs. The BFT associated to a graph is obtained using the following dictionary:

\medskip

\begin{itemize}
\item \textbf{\underline{Face}:} $U(N)$ symmetry group.\footnote{The case of general ranks, i.e.\ not equal for all faces, is extremely interesting. It is however not relevant for the questions discussed in this article, so we do not pursue it.}
\item \textbf{\underline{Edge}:} chiral multiplet $X_{i,j}$ in the bifundamental representation of the $U(N)_i \times U(N)_j$ symmetry groups corresponding to the faces on both sides of the edge. Introducing an orientation around nodes going clockwise around white nodes and counterclockwise around black ones, the fields transform in the fundamental representation of the head of the corresponding arrow and anti-fundamental representation of the tail. Chiral fields associated to external legs, i.e.\ edges connected to external nodes, are taken to be non-dynamical.
\item \textbf{\underline{Node}:} superpotential term given by the trace of the product of fields corresponding to edges terminating on the node. The superpotential term bears a positive sign for white nodes and negative sign for black nodes. External nodes, by definition, do not map to any superpotential term.
\end{itemize}

\medskip

In order to fully specify the BFT, it is also necessary to determine which symmetries are gauged. There are two natural choices \cite{Franco:2012wv}, which we now explain.\footnote{A related, and partially overlapping, class of theories was defined in \cite{Xie:2012mr}. Its string theory embedding was discussed in \cite{Heckman:2012jh}. Additional interesting works on BFTs and related topics can be found in \cite{Franco:2013pg,Baur:2013hwa}.} 

\bigskip

\paragraph{Gauging 1.}
In this case, the $U(N)$ symmetries associated to internal faces of the graph, namely faces whose perimeter does not involve any boundary,  are gauged. It is straightforward to see that bipartiteness guarantees that internal faces are even sided. This implies that they are anomaly free and can be consistently gauged. The remaining symmetry groups are global. We refer to the resulting class of gauge theories as $\text{BFT}_1$. The theories in this class are quiver gauge theories. Their quivers, including plaquettes representing the superpotential terms, are obtained by dualizing the bipartite graph \cite{Franco:2012mm}.

A particular sub-class of $\text{BFT}_1$'s has been the subject of intense activity in recent years. These theories are known as {\it  brane tilings} and correspond to BFTs on a 2-torus without boundaries \cite{Hanany:2005ve,Franco:2005rj,Franco:2005sm}. Brane tilings describe the theories on the worldvolume of D3-branes probing toric Calabi-Yau 3-folds and have played a key role in the identification of infinite families of explicit AdS/CFT dual pairs \cite{Franco:2005sm, Butti:2005sw}. More recently a physical realization in terms of D3 and D7-branes on toric Calabi-Yau 3-folds has been introduced for a more general class of BFTs, which includes graphs with boundaries \cite{Franco:2013ana}.

\bigskip

\begin{figure}[htb!]
\centering
\includegraphics[scale=0.6]{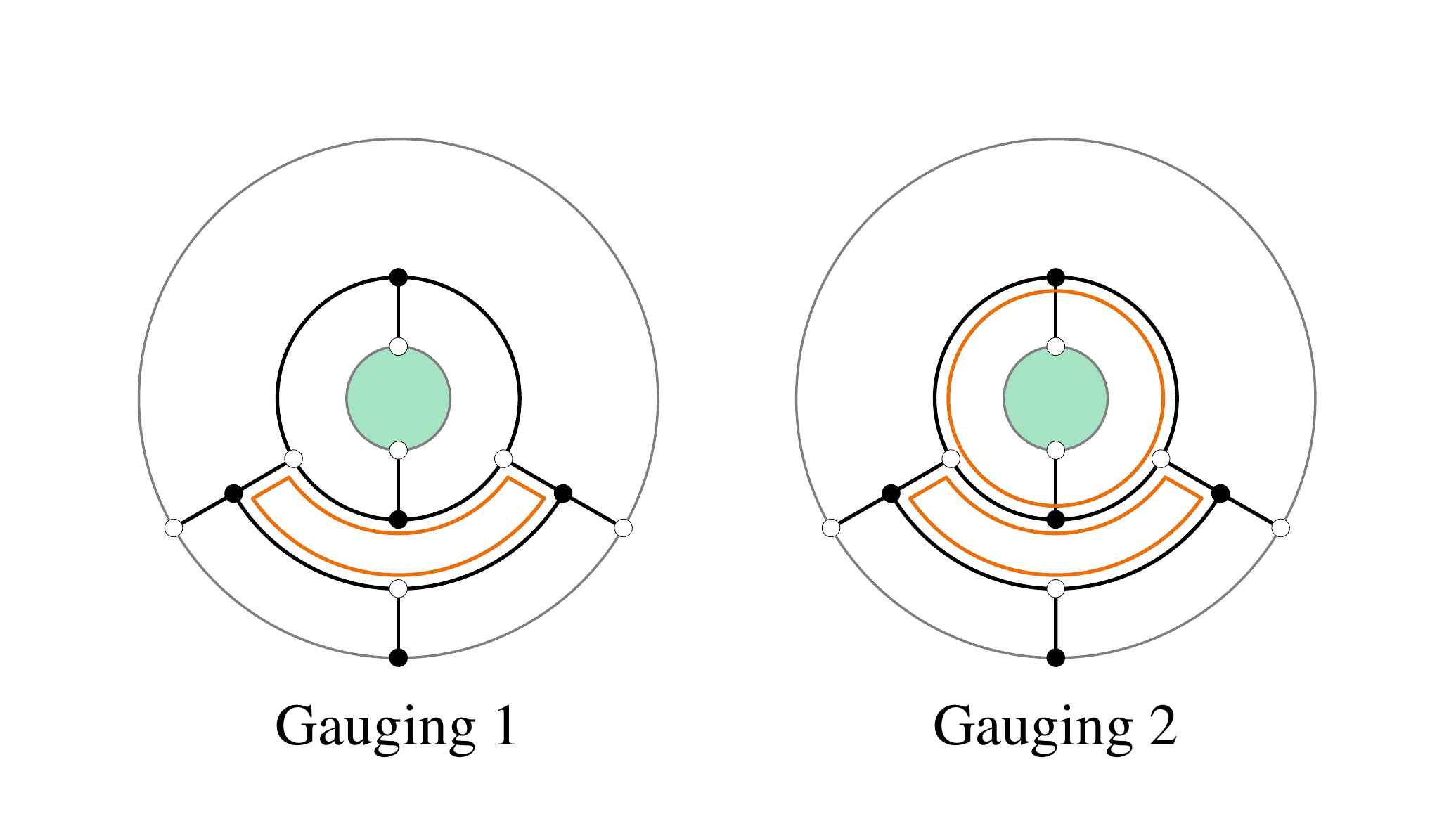}
\caption{Difference between $\text{BFT}_1$ and $\text{BFT}_2$ in an example with two boundaries. The orange loops are those which are gauged in each gauging. The surface has genus $g=0$ and hence there are no loops with non-trivial homology.}
\label{fig:gauge1and2}
\end{figure}

\paragraph{Gauging 2.}
Internal faces are not the only source of symmetries which are automatically anomaly-free. In fact any closed loop in the graph has this property. This leads us to a second class of BFTs, which we denote $\text{BFT}_2$, in which the symmetries associated to a basis of all closed loops are gauged. Gauging 2 is then an extension of gauging 1, where additional symmetries of the theory are gauged. Loops which cannot be expressed as faces or collection thereof, i.e.\ loops with a non-trivial homology around the $g>0$ Riemann surface, are identified with $U(1)$ gauged symmetries.\footnote{Whether and under what circumstances it is possible to promote some of these symmetries to non-Abelian is an interesting question that we will not pursue in this paper.} The difference between $\text{BFT}_1$ and $\text{BFT}_2$ is illustrated in \fref{fig:gauge1and2}.

For graphs on a disk there is no distinction between $\text{BFT}_1$ and $\text{BFT}_2$. The difference between the two gaugings arises in the presence of multiple boundaries and/or higher genus Riemann surfaces. For applications to on-shell diagrams, the relevant theories are classical Abelian BFTs, in which all symmetries are $U(1)$.\footnote{Since we focus on classical theories, we do not worry about issues concerning the UV completion of Abelian BFTs.} In this context it is also natural to focus on $\text{BFT}_2$'s, since the additional gauging makes the resulting theory independent of the underlying Riemann surface \cite{Franco:2012wv}.

\bigskip

\section{The Grassmannian and its Decompositions}

\label{section_Gr_decompositions}

In this section we review basic aspects about the Grassmannian and its
stratifications. We refer the interested reader to \cite{2006math09764P,Postnikovlectures,2012arXiv1210.5433T,oxley2006matroid,fink2010MatroidPolytopeSubdivisions} for more comprehensive discussions.

\subsection{Definition}
%
The Grassmannian \Gr($\mathbb{R}$) is the space of $k$-dimensional planes in $n$ dimensions that pass through the origin. 
Elements of \Gr($\mathbb{R}$) are typically represented by $k \times n$ matrices where the plane is the span of the $k$ $n$-dimensional row vectors. 
The action of $GL(k)$ on the basis vectors leaves the plane invariant, so the Grassmannian is the space of $k \times n$ matrices $C$ modulo $GL(k)$. The $GL(k)$ invariance can be used to fix any $k$ columns to form a $k\times k$ identity sub-matrix, e.g.\ for ${Gr_{2,4}}$ we can fix $C$ to the form
\begin{equation}
C=\begin{pmatrix}
1 & 0 & -c_3 & -c_4 \\
0 & 1 & c_1 & c_2 
\end{pmatrix} \quad ,
\end{equation}
where the signs have been introduced for later convenience. When mapping bipartite graphs to the Grassmannian, we will see that columns in this matrix correspond to all external nodes and rows correspond to those which are sources in a perfect orientation. From here on, we will always present elements of the Grassmannian in a form that has fixed the $GL(k)$ invariance.

\bigskip

\subsection{\pl Coordinates} 

The degrees of freedom of $C$ can alternatively be expressed by its $k\times k$ minors $\Delta_{I}$, where $I$ is a set with $k$ elements describing which columns participate in the minor; these are known as \pl coordinates. These minors are invariant under the action of $SL(k)$ and scale by a common factor under $GL(k)$. Since there are $\binom{n}{k}$ of these, it induces the \pl embedding of the Grassmannian \Gr$\hookrightarrow \mathbb{RP}^{\binom{n}{k}-1}$. The minors are not all independent, they satisfy relations known as the \pl relations
\begin{equation}
\sum_{i=1}^{k+1} (-1)^{i-1} \Delta_{J_1 \cup \,a_i} \, \Delta_{J_2\, \backslash \,a_i} = 0 ,
\end{equation}
where $J_1$ is any $(k-1)$-element subset of $[n]$, $J_2$ is any $(k+1)$-element subset of $[n]$ and $a_i$ is the $i$th element of $J_2$. In each term, $a_i$ is removed from $J_2$ and appended to the right of $J_1$. In this embedding, the Grassmannian is simply the subvariety described by the \pl relations.
For the example of ${Gr_{2,4}}$ above, we have 
\beq
\begin{array}{cclccclcccl}
\Delta_{12}&=&1 & \ \ \ \ & \Delta_{14} & = & c_2 & \ \ \ \ &  \Delta_{24}&=&c_4 \\
\Delta_{13}&=&c_1 & & \Delta_{23}&=&c_3 & & \Delta_{34}&=&c_1 c_4 - c_2 c_3
\end{array}
\eeq
and the single relation $ \Delta_{14} \Delta_{23} - \Delta_{13} \Delta_{24} + \Delta_{12} \Delta_{34} = 0$. The totally non-negative Grassmannian is given by those matrices $C$ with all $\Delta_I \geq 0$.

\bigskip

\subsection{Schubert Decomposition} 
\label{Schubert_dec}

There are many ways to decompose the Grassmannian into (possibly overlapping) sets, according to certain properties. Schubert cells\footnote{A cell is homeomorphic to an open ball and must have Euler number 1.} $\Omega_I$ are defined as those $C \in Gr_{k,n}$ where $\Delta_I$ is the first non-zero \pl coordinate, 
counted in lexicographic order\footnote{Lexicographic order is $1<2<3<4$, e.g.\ $1243<1324$, 
analogous to alphabetical order.}, 
i.e.\
\begin{equation}
\Omega_I=\{ C \in Gr_{k,n} \mid \Delta_I \text{ is the lexicographically minimal non-zero \pl coordinate} \}  .
\label{Schubertcell}
\end{equation}
For example,
\begin{equation}
\label{ex1}
C=\begin{pmatrix}
1 & 0 & 0 & -c_4 \\
0 & 1 & c_1 & c_2 
\end{pmatrix} \; \in \Omega_{12}  ,
\end{equation} 
because there is no other non-zero \pl coordinate with smaller lexicographic ordering than $I={12}$. The cyclically shifted Schubert cell $\Omega_I^{(i)}$ is defined similarly, but the lexicographic order is cyclically shifted to begin the counting at $i$, e.g.\ for the same example in \eref{ex1}, 
$C \in \Omega_{12}$ but also $C \in \Omega_{24}^{(2)}$ because the order is shifted to $2<3<4<1$, and since $\Delta_{23}=0$, the lexicographically smallest (with respect to the shifted order) non-zero $\Delta_I$ is now $I={24}$. Similarly, $C \in \Omega_{34}^{(3)}$ and $C \in \Omega_{41}^{(4)}$. 

Note that in each shifted Schubert cell $\Omega_I^{(i)}$ the \pl coordinates lexicographically larger 
(with respect to the shifted order $i$) than $I$ are free to be zero or non-zero.

The permuted Schubert cell $\Omega_I^w$ is defined as in \eref{Schubertcell} but with the lexicographic order being with respect to a permuted order $w(1)< w(1)< \cdots < w(n)$, where $w \in S_n$.

\bigskip

\subsection{Positroid Stratification} 
\label{sec:PositroidStrat}

The positroid stratification of the Grassmannian \Gr\  introduced by Postnikov \cite{2006math09764P} defines each stratum as
\begin{equation}
S_{\mathcal{I}} = \bigcap_{i=1}^n \Omega_{I_i}^{(i)} ,
\end{equation}
where $\mathcal{I}=\{I_1,\ldots,I_n\}$, and $I_i$ specifies which \pl coordinates are non-zero, only looking at those which are lexicographically minimal with respect to each shifted cyclic ordering starting at $i$. 
Note in particular that the \pl coordinates lexicographically smaller
with respect to each shifted order must be zero,
following the definition of the Schubert
decomposition.
For the example in \eref{ex1}, the non-zero \pl coordinates are $\Delta_{12}$, $\Delta_{13}$, $\Delta_{14}$, $\Delta_{24}$ and $\Delta_{34}$. With respect to the first order $i=1$, the lexicographically minimal one is $\Delta_{12}$; for $i=2$ the minimal one is $\Delta_{24}$; for $i=3$, $\Delta_{34}$; and finally for $i=4$, $\Delta_{41}=-\Delta_{14}$. Hence, this element of the Grassmannian is in the positroid stratum 
\begin{equation}
\label{stratex1}
S_{\mathcal{I}}=\{ C \in Gr_{2,4} \mid \Delta_{12} \neq 0 , \Delta_{24} \neq 0 , \Delta_{34} \neq 0 , \Delta_{14} \neq 0\} \quad .
\end{equation}
where $\Delta_{23}=0$ and we do not specify whether 
 $\Delta_{13}$ is vanishing or not.
Instead, consider the following stratum
\be
S_{\mathcal{I}} = \{ C \in Gr_{2,4} \mid \Delta_{14} \neq 0 , \Delta_{24} \neq 0 \} \quad .
\ee
This stratum contains those matrices for which lexicographically smaller \pl coordinates with respect to each shifted order are set to zero.
For the shifted order $i=1$, we note that $\Delta_{12}=0$ and $\Delta_{13}=0$ since they are lexicographically smaller than $\Delta_{14}$. For the shifted order $i=2$, $\Delta_{23}=0$ since it is 
lexicographically smaller than $\Delta_{24}$. For the shifted order $i=3$, we additionally have $\Delta_{34}=0$ since it is 
lexicographically smaller than $\Delta_{14}$ (along with $\Delta_{31}$ and $\Delta_{32}$). Finally  $\Delta_{41}\neq0$ is the
lexicographically smallest
with respect to the shifted order $i=4$.
So a matrix belonging to this positroid stratum is for instance
\begin{equation}
\label{ex2}
\begin{pmatrix}
c_1 & 1 & 0 & 0 \\
0 & 0 & 0 & 1 
\end{pmatrix} \; \in \; \; S_{\mathcal{I}} = \{ C \in Gr_{2,4} \mid \Delta_{14} \neq 0 , \Delta_{24} \neq 0 \} \quad .
\end{equation} 
Since a positroid stratum is in general more restricted than a Schubert cell, the positroid stratification refines the Schubert decomposition.

\bigskip

\subsection{Matroid Stratification}\label{matroidstra}

In order to describe this stratification, we have first to introduce the concept of 
\emph{matroids}.
The study of matroids is the analysis of an abstract theory of dependences.
We refer the interested reader to \cite{oxley2006matroid} for a comprehensive introduction,
here we review only some basic aspects.

\bigskip

\noindent \textbf{Definition of a Matroid.} A matroid of rank $k$ on a set $[n]$ is a non-empty collection $\mathcal{M} \subset \binom{[n]}{k} $
of $k$-element subsets in $[n]$, called \emph{bases} of $\mathcal{M}$, that satisfy the {\it exchange axiom}:

\medskip

{\it For any $I,J \in \mathcal{M}$ and $i \in I$, there exists a $j \in J$ such that $(I \setminus \{i \}) \cup \{ j\} \in \mathcal{M}$.}

\bigskip

\noindent \textbf{Matroid Polytope.} We can construct a polytope which efficiently encodes the linear dependencies among the bases of a matroid. Given a matroid $\mathcal{M}$ of rank $k$ on a set $[n]$, the matroid polytope
$\mathcal{P}(\mathcal{M})$ is the convex hull of the indicator vectors of the bases of $\mathcal{M}$
$$
\mathcal{P}(\mathcal{M})= \text{convex} \{ e_I ~:~ I \in \mathcal{M} \}
$$
where by $e_I$ we denote $e_I=\sum_{i \in I} e_i$ for any $I \in \mathcal{M}$, and
$\{e_1,\dots, e_n \}$ is the standard Euclidean basis of $\mathbb{R}^{n}$. Linear relations among matroid bases translate into linear relations between position vectors of points in the matroid. The construction of matroid polytopes is discussed in detail in \sref{section_graphs_polytopes_geometry}.

\bigskip

\noindent \textbf{Matroid Stratification.} Now we can discuss the matroid stratification of the Grassmannian $Gr_{k,n}$, which
further refines the positroid stratification. \\
Let $\mathcal{M} \subset \binom{[n]}{k} $ be a matroid. A matroid stratum is defined as follows
\begin{equation}
S_{\mathcal{M}}=\{ C \in Gr_{k,n} \mid \Delta_{I} \neq 0 \text{ if and only if } I \in \mathcal{M} \} .
\end{equation}
Note that each stratum is defined by which \pl coordinates are non-zero and which ones are zero; here all \pl coordinates are specified.
This stratification can also be expressed as the common refinement of the $n!$ permuted Schubert cells $\Omega_I^w$.

For the example of $Gr_{2,4}$, 
$\{ 12, 24,  34, 14 \}$ is a matroid, which corresponds to elements $C \in Gr_{2,4} $ with $\{ \Delta_{12} \neq 0 , \Delta_{24} \neq 0 , \Delta_{34} \neq 0 , \Delta_{14} \neq 0 , \Delta_{13} = 0 , \Delta_{23} = 0\}$.
In the positroid stratum specified in \eref{stratex1} there is exactly one more matroid stratum: that which contains $C \in Gr_{2,4} $ with $\{ \Delta_{12} \neq 0 , \Delta_{24} \neq 0 , \Delta_{34} \neq 0 , \Delta_{14} \neq 0 , \Delta_{13} \neq 0 , \Delta_{23} = 0\}$. 
The matrix \eref{ex1} belongs to this matroid stratum.

\bigskip

\subsection{Positroid Cells} 

\label{section_positroid_cells}

Postnikov showed that intersecting the matroid stratification with the totally non-negative Grassmannian $Gr_{k,n}^{\geq 0}$ gives a cell decomposition of $Gr_{k,n}^{\geq 0}$ \cite{2006math09764P}. Only one matroid stratum in each positroid stratum has a non-empty intersection with $Gr_{k,n}^{\geq 0}$, and it is this intersection which is the positroid cell.\footnote{These are called cells since they are homeomorphic
to an open ball of appropriate dimensions.} Equivalently, the positroid cell decomposition of $Gr_{k,n}^{\geq 0}$ can be obtained as the intersection of the positroid stratification with the totally non-negative Grassmannian $Gr_{k,n}^{\geq 0}$. This cell is the only one for which non-negative \pl coordinates are compatible with the \pl relations.

The positroid cell whose \pl coordinates are all different from zero (and positive) is the top-dimensional cell, which we refer to as the top-cell. 
Postnikov showed that the positroid cells are indexed by \Le \ diagrams and planar bipartite graphs \cite{2006math09764P}.

\bigskip 

\subsubsection{Deodhar Decomposition} 

The Deodhar decomposition is a refinement of the positroid stratification, but in turn it is refined by the matroid stratification, i.e.\ in general there are several Deodhar components in each positroid stratum, but each Deodhar component contains several matroid strata. For example, the refinement of the positroid stratum $\{ \Delta_{12} \neq 0 , \Delta_{23} \neq 0 , \Delta_{34} \neq 0 , \Delta_{14} \neq 0 \}$ is:

\begin{center}
\bigskip 
\begin{tabular}{|c|c|c|c|c|}
\hline
Positroid stratum & \multicolumn{4}{|c|}{$\Delta_{12}\neq 0 , \Delta_{23} \neq 0 , \Delta_{34} \neq 0 , \Delta_{14} \neq 0 $} \\
\hline
\ \ Deodhar components \ \ & \multicolumn{2}{|c|}{$\Delta_{13} \neq 0$} & \multicolumn{2}{c|}{$\Delta_{13} = 0$} \\
\hline
Matroid strata & $\Delta_{24} \neq 0$ & $\Delta_{24} = 0$ & $\Delta_{24} \neq 0$ & $\Delta_{24} = 0$  \\
\hline
\end{tabular}
\bigskip
\end{center}

Each Deodhar component was shown to be indexed by so-called Go-diagrams \cite{2012arXiv1204.6446K} and subsequently by (generally non-planar) networks \cite{2012arXiv1210.5433T}, which have a direct mapping to elements of the Grassmannian. The graph that represents a Deodhar component actually is in a specific matroid stratum, but each Deodhar component will have only one representative. As a result, these representatives can be chosen to represent the entire Deodhar component.

\bigskip

\section{Bipartite Graphs and the Grassmannian, a First Encounter}

\label{sec:planarboundmeas}

In this section we review the map between planar bipartite graphs and
the Grassmannian introduced by Postnikov in \cite{2006math09764P} and begin its generalization to arbitrary bipartite graphs. Further details of the generalization are developed in \sref{section_boundary_measurement_non-planar}. This map is known as the {\it boundary measurement}, and maps a bipartite graph with $k$ sources and $n$ external vertices to an element of \Gr.

The boundary measurement is an important ingredient in the study of on-shell diagrams. As we review in \sref{singularities}, the corresponding integrand is determined by the Grassmannian element associated to the graph.

Given a bipartite graph, the boundary measurement is constructed as follows:

\begin{itemize}
\item[{\bf 1)}] Choose an arbitrary perfect orientation of the diagram. This determines a source set. We denote the number of external vertices by $n$, and the number of sources by $k$.
\item[{\bf 2)}] Construct the $n_v \times n_v$ {\it path matrix} $\mathcal{M}$, where $n_v$ is the total number of nodes in the graph. Each matrix entry $\mathcal{M}_{i,j}$ entry contains the weights of the oriented paths in the perfect orientation connecting node $i$ and node $j$. An efficient way for constructing $\mathcal{M}$ is presented in Appendix \ref{PathM}.
\item[{\bf 3)}] Construct the $k\times n$ dimensional matrix $\mathcal{M}^C$. This is a sub-matrix of $\mathcal{M}$ in which columns are given by all external nodes and rows correspond to external nodes which are sources of the perfect orientation.
\item[{\bf 4)}] Modify signs in the entries of $\mathcal{M}^C$. We will discuss below the reasons for introducing such signs and introduce a systematic prescription for their determination.
\end{itemize}
The discussion above is completely general and applies to arbitrary bipartite graphs.

There are three different kind of entries in $\mathcal{M}^C$. The entries which contain paths that go from a source to the same source are always equal to 1. Some entries are 0, representing the fact that sometimes it is impossible to flow from a source to a given external node. In particular, there are no oriented flows between two sources. The paths contributing to entries in $\mathcal{M}^C$ can be identified with {\it single component} flows, which in addition take the form $\mathfrak{p}_\mu = \tilde{p}_\mu / \tilde{p}_{\text{ref}}$ for some oriented perfect matching $\tilde{p}_\mu$.\footnote{In the presence of loops, entries will in general have the form $\frac{\mathfrak{p}_i}{1-\mathfrak{p}_{\text{loop}}}$.}

The matrix $\mathcal{M}^C$ is already extremely useful for some applications, which do not require a precise knowledge of the sign assignments that take us to the boundary measurement $C$. By studying the entries of the matrix, it is possible to determine the connectivity of external nodes. This fact will be heavily used in \sref{section_graphs_polytopes_geometry}. Similarly, we can use it for determining the number of its degrees of freedom: it is the number of non-zero minors minus the number of relations between $k \times k$ minors, minus 1. This is equal to the number of degrees of freedom of $C$, which is the dimensionality of the associated element of the Grassmannian.

\bigskip

\subsection*{Sign Prescription}

We are ready to discuss the sign prescription, to finally map $\mathcal{M}^C \mapsto C \in Gr_{k,n}$. Here we will focus on the case of planar graphs, i.e.\ graphs on a disk, and follow \cite{2006math09764P}. The implementation of signs for non-planar graphs will be the topic of \sref{section_boundary_measurement_non-planar}.
 
For planar graphs, the signs in the boundary measurement are chosen such that two nice properties are {\it simultaneously} achieved: all maximal minors of $C$ are non-negative for non-negative edge weights and, moreover, these minors are simply sums of products of flows. In addition, we will pick signs such that denominators cannot vanish for strictly positive edge weights.\footnote{Here we consider the analytic continuation of the geometric series giving rise to a non-trivial denominator.} Such potentially vanishing denominators arise when formally summing the geometric series that arise in the presence of closed oriented loops.\footnote{Another natural choice for which all minors are sums of flows corresponds to not introducing any signs to $\mathcal{M}^C$ \cite{2012arXiv1202.3128T}. However non-trivial signs have to be delicately chosen in order to simultaneously achieve the other two properties mentioned in this paragraph.}

In order to construct a matrix with definite non negative minors, we have to modify some signs in the entries of $\mathcal{M}^C$. The prescription consists in first introducing a sign $(-1)^{s(i,j)}$ to the entry $\mathcal{M}^C_{i,j}$, where $s(i,j)$ is the number of sources strictly between $i$ and $j$, neglecting periodicity. Secondly, one has also to introduce a $(-1)$ factor to every loop. These two modifications conspire in such a way to
obtain a matrix $C$ whose minors are all non-negative, and moreover such that its minors remain simple sums of flows.

\bigskip

\begin{figure}[h]
\begin{center}
\includegraphics[width=5cm]{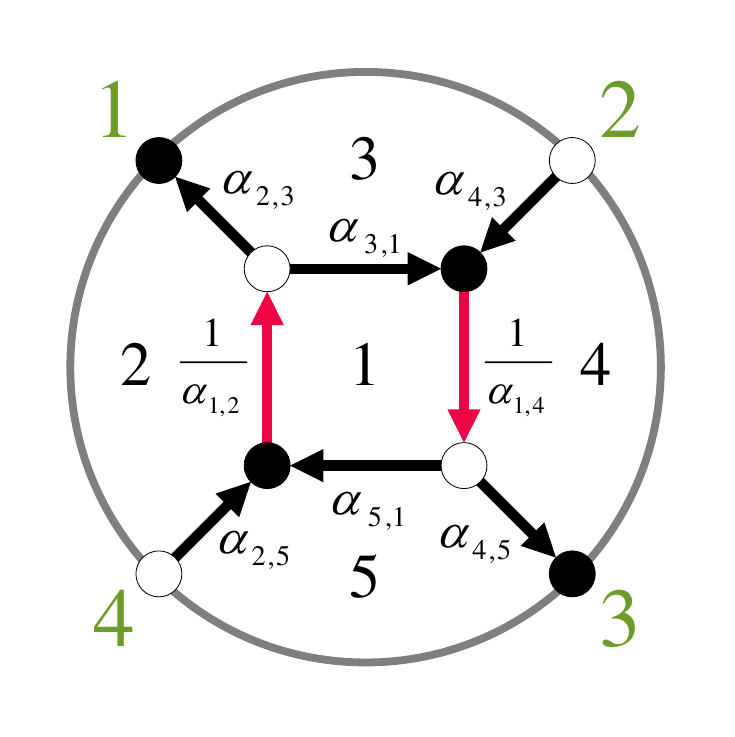}
\caption{Bipartite graph for the top-cell of $Gr_{2,4}$. The reference perfect matching is shown in red. Arrows indicate the corresponding perfect orientation.}
\label{squarebox1}
\end{center}
\end{figure}

\paragraph{Example.} We now provide an example to illustrate this method. We begin with the diagram displayed in \fref{squarebox1}, and the perfect orientation associated to the reference matching consisting of edges $X_{1,2}, X_{1,4}$. The relevant subset of the path matrix, choosing the clockwise ordering starting at the edge $X_{2,3}$, is

\begin{equation}
\mathcal{M}^C =
\left(
\begin{array}{c|cccc}
 & 1 & 2 & 3 & 4 \\
\hline
2 \ \ & \frac{\alpha_{2,3} \alpha_{4,3} \alpha_{5,1}}{\alpha_{1,2} \alpha_{1,4} \left(1-\frac{\alpha_{3,1} \alpha_{5,1}}{\alpha_{1,2} \alpha_{1,4}}\right)} & 1 & \frac{\alpha_{4,3} \alpha_{4,5}}{\alpha_{1,4} \left(1-\frac{\alpha_{3,1} \alpha_{5,1}}{\alpha_{1,2} \alpha_{1,4}}\right)} & 0 \\
4 \ \ & \frac{\alpha_{2,3} \alpha_{2,5}}{\alpha_{1,2} \left(1-\frac{\alpha_{3,1} \alpha_{5,1}}{\alpha_{1,2} \alpha_{1,4}}\right)} & 0 & \frac{\alpha_{2,5} \alpha_{3,1} \alpha_{4,5}}{\alpha_{1,2} \alpha_{1,4} \left(1-\frac{\alpha_{3,1} \alpha_{5,1}}{\alpha_{1,2} \alpha_{1,4}}\right)} & 1
\end{array}
\right) = {\large \left(
\begin{array}{c|cccc}
& \mbox{\normalsize{1}} & \mbox{\normalsize{2}} & \mbox{\normalsize{3}} & \mbox{\normalsize{4}} \\ \hline
\mbox{\normalsize{2}} \ \ & \frac{\mathfrak{p}_4}{1-\mathfrak{p}_7} & 1 & \frac{\mathfrak{p}_2}{1-\mathfrak{p}_7} & 0 \\
\mbox{\normalsize{4}} \ \ & \frac{\mathfrak{p}_3}{1-\mathfrak{p}_7} & 0 & \frac{\mathfrak{p}_5}{1-\mathfrak{p}_7} & 1
\end{array}
\right)},
\end{equation}

\noindent where the labeling of perfect matchings follows that of \fref{fig:sqbpms}. Once the signs are introduced, this is associated with the top-cell of $Gr_{2,4}$, since all entries which can be non-zero are generically non-zero. This example has a loop in the perfect orientation, which manifests itself as several terms in the denominator, as explained in Appendix \ref{PathM}. The minors of this matrix take on a very simple form:
\begin{equation}
\begin{array}{lcl}
 m_{12} = -\frac{\alpha_{1,4} \alpha_{2,3} \alpha_{2,5}}{\alpha_{1,2} \alpha_{1,4}-\alpha_{3,1} \alpha_{5,1}} & \text{\phantom{space}} & m_{23} = \frac{\alpha_{2,5} \alpha_{3,1} \alpha_{4,5}}{\alpha_{1,2} \alpha_{1,4}-\alpha_{3,1} \alpha_{5,1}} \\
 m_{13} = -\frac{\alpha_{2,3} \alpha_{2,5} \alpha_{4,3} \alpha_{4,5}}{\alpha_{1,2} \alpha_{1,4}-\alpha_{3,1} \alpha_{5,1}} & \text{\phantom{space}} &  m_{24} = 1 \\
 m_{14} = \frac{\alpha_{2,3} \alpha_{4,3} \alpha_{5,1}}{\alpha_{1,2} \alpha_{1,4}-\alpha_{3,1} \alpha_{5,1}} & \text{\phantom{space}} & m_{34} = \frac{\alpha_{1,2} \alpha_{4,3} \alpha_{4,5}}{\alpha_{1,2} \alpha_{1,4}-\alpha_{3,1} \alpha_{5,1}} \\
\end{array} \quad .
\end{equation}
Several remarks are in order. First, all the minors of $\mathcal{M}^C$ have the form of sums of flows, divided by possible loops, thanks to non-trivial cancellations. Secondly, all minors are non-zero, reflecting the fact that the element of the Grassmannian associated to $\mathcal{M}^C$ has maximal dimension. Thirdly, some of the minors are negative, for positive edge weights.

We finally proceed in modifying the signs of the matrix $\mathcal{M}^C$ to obtain the element of the totally non negative Grassmannian. The $(-1)^{s(i,j)}$ factor implies that we have to multiply the entry $\mathcal{M}^C_{2,1}$ by $(-1)$. The $(-1)$ factor for loops amounts to replacing $\mathfrak{p}_7 \to -\mathfrak{p}_7$. 
These two operations map $\mathcal{M}^C$ into the relevant element of the Grassmannian $C \in Gr_{2,4}$:

\begin{equation}
\label{C24}
C=
\left(
\begin{array}{c|cccc}
& 1 & 2 & 3 & 4 \\ \hline
1 \ \ & \frac{\alpha_{2,3} \alpha_{4,3} \alpha_{5,1}}{\alpha_{1,2} \alpha_{1,4} \left(1+\frac{\alpha_{3,1} \alpha_{5,1}}{\alpha_{1,2} \alpha_{1,4}}\right)} & 1 & \frac{\alpha_{4,3} \alpha_{4,5}}{\alpha_{1,4} \left(1+\frac{\alpha_{3,1} \alpha_{5,1}}{\alpha_{1,2} \alpha_{1,4}}\right)} & 0 \\
2 \ \ &  -\frac{\alpha_{2,3} \alpha_{2,5}}{\alpha_{1,2} \left(1+\frac{\alpha_{3,1} \alpha_{5,1}}{\alpha_{1,2} \alpha_{1,4}}\right)} & 0 & \frac{\alpha_{2,5} \alpha_{3,1} \alpha_{4,5}}{\alpha_{1,2} \alpha_{1,4} \left(1+\frac{\alpha_{3,1} \alpha_{5,1}}{\alpha_{1,2} \alpha_{1,4}}\right)} & 1
\end{array}
\right)  .
\end{equation}
The maximal minors of $C \in Gr_{k,n}$ are the \pl coordinates $\Delta_I$. For the example above, the \pl coordinates are:

\begin{equation}
\begin{array}{lcl}
 \Delta _{12} = \frac{\alpha_{1,4} \alpha_{2,3} \alpha_{2,5}}{\alpha_{1,2} \alpha_{1,4}+\alpha_{3,1} \alpha_{5,1}} = \frac{\mathfrak{p}_3}{1+\mathfrak{p}_7} & \text{\phantom{space}}& \Delta _{23} = \frac{\alpha_{2,5} \alpha_{3,1} \alpha_{4,5}}{\alpha_{1,2} \alpha_{1,4}+\alpha_{3,1} \alpha_{5,1}} = \frac{\mathfrak{p}_5}{1+\mathfrak{p}_7} \\
 \Delta _{13} = \frac{\alpha_{2,3} \alpha_{2,5} \alpha_{4,3} \alpha_{4,5}}{\alpha_{1,2} \alpha_{1,4}+\alpha_{3,1} \alpha_{5,1}} = \frac{\mathfrak{p}_1}{1+\mathfrak{p}_7} & \text{\phantom{space}}& \Delta _{24} = 1 \\
 \Delta _{14} = \frac{\alpha_{2,3} \alpha_{4,3} \alpha_{5,1}}{\alpha_{1,2} \alpha_{1,4}+\alpha_{3,1} \alpha_{5,1}} = \frac{\mathfrak{p}_4}{1+\mathfrak{p}_7} &  \text{\phantom{space}}& \Delta _{34} = \frac{\alpha_{1,2} \alpha_{4,3} \alpha_{4,5}}{\alpha_{1,2} \alpha_{1,4}+\alpha_{3,1} \alpha_{5,1}} = \frac{\mathfrak{p}_2}{1+\mathfrak{p}_7} \\
\end{array}
\end{equation}

\noindent which are manifestly positive, for positive edge weights.

\bigskip

\section{Stratification and Singularity Structure of On-Shell Diagrams}

\label{singularities}

In \sref{onshellDiag} 
we discussed the connection between on-shell diagrams 
of $\mathcal{N}=4$ SYM, the Grassmannian,
and bipartite graphs.
The authors 
of \cite{ArkaniHamed:2012nw}
explained how to construct the integrand associated to a planar on-shell diagram
using twistor space variables, in terms of data associated to the bipartite graph.
In the previous section we have reviewed how to associate to on-shell diagrams
the edge weights $\alpha_{ij}$ and the boundary measurement matrix $C(\alpha_{ij}) \in 
Gr_{k,n}$ of the Grassmannian.

The edge weight parameterization of the Grassmannian is
redundant.
The independent degrees of freedom are a subset $\beta_i$ of the edge weights of 
dimension $d=F-1$,
where $F$ is the number of faces of the graph.

The differential form associated to an on-shell diagram is \cite{ArkaniHamed:2012nw}
\be
 \frac{d \beta_1}{ \beta_1} \wedge \dots \wedge \frac{d \beta_d}{ \beta_d} ~ \delta^{k \times 4}(C \cdot \tilde \eta)
\delta^{k \times 2} (C \cdot \tilde \lambda) \delta^{2 \times (n-k)} (\lambda \cdot C^{\perp}),
\ee
where ($\tilde \eta, \tilde \lambda, \lambda$) are the kinematical variables of the scattered particles,
in $\mathcal{N}=4$ twistor space.
The delta functions provide $2 n-4$ constraints. Hence, depending on the degrees of freedom
in the matrix $C$, i.e.\ on $d$, different situations arise.
If $d=2 n-4$ the integral over the differential form
is fully localized, and the result is an ordinary function of the external data;
this is the so-called leading singularity. If $d < 2n -4$ we have more constraints than degrees
of freedom $\beta_i$, so the leftover constraints impose conditions on the external data; this is a singularity. 
If $d>2n-4$ there are some degrees of freedom left unfixed by the delta functions
which can be integrated over.
Moreover,
in some cases the differential form can be such that 
some of the $\frac{d \beta_i}{ \beta_i}$ integrations factorize, leaving externals
$\log \beta_i$ factors. 
This happens when the corresponding graph is reducible. We will discuss the notion of graph reducibility in \sref{section_graph_equivalence_and_reduction}.

Understanding the singularity structure of the differential forms associated to on-shell diagrams is of great physical interest.
For instance, in the case of planar $\mathcal{N}=4$ SYM, 
the study of such singularities is connected to a generalization of the BCFW recursion relation
which fully determines the scattering amplitudes to all loop orders \cite{Britto:2004ap,Britto:2005fq,ArkaniHamed:2012nw}.

Given a differential form related to an on-shell diagram,
the singularity structure contains the information of the residues 
at the poles of the differential form, which are generically located at some
$\beta_i=0$.\footnote{Many coordinate charts $\vec \beta$ are necessary to cover all the poles of the differential form.} 
These singularities correspond to elements in the Grassmannian where the 
number of degrees of freedom in the matrix $C$ has been reduced, 
 by turning off some $\beta_i$.

The singularity pattern can be organized in a layered partially ordered set (poset).
At the top level we have the original diagram and the associated differential form. At the next level, there are the differential forms obtained at the poles of the original one, with one less degrees of freedom, and so on.
This procedure continues until it reaches the trivial configuration with no 
poles left. We provide graphical realizations of this is \sref{Strat_NewRegionNewMethod}, e.g.\ \fref{G24posi}.

In terms of the Grassmannian element determining the differential form, 
the number of degrees of freedom in $C$ is reduced by one when going from one level of the poset to the next one.
In terms of the bipartite graph, 
each step coincides with the removal of so-called \emph{removable edges},
which are defined as those which yield
subgraphs where 
$d \to d-1$. The precise notion of \emph{removable edges} and how to identify them will be discussed in \sref{section_efficient_reducibility}.

In summary,
given a differential form related to an on-shell diagram, its singularity structure can be understood
from the corresponding bipartite graph by decomposing the graph into subgraphs by removing only
removable edges. This provides a lattice of subgraphs, whose corresponding differential
forms are the singularities of the original differential form, 
organized by number of degrees of freedom.

In the planar case, if the original graph is top-dimensional, this graph decomposition 
is equivalent to the positroid stratification of the associated Grassmannian. 
In \sref{sec:NonplDecomp}, we will introduce a natural generalization of this decomposition which also applies to the non-planar graphs.

\bigskip

\section{From Bipartite Graphs to Polytopes and Toric Geometry} 

\label{section_graphs_polytopes_geometry}

In this section we will associate bipartite graphs to {\it matching} and {\it matroid polytopes}, which will play a prominent role in the rest of the paper.\footnote{Throughout this article, $d$-dimensional polytopes should be regarded modulo $SL(d,\mathbb{Z})$ transformations.} Equivalently, these convex polytopes can be interpreted as the toric diagrams defining certain toric, non-compact Calabi-Yau (CY) manifolds which we denote {\it master} and {\it moduli} spaces for their relation to BFTs. We will present various alternative approaches to these objects:

\begin{itemize}
\item Classifying matroid elements and their relations (perfect orientations).
\item Giving a geometric description of flows (flows). 
\item As master and moduli spaces of BFTs (perfect matchings).
\end{itemize}

Interestingly, each viewpoint naturally emphasizes different objects, listed above in parentheses. However, all of them are equivalent, as explained in \sref{RelationFlowPerfOrientMatch}. It is important to have multiple perspectives on the same objects, since they are best suited for addressing different questions.

Part of the material presented in this section has previously appeared in the literature, in some cases only for the case of planar graphs \cite{2007arXiv0706.2501P,2008arXiv0801.4822T,Amariti:2013ija}. 
A key point of this article is that these polytopes are also extremely useful beyond planar graphs. 

\begin{figure}[h]
\begin{center}
\includegraphics[width=5.5cm]{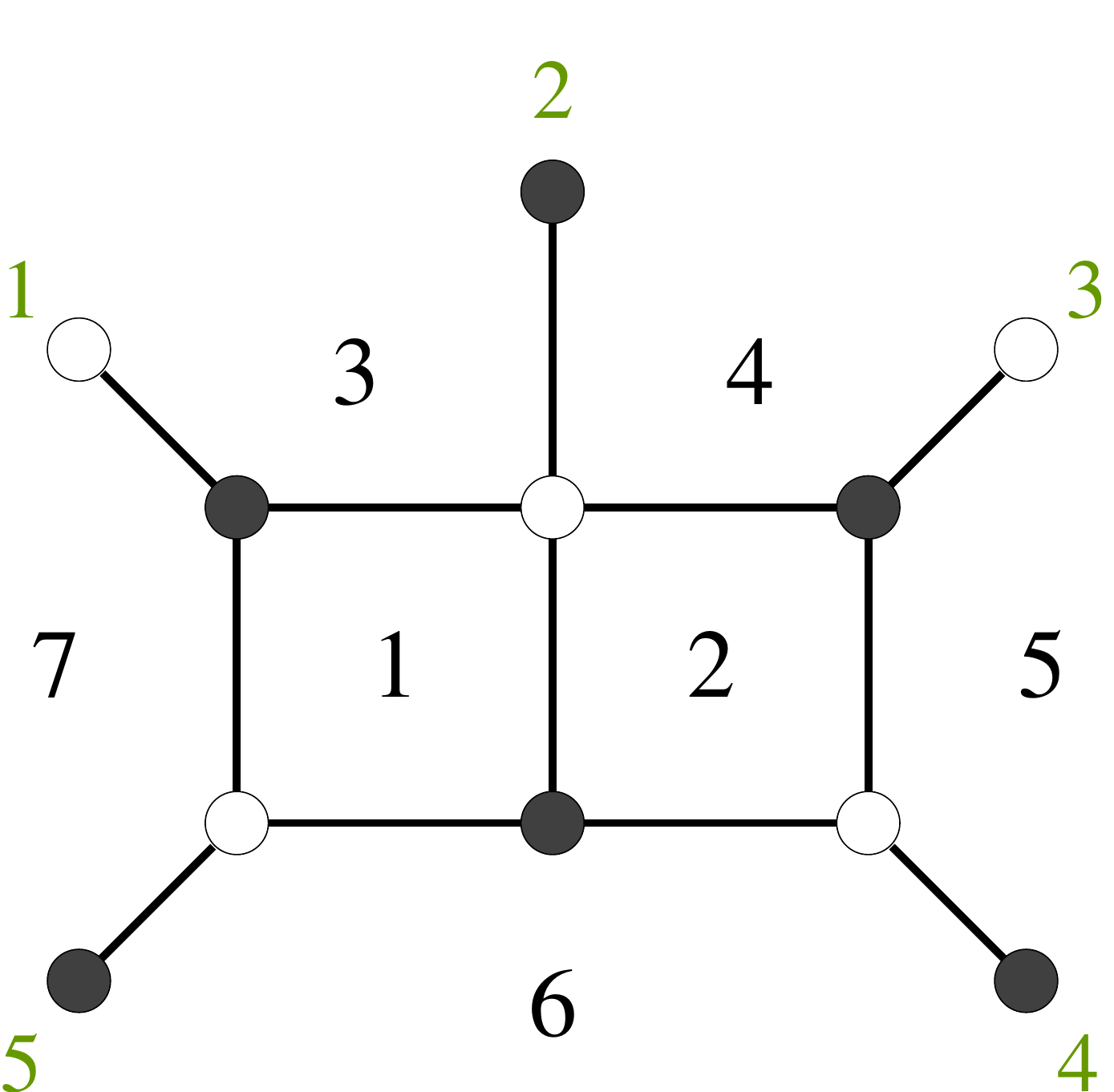}
\caption{An on-shell diagram for the top dimensional cell of $Gr_{2,5}$.}
\label{G25_graph}
\end{center}
\end{figure}

We will use the explicit example in \fref{G25_graph} for illustrating our ideas. This is an on-shell diagram associated to the top dimensional cell of $Gr_{2,5}$. This example is chosen because it exhibits more richness than the simpler $Gr_{2,4}$ considered so far. In general, the polytopes we will define live in high dimensional integer lattices. It is thus typically impractical to provide a graphical representation of them.  Instead, we will describe them in terms of matrices giving the position vectors of points in them. 

\bigskip

\subsection{Polytopes from Matroids}

\label{section_polytopes_from_matroids}

Here we introduce the polytopes we want to study and a first perspective on them.

\bigskip

\paragraph{Matching Polytope.} The first polytope we will construct encodes the map between edges and perfect matchings. Given a bipartite graph with $E$ edges $X_i$, $i=1,\ldots, E$ and $c$ perfect matchings $p_\mu$, $\mu=1,\ldots, c$, we define the $(E\times c)$-dimensional  {\it perfect matching matrix} $P$ as follows:
\beq
P_{i\mu}=\left\{ \begin{array}{ccccc} 1 & \rm{ if } & X_i  & \in & p_\mu \\
0 & \rm{ if } & X_i  & \notin & p_\mu
\end{array}\right.
\label{P_matrix}
\eeq
This matrix can be interpreted as defining the {\it matching polytope}, in which there is a distinct point for every perfect matching, with a position vector in $\mathbb{Z}^E$ given by the corresponding column vector \cite{2007arXiv0706.2501P,Franco:2012mm}.\footnote{Strictly speaking, we have not defined the matching polytope in terms of matroids. The connection to the matroid polytope, which we introduce below, will become clear in coming subsections once we develop other viewpoints on these objects.}

\begin{figure}[h]
\begin{center}
\includegraphics[width=15cm]{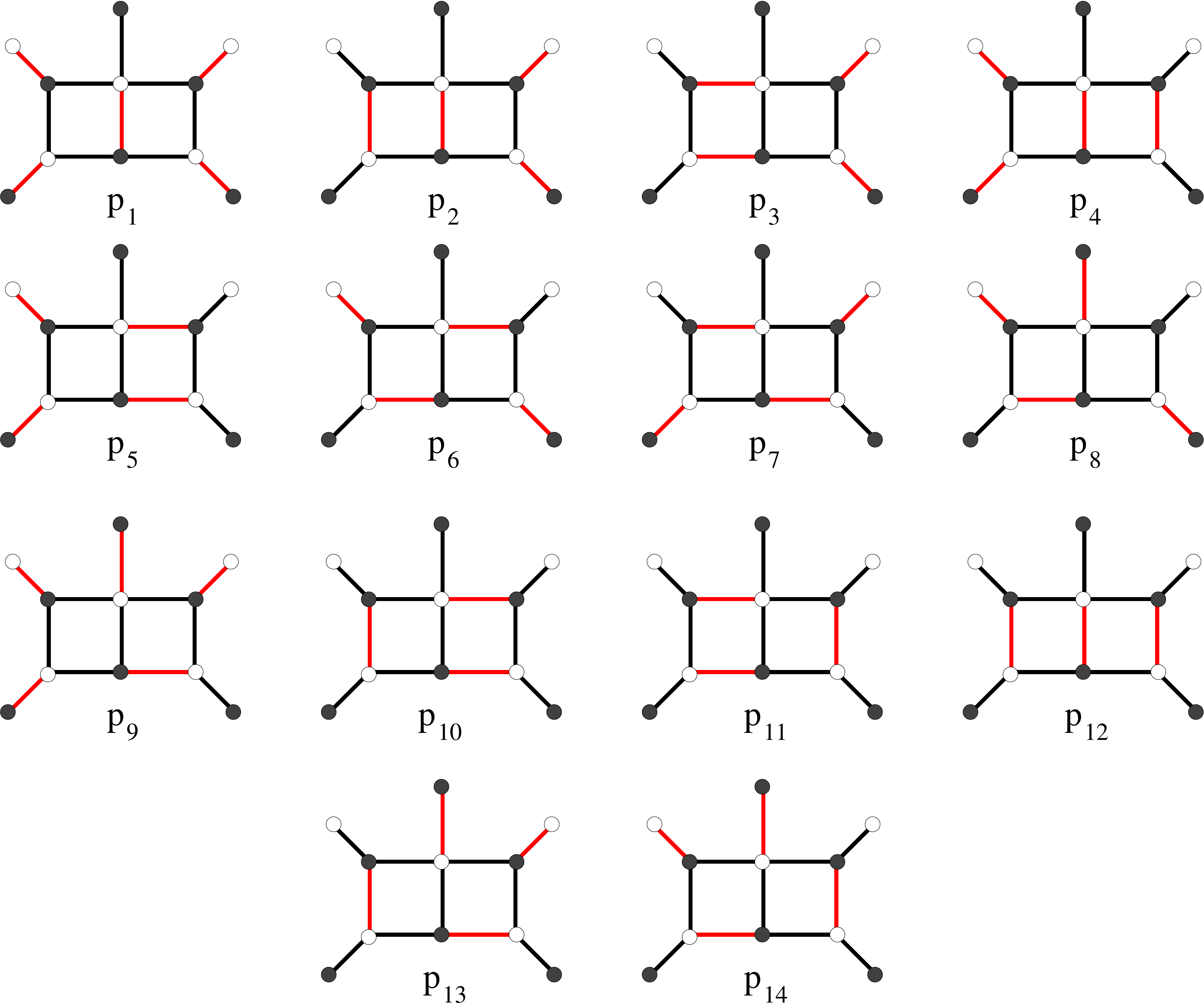}
\caption{The 14 perfect matchings for the bipartite graph in \fref{G25_graph}.}
\label{G25_pms}
\end{center}
\end{figure}

Let us construct the matching polytope for the explicit example at hand. The graph in \fref{G25_graph} has 14 perfect matchings, which can be determined using \eref{PM_Kastel}. They are shown in \fref{G25_pms}. The perfect matching matrix thus becomes:

{\footnotesize
\beq
P=\left(
\begin{array}{c|cccccccccccccc}
& \ p_1 \ & \ p_2 \ & \ p_3 \ & \ p_4 \ & \ p_5 \ & \ p_6 \ & \ p_7 \ & \ p_8 \ & \ p_9 \ & p_{10} & p_{11} & p_{12} & p_{13} & p_{14} \\
\hline
X_{1,3} & 0 & 0 & 1 & 0 & 0 & 0 & 1 & 0 & 0 & 0 & 1 & 0 & 0 & 0 \\
X_{4,2} &  0 & 0 & 0 & 0 & 1 & 1 & 0 & 0 & 0 & 1 & 0 & 0 & 0 & 0 \\
X_{2,5} &  0 & 0 & 0 & 1 & 0 & 0 & 0 & 0 & 0 & 0 & 1 & 1 & 0 & 1 \\
X_{6,2} &  0 & 0 & 0 & 0 & 1 & 0 & 1 & 0 & 1 & 1 & 0 & 0 & 1 & 0 \\
X_{1,6} &  0 & 0 & 1 & 0 & 0 & 1 & 0 & 1 & 0 & 0 & 1 & 0 & 0 & 1 \\
X_{7,1} &  0 & 1 & 0 & 0 & 0 & 0 & 0 & 0 & 0 & 1 & 0 & 1 & 1 & 0 \\
X_{2,1} &  1 & 1 & 0 & 1 & 0 & 0 & 0 & 0 & 0 & 0 & 0 & 1 & 0 & 0 \\
X_{3,4} &  0 & 0 & 0 & 0 & 0 & 0 & 0 & 1 & 1 & 0 & 0 & 0 & 1 & 1 \\
X_{5,4} &  1 & 1 & 1 & 0 & 0 & 0 & 1 & 1 & 1 & 0 & 0 & 0 & 1 & 0 \\
X_{5,6} &  1 & 1 & 1 & 0 & 0 & 1 & 0 & 1 & 0 & 0 & 0 & 0 & 0 & 0 \\
X_{6,7} &  1 & 0 & 0 & 1 & 1 & 0 & 1 & 0 & 1 & 0 & 0 & 0 & 0 & 0 \\
X_{3,7} &  1 & 0 & 0 & 1 & 1 & 1 & 0 & 1 & 1 & 0 & 0 & 0 & 0 & 1 \\
\end{array}
\right)
\label{P_matrix_G25}
\eeq}

Generically, the matching polytope lives in a lower dimensional subspace of $\mathbb{Z}^E$. This fact can be made explicit by row-reducing $P$, which for \eref{P_matrix_G25} results in the following matrix:

{\footnotesize
\beq
G_{\text{matching}}=\left(
\begin{array}{cccccccccccccc}
\ p_1 \ & \ p_2 \ & \ p_3 \ & \ p_4 \ & \ p_5 \ & \ p_6 \ & \ p_7 \ & \ p_8 \ & \ p_9 \ & p_{10} & p_{11} & p_{12} & p_{13} & p_{14} \\
\hline
 1 & 0 & 0 & 0 & 0 & 0 & 0 & 0 & 0 & -1 & -1 & -1 & -1 & -1 \\
 0 & 1 & 0 & 0 & 0 & 0 & 0 & 0 & 0 & 1 & 0 & 1 & 1 & 0 \\
 0 & 0 & 1 & 0 & 0 & 0 & 1 & 0 & 0 & 0 & 1 & 0 & 0 & 0 \\
 0 & 0 & 0 & 1 & 0 & 0 & 0 & 0 & 0 & 0 & 1 & 1 & 0 & 1 \\
 0 & 0 & 0 & 0 & 1 & 0 & 1 & 0 & 1 & 1 & 0 & 0 & 1 & 0 \\
 0 & 0 & 0 & 0 & 0 & 1 & -1 & 0 & -1 & 0 & 0 & 0 & -1 & 0 \\
 0 & 0 & 0 & 0 & 0 & 0 & 0 & 1 & 1 & 0 & 0 & 0 & 1 & 1 
\end{array}
\right)
\label{G_matching_G25_matroids}
\eeq}

It is straightforward to verify that the points defined by the previous matrix actually live in a 6d hyperplane at unit distance from the origin, and hence one of the dimensions in \eref{G_matching_G25_matroids} can be projected out. It is thus possible to neglect one dimension, by e.g.\ discarding a row in $G$. From now on we refer to the \textit{dimension of the matching polytope} as the dimension of the hyperplane on which the points lie; in the example above this is 6 dimensions. Thus, for planar graphs the dimensionality of the matching polytope is equal to the total number of faces minus one, i.e.\ $F-1$. The dimensionality and how it generalizes to non-planar graphs are best understood in terms of flows in a perfect orientation. This will be discussed in \sref{section_matching_from_flows}.

\bigskip

\paragraph{Matroid Polytope.} The matroid polytope was introduced in \sref{matroidstra} to encode the elements of a matroid and their relations. The source sets $I_\mu$, $\mu=1,\ldots c$, of perfect orientations in a planar graph are in one-to-one correspondence with elements of a matroid. A central theme of the current paper is the extension of notions such as the matroid polytope to non-planar graphs. Additional details of such generalizations will be given in later sections. The discussion in this section will thus continue under the assumption of completely general bipartite graphs, i.e.\ our matroid polytopes should be regarded as the ones usually defined for planar graphs.

Matroid bases are in one-to-one correspondence with source sets of perfect orientations. Given the external nodes $n_i^{(e)}$, $i=1,\ldots,n$ and source sets $I_{\mu}$ of perfect orientations, the matroid polytope is defined as follows:
\beq
G_{\text{matroid},i\mu}=\left\{ \begin{array}{ccccc} 1 & \rm{ if } & n^{(e)}_i  & \in & I_\mu \\
0 & \rm{ if } & n^{(e)}_i  & \notin & I_\mu
\end{array}\right.
\label{matroid_polytope}
\eeq
where column vectors give the positions of points in the polytope. At this point, it is important to emphasize a phenomenon which will later reappear in multiple incarnations. In general, different perfect orientations can share the same source set, which in turn implies they are mapped to the same point in the matching polytope. The precise sense in which such perfect orientations imply multiple ``contributions" to a given matroid element will be clarified in \sref{section_Plucker_coordinates} in terms of \pl coordinates.

Similarly to the matching polytope, the 
matroid polytope lies in a hyperplane at unit distance from the origin, i.e.\ it has $F_e-1$ independent dimensions. 
Interestingly, since the dimensionality of the matroid polytope is only controlled by external nodes, 
it remains equal to $F_e-1$ in the non-planar case. We present a further discussion of this point in \sref{Matroid_Polytope_Flows}.

\begin{figure}[h]
\begin{center}
\includegraphics[width=15cm]{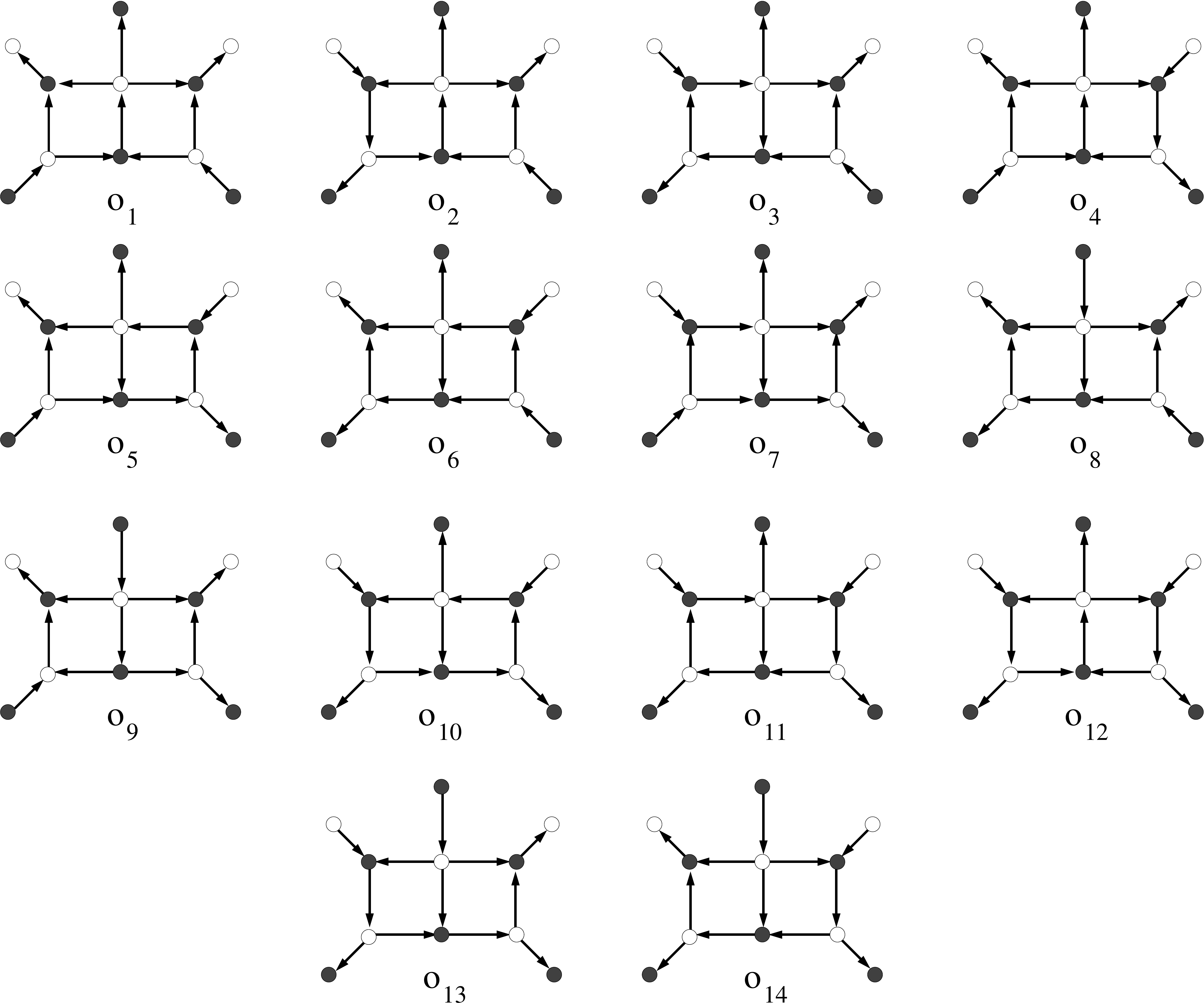}
\caption{Perfect orientations for the perfect matchings in \fref{G25_pms}.}
\label{G25_perfect_orientations}
\end{center}
\end{figure}

Returning to our explicit example, \fref{G25_perfect_orientations} gives the 14 perfect orientations associated to the perfect matchings in \fref{G25_pms}. We denote $o_\mu$ the perfect orientation corresponding to a perfect matching $p_\mu$.

There are 10 possible source sets in this case, i.e.\ 10 matroid elements, and \eref{matroid_polytope} becomes:

{\footnotesize
\beq
G=\left(
\begin{array}{r|c|cc|cc|c|c|c|c|ccc|c|c}
& \{45\} & \multicolumn{2}{c|}{\{14\}} & \multicolumn{2}{c|}{\{35\}} & \{34\} & \{15\} & \{24\} & \{25\} & \multicolumn{3}{c|}{\{13\}} & \{12\} & \{23\} \\ \hline
& \ o_1 \ & \ o_2 \ & \ o_3 \ & \ o_4 \ & \ o_5 \ & \ o_6 \ & \ o_7 \ & \ o_8 \ & \ o_9 \ & o_{10} & o_{11} & o_{12} & o_{13} & o_{14} \\ \hline
1 \ \ & 0 & 1 & 1 & 0 & 0 & 0 & 1 & 0 & 0 & 1 & 1 & 1 & 1 & 1 \\
2 \ \ & 0 & 0 & 0 & 0 & 0 & 0 & 0 & 1 & 1 & 0 & 0 & 0 & 1 & 0 \\
3 \ \ & 0 & 0 & 0 & 1 & 1 & 1 & 0 & 0 & 0 & 1 & 1 & 1 & 0 & 1 \\
4 \ \ & 1 & 1 & 1 & 0 & 0 & 1 & 0 & 1 & 0 & 0 & 0 & 0 & 0 & 0 \\
5 \ \ & 1 & 0 & 0 & 1 & 1 & 0 & 1 & 0 & 1 & 0 & 0 & 0 & 0 & 0 
\end{array}
\right)
\label{G_matroid_G25}
\eeq}

\smallskip

\noindent This example explicitly shows how source sets can be shared by more than one perfect orientation. For example ${\{14\}}$ corresponds to both $p_2$ and $p_3$. Similarly, $\{35\}$ and $\{13\}$ arise from multiple perfect orientations.

It is convenient to introduce a more compact version of this matrix, which only provides the positions of points in the matroid polytope and the multiplicities of perfect orientations contributing to each of them. For \eref{G_matroid_G25}, we have:
{\footnotesize
\beq
G_{\text{matroid}}=\left(
\begin{array}{cccccccccc}
\{45\} & \{14\} & \{35\} & \{34\} & \{15\} & \{24\} & \{25\} & \{13\} & \{12\} & \{23\} \\
\hline
  0 & 1 & 0 & 0 & 1 & 0 & 0 & 1 & 1 & 1 \\
 0 & 0 & 0 & 0 & 0 & 1 & 1 & 0 & 1 & 0 \\
 0 & 0 & 1 & 1 & 0 & 0 & 0 & 1 & 0 & 1 \\
 1 & 1 & 0 & 1 & 0 & 1 & 0 & 0 & 0 & 0 \\
 1 & 0 & 1 & 0 & 1 & 0 & 1 & 0 & 0 & 0 \\ \hline
\ {\bf 1} \ & \ {\bf 2} \ & \ {\bf 2} \ & \ {\bf 1} \ & \ {\bf 1} \ & \ {\bf 1} \ & \ {\bf 1} \ & \ {\bf 3} \ & \ {\bf 1} \ & \ {\bf 1} \
\end{array}
\right) ,
\label{G_matroid_G25_from_matroids}
\eeq}

\smallskip

\noindent where the last row indicates the multiplicities of perfect orientations. The polytope lives on a 4d hyperplane.

\bigskip

\subsection{Polytopes from Flows}

\label{section_polytopes_from_flows}

Here we introduce a second route to matching and matroid polytopes, based on a geometric description of flows. The thoughts in this section are a continuation of the ones introduced in \cite{Franco:2012wv} and related ideas, albeit emphasizing slightly different issues, can be found in \cite{Amariti:2013ija}. Similar descriptions of flows have appeared earlier in the literature, see e.g.\ \cite{2008arXiv0801.4822T}.

The first step in order to discuss flows is to pick an underlying perfect orientation. Alternative choices of the reference perfect orientation lead to trivial modifications of the polytopes. 

\begin{figure}[h]
\begin{center}
\includegraphics[width=5cm]{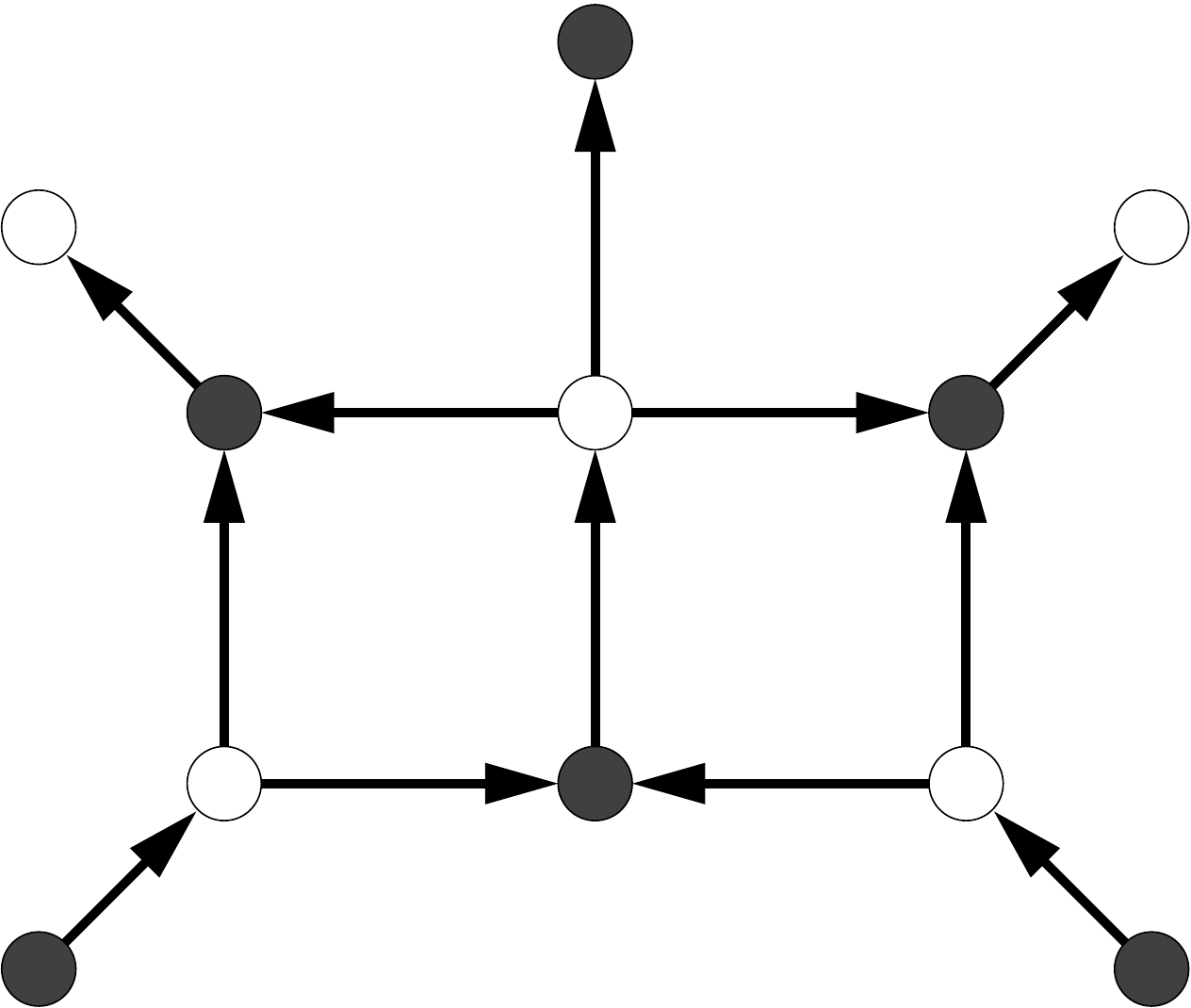}
\caption{An on-shell diagram for the top dimensional cell of $Gr_{2,5}$ and a choice of perfect orientation.}
\label{G25_orientation}
\end{center}
\end{figure}

\begin{figure}[h]
\begin{center}
\psfrag{p1}{hola}
\includegraphics[width=15cm]{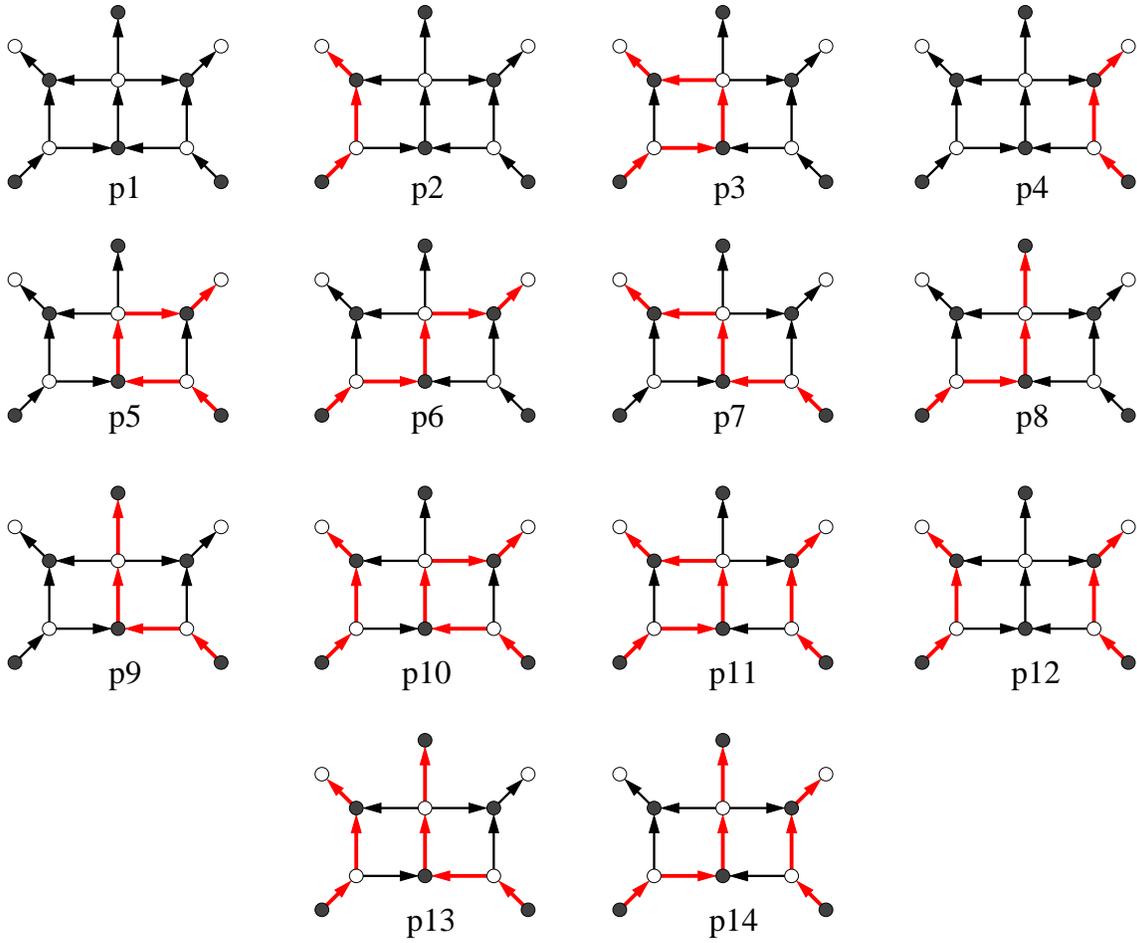}
\caption{All flows corresponding to the perfect orientation in \fref{G25_orientation}.}
\label{G25_loops}
\end{center}
\end{figure}

For the example at hand, let us focus on the perfect orientation $o_1$, which we reproduce in \fref{G25_orientation}. \fref{G25_loops} shows all flows in it. As previously discussed, flows can be open, closed or a combination of disjoint components.

\bigskip

\subsubsection{Matching Polytopes: a Fully Refined Description of Flows}

\label{section_matching_from_flows}

Flows in a perfect orientation can be fully specified by expanding them in terms of a basis. For graphs on a disk, a convenient basis is given by the loops circling clockwise around faces, both internal and external. It is indeed useful to distinguish between the two types of faces. We call the internal faces  $w_i$, $i=1,\ldots,F_i$, and the external ones $x_j$, $j=1,\ldots,F_e$, with $F_i+F_e=F$. These variables are subject to the constraint $\prod_{i=1}^{F_e} w_i \prod_{j=1}^{F_i} x_j=1$. This implies that one of them is actually redundant which, without loss of generality, we can take it to be one of the external faces. This is the manifestation, in the language of flows, of the extra coordinate we discussed in the previous section. Flows $\mathfrak{p}_\mu$ are thus mapped to points in an $(F-1)$-dimensional space with integer coordinates, according to: 

\beq
\mathfrak{p}_\mu=\prod_{i=1}^{F_e-1} w_i^{a_{i,\mu}} \prod_{j=1}^{F_i} x_j^{b_{j,\mu}} \ \ \ \ \mapsto \ \ \ \ \begin{array}{c}{\rm \underline{Coordinates}:} \\ (a_{1,\mu},\ldots,n_{F_i-1,\mu},b_{1,\mu},\ldots,b_{F_e,\mu}) \end{array}
\label{flows_from_loops}
\eeq
Since these coordinates allow a full identification between flows, each of them is mapped to a distinct point. The resulting polytope is indeed the matching polytope. 

For the flows in \fref{G25_loops}, the points in the polytope can be summarized as the column vectors of the following matrix:

{\footnotesize
\beq
G_{\text{matching}}=\left(
\begin{array}{c|cccccccccccccc}
& \ \mathfrak{p}_1 \ & \ \mathfrak{p}_2 \ & \ \mathfrak{p}_3 \ & \ \mathfrak{p}_4 \ & \ \mathfrak{p}_5 \ & \ \mathfrak{p}_6 \ & \ \mathfrak{p}_7 \ & \ \mathfrak{p}_8 \ & \ \mathfrak{p}_9 \ & \mathfrak{p}_{10} & \mathfrak{p}_{11} & \mathfrak{p}_{12} & \mathfrak{p}_{13} & \mathfrak{p}_{14} \\ \hline
a_1 & 0 & 0 & -1 & 0 & 0 & -1 & 0 & -1 & 0 & 0 & -1 & 0 & 0 & -1 \\
a_2 & 0 & 0 & 0 & 0 & 1 & 0 & 1 & 0 & 1 & 1 & 0 & 0 & 1 & 0 \\ \hline
b_1 & 0 & -1 & -1 & 0 & 0 & -1 & 0 & -1 & 0 & -1 & -1 & -1 & -1 & -1 \\
b_2 & 0 & 0 & 0 & 0 & 0 & -1 & 1 & -1 & 0 & 0 & 0 & 0 & 0 & -1 \\
b_3 & 0 & 0 & 0 & 0 & 0 & -1 & 1 & 0 & 1 & 0 & 0 & 0 & -1 & 0 \\
b_4 & 0 & 0 & 0 & 1 & 1 & 0 & 1 & 0 & 1 & 1 & 1 & 1 & -1 & 1 
\end{array}
\right) .
\label{G_matching_G25_from_flows}
\eeq}
\smallskip

\noindent This result coincides with \eref{G_matching_G25_matroids}. 

Bipartite graphs with higher genus and zero or multiple boundary components can be treated similarly. In such cases, the basis of cycles needs to be appropriately extended as follows \cite{Franco:2012wv}:

\bigskip

\begin{itemize} 
\item {\bf Higher genus:} include $\alpha_i$ and $\beta_i$, $i=1,\ldots,g$ pairs of fundamental cycles for a genus $g$ Riemann surface.
\item {\bf Boundaries:} for $B\geq 1$ boundaries, it is necessary to include paths $B-1$ independent paths connecting the different boundary components.
\end{itemize}

\bigskip

For clarity, the discussion that follows is centered on the case of the disk. Extending it to general graphs along the lines just mentioned is straightforward.

\bigskip

\subsubsection{Matroid Polytope: Keeping Partial Information About Paths}
\label{Matroid_Polytope_Flows}

For certain questions, having a full specification of flows, such as the one given in \sref{section_matching_from_flows}, is more than it is necessary. For example, in order to determine which entries in the boundary measurement are non-vanishing, knowledge of which external nodes are connected by a given flow is sufficient.\footnote{Recall that determining the non-vanishing entries of $C$ is equivalent to finding them for $\mathcal{M}^C$.} The detailed trajectories of flows along the bulk of the graph are unimportant. It is sufficient to identify the edges through which they enter and exit the graph. In terms of the loop coordinates defined in \sref{section_matching_from_flows}, this is fully determined by keeping only those coordinates associated to the $F_e-1$ independent external faces. For planar graphs we drop the coordinates associated to internal faces. In more general cases, we also discard those coordinates associated to paths between different boundary components and fundamental cycles on higher genus Riemann surfaces.

\begin{figure}[h]
\begin{center}
\includegraphics[width=5.25cm]{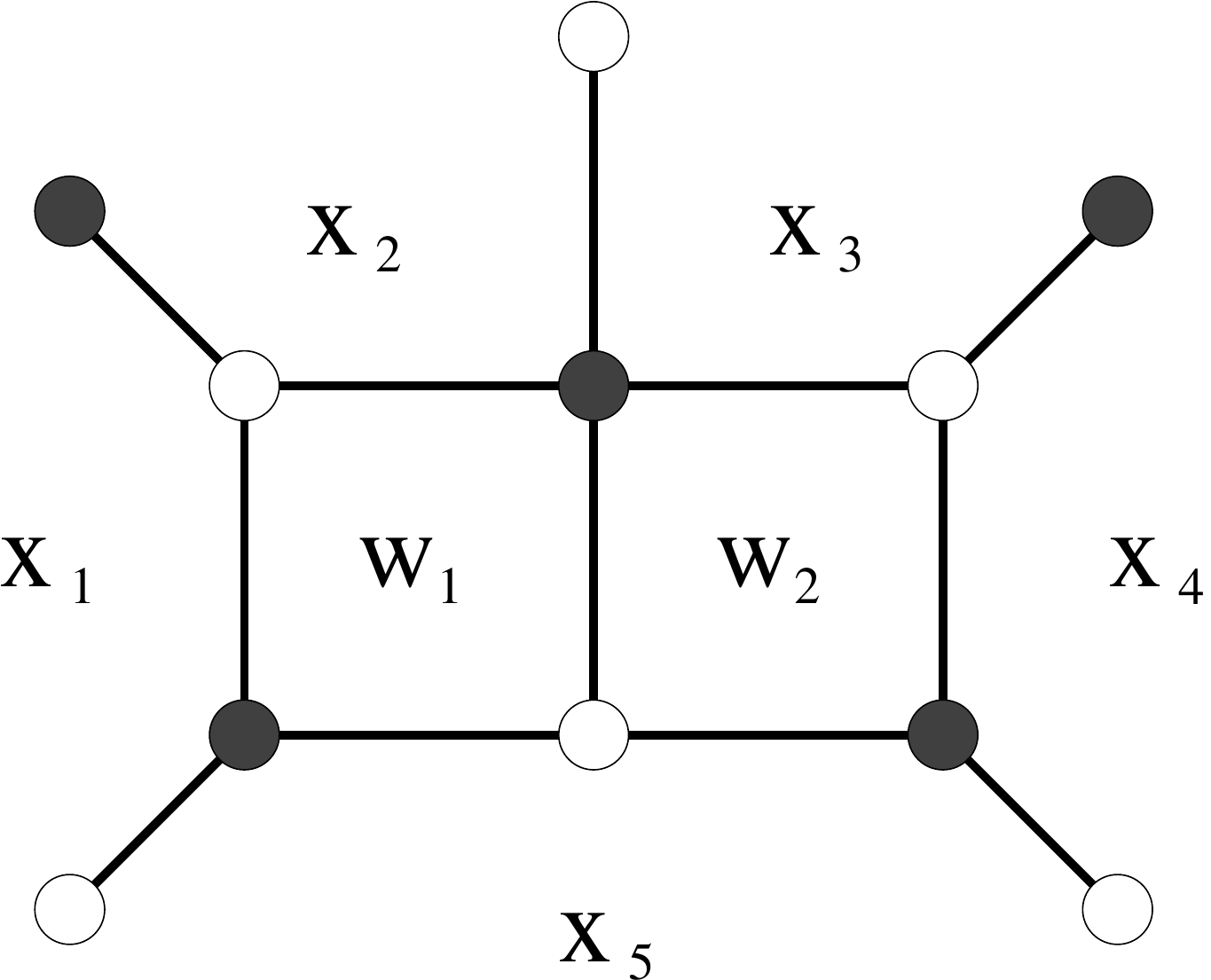}
\caption{An on-shell diagram for the top dimensional cell of $Gr_{2,5}$, with a new labeling of faces that is suitable for the analysis in this section.}
\label{G25_face_labels}
\end{center}
\end{figure}

Let us consider the $Gr_{2,5}$ example. Keeping the $b_{j,\mu}$ coordinates and discarding the two $a_{i,\mu}$ associated to the internal faces, \eref{G_matching_G25_from_flows} reduces as follows:

{\footnotesize
\beq
\begin{array}{c}
G_{\text{matching}}=\left(
\begin{array}{c|c|cc|cc|c|c|c|c|ccc|c|c}
& \ \mathfrak{p}_1 \ & \ \mathfrak{p}_2 \ & \ \mathfrak{p}_3 \ & \ \mathfrak{p}_4 \ & \ \mathfrak{p}_5 \ & \ \mathfrak{p}_6 \ & \ \mathfrak{p}_7 \ & \ \mathfrak{p}_8 \ & \ \mathfrak{p}_9 \ & \mathfrak{p}_{10} & \mathfrak{p}_{11} & \mathfrak{p}_{12} & \mathfrak{p}_{13} & \mathfrak{p}_{14} \\ \hline
a_1 & 0 & 0 & -1 & 0 & 0 & -1 & 0 & -1 & 0 & 0 & -1 & 0 & 0 & -1 \\
a_2 & 0 & 0 & 0 & 0 & 1 & 0 & 1 & 0 & 1 & 1 & 0 & 0 & 1 & 0 \\  \hline
b_1 & 0 & -1 & -1 & 0 & 0 & -1 & 0 & -1 & 0 & -1 & -1 & -1 & -1 & -1 \\
b_2 & 0 & 0 & 0 & 0 & 0 & -1 & 1 & -1 & 0 & 0 & 0 & 0 & 0 & -1 \\
b_3 & 0 & 0 & 0 & 0 & 0 & -1 & 1 & 0 & 1 & 0 & 0 & 0 & -1 & 0 \\
b_4 & 0 & 0 & 0 & 1 & 1 & 0 & 1 & 0 & 1 & 1 & 1 & 1 & -1 & 1
\end{array}
\right) \\ \\
\downarrow \\ \\
G_{\text{matroid}}=\left(
\begin{array}{c|cccccccccc}
& \ \pi_1 \ & \ \pi_2 \ & \ \pi_3 \ & \ \pi_4 \ & \ \pi_5 \ & \ \pi_6 \ & \ \pi_7 \ & \ \pi_8 \ & \ \pi_9 \ & \pi_{10} \\ \hline
b_1 & 0 & -1 & 0 & -1 & 0 & -1 & 0 & -1 & -1 & -1 \\
b_2 & 0 & 0 & 0 & -1 & 1 & -1 & 0 & 0 & 0 & -1 \\
b_3 & 0 & 0 & 0 & -1 & 1 & 0 & 1 & 0 & -1 & 0 \\
b_4 & 0 & 0 & 1 & 0 & 1 & 0 & 1 & 1 & -1 & 1 \\ \hline
 & \ \ {\bf 1} \ \ & \ \ {\bf 2} \ \ & \ \ {\bf 2} \ \ & \ \ {\bf 1} \ \ & \ \ {\bf 1} \ \ & \ \ {\bf 1} \ \ & \ \ {\bf 1} \ \ & \ \ {\bf 3} \ \ & \ \ {\bf 1} \ \ & \ \ {\bf 1} \ \
\end{array}
\right)
\end{array}
\label{G_matroid_G25_from_loops}
\eeq}
where the $\pi_i$ are the vertices obtained by only keeping the $b$ coordinates. This is precisely the matroid polytope given in \eref{G_matroid_G25_from_matroids}, after projecting out a redundant dimension.

Flows provide an alternative perspective on the emergence of the non-trivial multiplicities for points in the matroid polytope. Such multiplicities arise because paths that coincide on external legs but differ in the interior of the graph are projected down to the same point after eliminating the extra coordinates. 

A corollary of the discussion in this section is that the matroid polytope encodes the connectivity between external legs in a perfect orientation, i.e.\ it specifies which entries in the boundary measurement are non-zero.

\bigskip

\subsubsection{Perfect Matchings and \pl Coordinates}

\label{section_Plucker_coordinates}

In \sref{section_polytopes_from_matroids} we observed that different perfect matchings can give rise to perfect orientations with the same source set and hence provide multiple ``contributions" to a given matroid element. This phenomenon manifests as non-trivial multiplicities for points in the matroid polytope. We are now ready to explain in what sense these objects contribute to the same matroid element in more detail. 

Matroid elements $\{i_1 \ldots i_k\}$ are in one-to-one correspondence with \pl coordinates $\Delta_{i_1\ldots i_k}$ which, in turn, are given by minors of the boundary measurement matrix. All flows associated to a given point in the matroid polytope contribute to the same entries in the boundary measurement matrix. As a result, every perfect matching is mapped to a specific \pl coordinate \cite{2006math09764P,2007arXiv0706.2501P,2008arXiv0801.4822T,Amariti:2013ija}. In summary, each point in the matroid polytope is associated with a single \pl coordinate, but may get contributions from multiple perfect matchings.

For the example in this section, \eref{G_matroid_G25} implies the following relation between perfect matchings and \pl coordinates:

\beq
\begin{array}{|c|c|cc|cc|c|c|c|c|ccc|c|c|} 
\hline
\ \mbox{ \pl coordinate } \ & \ \Delta_{45} \ & \multicolumn{2}{c|}{\Delta_{14}} & \multicolumn{2}{c|}{\Delta_{35}} & \ \Delta_{34} \ & \ \Delta_{15} \ & \ \Delta_{24} \ & \ \Delta_{25} \ & \multicolumn{3}{c|}{\Delta_{13}} & \ \Delta_{12} \ & \ \Delta_{23} \ \\ \hline
\mbox{ PM } & \ p_1 \ & \ p_2 \ & \ p_3 \ & \ p_4 \ & \ p_5 \ & \ p_6 \ & \ p_7 \ & \ p_8 \ & \ p_9 \ & p_{10} & p_{11} & p_{12} & p_{13} & p_{14} \\ \hline
\end{array}
\nonumber
\eeq

\bigskip

\subsection{Polytopes from BFTs}

\label{section_polytopes_from_BFTs}

Interpreting bipartite graphs in terms of the corresponding BFTs, the matching and matroid polytopes become two very natural geometries for a quantum field theorist. With the goal of obtaining these geometries, we focus on classical Abelian BFTs. By this we mean BFTs in which all symmetry groups are $U(1)$ and gauge couplings are fixed and finite, with no quantum RG running.\footnote{A full investigation of the quantum behavior of BFTs with general ranks is certainly a well-motivated and interesting problem, but it is beyond the focus of this article.} Our discussion will be brief, and we refer the reader to 
\cite{Franco:2012mm,Franco:2012wv} for a detailed presentation.

\bigskip

\subsubsection{The Matching Polytope from the Master Space}

The {\it master space} of 4d $\mathcal{N}=1$ is defined as the space of solutions to vanishing F-term equations  \cite{Forcella:2008bb}. The special structure of BFT superpotentials, which are determined by bipartite graphs, reduces the determination of the master space to a combinatorial problem. F-terms automatically vanish with the following change of variables

\beq
X_i=\prod_\mu p_\mu^{P_{i \mu}} ,
\label{map_X_from_pm}
\eeq
where $X_i$ are the scalar components of chiral multiplets associated to edges, $p_\mu$ are new fields that are in one-to-one correspondence with perfect matchings and $P$ is the perfect matching matrix defined in \eref{P_matrix}.\footnote{It is important to emphasize the difference between \eref{map_X_from_pm} and the definition of oriented perfect matchings introduced in \sref{orientedge}, which are given by $\tilde{p}_\mu =\prod_i \alpha_i^{P_{i \mu}}$. While edge weights are naturally interpreted as products of perfect matchings for solving F-term equations, oriented perfect matchings should be thought as the product of oriented edge weights. In both cases, the object controlling the map is the $P$ matrix. Avoiding inconsistencies associated with this subtle difference was one of the main reasons for introducing the concepts of oriented perfect matchings and edge weights.} Perfect matchings can thus be interpreted as GLSM fields parametrizing the master space. The master space of a BFT is toric CY manifold whose toric diagram is the matching polytope \cite{Franco:2012mm}. The positions of perfect matchings in the matching polytope encode linear relations between the $p_\mu$ variables associated to F-term equations.

\bigskip

\subsubsection{The Matroid Polytope from the Moduli Space}

The moduli space of the BFT is obtained from its master space, by further demanding vanishing of D-terms. In order to do so, it is necessary assign charges under all $U(1)$ gauge groups to the $p_\mu$ fields.  These charges are deduced from those of the edge fields via the map \eref{map_X_from_pm}. For every $U(1)^{(\alpha)}$ factor of the gauge group and every edge chiral multiplet $X_i$ associated to an edge, we have:
\beq
Q^{(\alpha)}(X_i)=\sum_{\mu=1}^c  P_{i\mu} \, Q^{(\alpha)}(p_\mu).
\eeq
This set of equations can be used to determine an assignation of $Q^{(\alpha)}(p_\mu)$ charges. Since the system is not invertible, the resulting charges are generically not uniquely determined. The moduli space is however independent of the chosen solution.  It is obtained by projecting the master space on the space of gauge invariants. The moduli space is also a toric CY manifold and its toric diagram is obtained from the one of the master space by projecting it onto the null space of the matrix of gauge charges of the perfect matchings.

The previous discussion holds in general. However, the specific toric CY obtained as a result depends on whether the BFT is defined with gauging 1 or 2. When computed in gauging 2, the toric diagram of the moduli space is the matroid polytope \cite{Franco:2012wv}.\footnote{The BFTs resulting from gauging 1 and the associated moduli spaces are interesting in their own right. Given the questions we want to address in this paper, we will strictly focus on gauging 2.}

Making contact with the discussion in \sref{section_polytopes_from_flows} in terms of a geometric description of flows, eliminating a coordinate is physically achieved by gauging the corresponding $U(1)$ symmetry group in the BFT context. Gauging 2, the maximal gauging, corresponds to keeping only the $b_{i,\mu}$ coordinates.

\bigskip

\subsection{A Fast Algorithm for Finding the Matroid Polytope}
\label{Fast_Matroid}

Here we introduce a practical implementation of the ideas in previous sections leading to an efficient algorithm for the determination of the matroid polytope of a bipartite graph.

There exists a one-to-one correspondence between external faces and external legs in a bipartite graph. This correspondence underlies the identification of flow connectivity in terms of external faces of \sref{section_matching_from_flows}. Without loss of generality, in the case of a single boundary, every external face can be traded by the external leg separating it from the consecutive external face when going around the boundary clockwise. It is straightforward to extend this map to graphs with multiple boundaries.

In analogy to the matching polytope, this correspondence implies the matroid polytope is given by a {\it reduced perfect matching matrix}, with columns given by perfect matchings but rows only associated to external legs. Denoting external edges by $X^{(e)}_i$ and perfect matchings by $p_\mu$, we have:

\beq
G_{\text{matroid},i\mu}=\left\{ \begin{array}{ccccc} 1 & \rm{ if } & X^{(e)}_i  & \in & p_\mu \\
0 & \rm{ if } & X^{(e)}_i  & \notin & p_\mu
\end{array}\right. .
\label{reduced_P_matrix}
\eeq
This method for determining matroid polytopes is almost identical and trivially related to the one given by \eref{matroid_polytope}, based on perfect orientations. In our opinion, \eref{reduced_P_matrix} is even simpler to implement computationally, since it is written directly in terms of perfect matchings, which can be straightforwardly found via reduced Kasteleyn matrices.

\bigskip

\section{Graph Equivalence and Reduction}

\label{section_graph_equivalence_and_reduction}
 
In this section we introduce the notions of graph {\it equivalence} and {\it reducibility}, which concern the possibility of using different graphs for describing the same element in the Grassmannian.

\bigskip

\paragraph{{\bf Equivalence.}} Two graphs are {\it equivalent} if they have the same matroid polytope, modulo $SL$ transformations and multiplicities. Following \sref{section_graphs_polytopes_geometry}, equivalent graphs cover the same regions of the Grassmannian. They lead to the same set of generically non-zero entries in the boundary measurement, and to the same set of non-zero \pl coordinates. This notion of equivalence is also well-motivated in the BFT interpretation, since it implies that the corresponding theories have the same moduli space.\footnote{In the non-Abelian case the equality of moduli spaces is a necessary condition for two theories to be Seiberg dual \cite{Seiberg:1994pq,Feng:2000mi,Feng:2001xr,Feng:2001bn,Beasley:2001zp,Feng:2002zw}. Strictly speaking, the duality does not exist for Abelian theories, to which we restrict in this paper, since the theories are not asymptotically free. The matching of moduli spaces is however a well-defined mathematical question regarding natural geometric objects in the field theory.} Integrating out 2-valent nodes, square moves and bubble reductions lead to equivalent theories. We refer the reader to \cite{Franco:2012mm} for a detailed description of these graph transformations. In some cases, edge removal can also lead to equivalent theories. In the specific case of planar bipartite graphs, there is a one-to-one correspondence between equivalence classes of graphs and positroid cells of the Grassmannian.

\bigskip
 
\paragraph{{\bf Reducibility.}} A graph is {\it reduced} or {\it irreducible} if it has the minimum number of independent closed paths within a given equivalence class.\footnote{The notion of independent closed paths generalizes the one of internal faces, which is typically used for planar graphs.} Being defined up to equivalence transformations, reduced graphs are clearly not unique. More practically, a graph is reducible if it is possible to remove edges without changing its matroid polytope, modulo multiplicities.\footnote{We will assume this definition is equivalent to the one of irreducible graphs. This assumes that all reductions can be implemented by edge removals. It would be interesting to prove rigorously that this is the most general type of reduction, i.e.\ including those associated to bubble reductions and excluding any other exotic possibility.}

There are various alternative interpretations of graph reducibility. From the perspective of \sref{section_polytopes_from_flows} we see that, given a perfect orientation, reducibility translates into {\it redundant connectivity} between external legs of a graph. A graph is reducible if it is possible to remove edges, which results in the disruption of some oriented paths, such that every originally connected pair of external nodes remains so after the removal. Following \sref{section_Plucker_coordinates}, reducibility can also be thought of as the ability to eliminate edges of the graph while keeping contributions to all \pl coordinates, i.e.\ without setting any of them to zero.  

Roughly speaking, reduced graphs possess the minimal amount of structure necessary for describing the elements in the Grassmannian associated to the corresponding equivalence class.

\bigskip

\section{Quantifying Graph Reducibility}
\label{sec:QuantGraphRed}

Heuristically, the more flows connecting external nodes that exist, the more likely connectivity is preserved after removing an edge. In other words, the degree of reducibility of a graph is correlated with the multiplicities of perfect matchings associated to the same points in the matroid polytope. These multiplicities can thus be used as indicators of (relative) reducibility.\footnote{These multiplicities have been extensively studied for dimer models, i.e.\ bipartite graphs on a 2-torus, particularly in relation to Seiberg duality in the corresponding BFTs, see e.g.\ \cite{Feng:2000mi,Feng:2001xr,Feng:2002zw,Hanany:2005ve}.}

It is important to emphasize that multiplicities greater than one do not imply that a graph is reducible. An efficient method for addressing this question will be introduced in \sref{section_efficient_reducibility}.

\begin{figure}[h]
\begin{center}
\includegraphics[width=6.5cm]{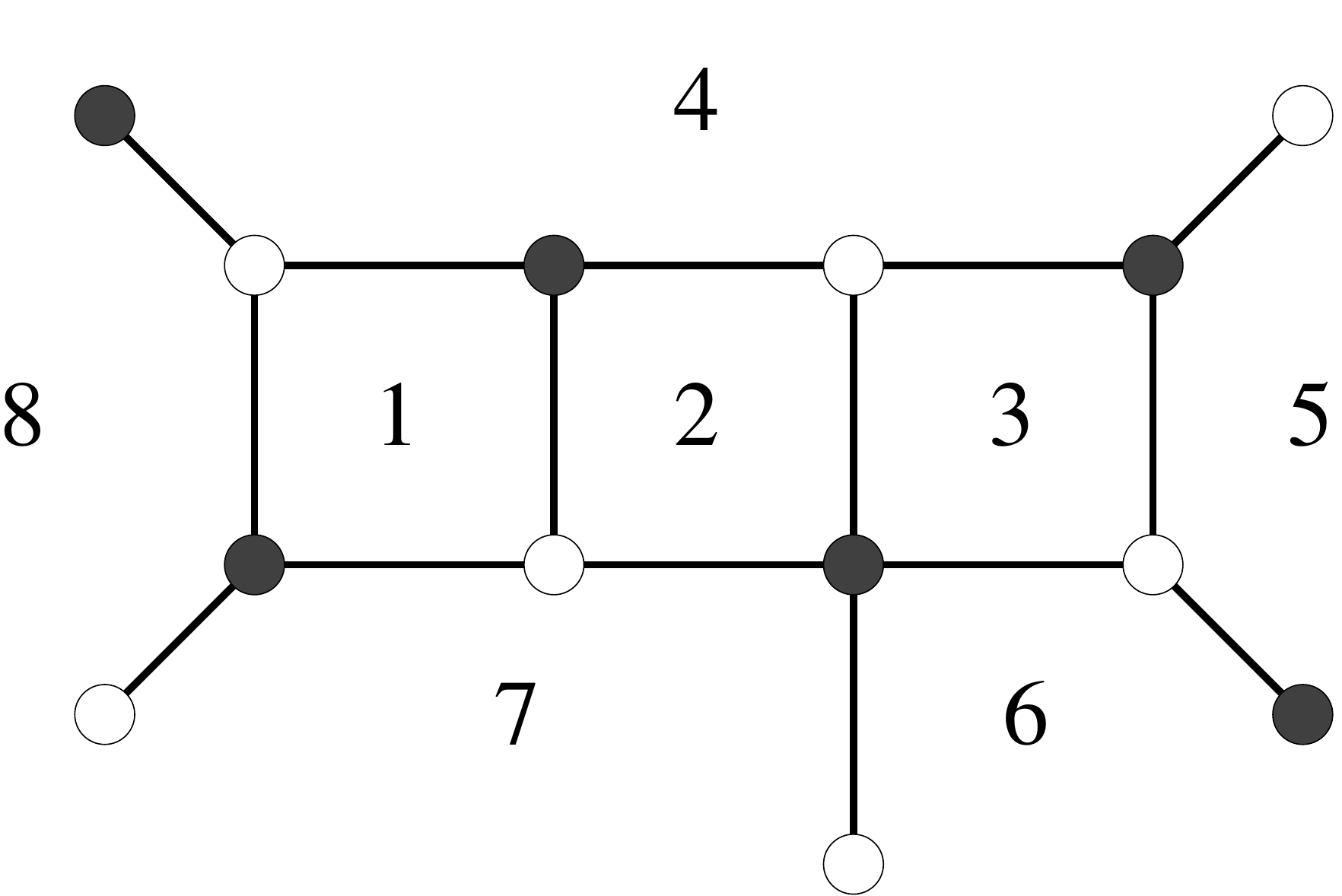}
\caption{A reducible bipartite graph corresponding to the top-dimensional cell of $Gr_{2,5}$.}
\label{G25_reducible}
\end{center}
\end{figure}

In order to illustrate these ideas, let us consider the graph in \fref{G25_reducible}, which is related to \fref{G25_graph} by reduction. The matroid polytope is given by the following matrix:

{\footnotesize
\beq
G_{\text{matroid}}=\left(
\begin{array}{cccccccccc}
 0 & 1 & 0 & 0 & 1 & 0 & 1 & 1 & 0 & 0 \\
 0 & 0 & 1 & 0 & 0 & 0 & 0 & -1 & 1 & 1 \\
 0 & 0 & 0 & 1 & 1 & 0 & 0 & 0 & 1 & 1 \\
 0 & 0 & 0 & 0 & 0 & 1 & 1 & 1 & -1 & 0 \\
 0 & 0 & 0 & 0 & 0 & 0 & 0 & 0 & 0 & 0 \\ \hline
\ \ {\bf 5} \ \ & \ \ {\bf 3} \ \ & \ \ {\bf 2} \ \ & \ \ {\bf 3} \ \ & \ \ {\bf 2} \ \ & \ \ {\bf 1} \ \ & \ \ {\bf 1} \ \ & \ \ {\bf 1} \ \ & \ \ {\bf 1} \ \ & \ \ {\bf 1} \ \ 
\end{array}
\right) .
\eeq}
This polytope coincides with the one given by \eref{G_matroid_G25_from_loops}, but the new graph has 20 perfect matchings and multiplicities are hence increased. As explained, this is a manifestation of the redundant connectivity associated to reducibility. Similar examples were presented in \cite{Franco:2012mm}.

\bigskip

\subsection{An Efficient Approach to Reducibility}\label{Algoredu}

\label{section_efficient_reducibility}

Determining whether a bipartite graph is reduced is an important question for various applications. For planar graphs, there is a combinatorial diagnostic for reducibility based on zig-zag paths (see e.g.\ \cite{ArkaniHamed:2012nw} and references therein). Determining zig-zags and their properties can however be rather impractical. Furthermore, whether and how this method generalizes to non-planar graphs is currently unknown. In this section we introduce an alternative test for reducibility with two salient features: it is straightforward to implement and it applies to both planar and non-planar graphs.

The discussion in \sref{section_graph_equivalence_and_reduction} makes it clear that the matroid polytope is the central player for determining graph equivalence and hence reducibility, which can be formulated as follows:

\bigskip

{\it A graph is irreducible if it is impossible to remove any edge without deleting points in the matroid polytope, i.e.\ without at least one perfect matching surviving for each of them.} 

\bigskip

This approach, originally advocated in \cite{Franco:2012mm}, leads to a practical procedure for determining whether a graph is irreducible. 

\begin{itemize}
\item[{\bf 1)}] Define $E_\alpha$ to be the set of edges that are present in {\it all} perfect matchings corresponding to a point $\alpha$ in the matroid polytope, $\alpha=1,\ldots, n_p$.
\item[{\bf 2)}] Combine them to form the set of edges that cannot be deleted $E_{\rm{und}}=\cup_\alpha E_\alpha$. In particular, $E_{\rm{und}}$ contains all edges in perfect matchings associated to multiplicity one points in the matroid polytope. 
\end{itemize}

Then, graph is reduced if and only if $E_{\rm{und}}$ is equal to the set $E_{\rm{tot}}$ of all edges in the graph. If $E_{\rm{und}} \nsubseteq E_{\rm{tot}}$, removing any single edge in $E_{\rm{tot}} - E_{\rm{und}}$ results in a reduction of the graph. Notice however that, in general, it is not possible to simultaneously remove more than one edge $E_{\rm{tot}} - E_{\rm{und}}$ without eliminating points from the matroid polytope.

\bigskip

\subsubsection*{Matrix Implementation}

The previous procedure can be nicely implemented in matrix language. Let us consider the perfect matching matrix $P$ in terms of which, as seen in \eref{map_X_from_pm}, edge removal is very transparent. When an edge $X_i$ is deleted, the perfect matchings $p_\mu$ for which $P_{i\mu}=1$ disappear.

Our main goal is to identify which edges, if any, can be deleted while keeping at least one perfect matching per point in the matroid polytope. For this purpose, it is natural to define a new matrix $ \mathscr{P}$, by multiplying the entries of $P$ associated to each point $\pi_\alpha$ in the matroid polytope as follows:

\begin{equation}
\mathscr{P}_{i \alpha} \equiv \prod_{p_\mu \in \pi_\alpha} P_{i \mu} .
\label{P_ialpha}
\end{equation} 
This results in a new $m \times n_p$ matrix $\mathscr{P}$, where $m$ is the number of edges, as it is for $P$, and $n_p$ is the number of distinct points in the matroid polytope. 

A vanishing entry $\mathscr{P}_{i \alpha}=0$ implies that removal of the edge $X_i$ preserves the point $\alpha$ in the matroid polytope, albeit not necessarily its multiplicity. Similarly $\mathscr{P}_{i \alpha}=1$ signifies that the removal of $X_i$ kills all perfect matchings at point $\pi_\alpha$. The construction of $\mathscr{P}$ is very efficient given $P$ and immediately displays the reducibility of a graph: if $\mathscr{P}$ has a row of zeroes, the graph is reducible since it is possible to remove the corresponding edge while preserving all points in the matroid polytope. 

Let us illustrate this construction for the example in \fref{G25_reducible}, for which we obtain

{\footnotesize
\beq
P=\left(
\begin{array}{l|ccccc|ccc|cc|ccc|cc|c|c|c|c|c}
& \multicolumn{5}{c|}{\pi_1} & \multicolumn{3}{c|}{\pi_2} & \multicolumn{2}{c|}{\pi_3} & \multicolumn{3}{c|}{\pi_4} & \multicolumn{2}{c|}{\pi_5} & \multicolumn{1}{c|}{\pi_6} & \multicolumn{1}{c|}{\pi_7} & \multicolumn{1}{c|}{\pi_8} & \multicolumn{1}{c|}{\pi_9} & \multicolumn{1}{c}{\pi_{10}} \\ \hline
& \ p_{1} \ & \ p_{2} \ & \ p_{3} \ & \ p_{4} \ & \ p_{5} \ & \ p_{6} \ & \ p_{7} \ & \ p_{8} \ & \ p_{9} \ & p_{10} & p_{11} & p_{12} & p_{13} & p_{14} & p_{15} & p_{16} & p_{17} & p_{18} & p_{19} & p_{20} \\ \hline
X_{1,2} \ \ & 0 & 1 & 1 & 0 & 0 & 0 & 1 & 0 & 1 & 0 & 0 & 1 & 1 & 0 & 1 & 0 & 0 & 0 & 0 & 1 \\
X_{1,8} & 1 & 1 & 1 & 0 & 0 & 1 & 1 & 0 & 1 & 0 & 0 & 0 & 0 & 0 & 0 & 0 & 0 & 0 & 0 & 0 \\
X_{2,4} & 1 & 0 & 0 & 0 & 0 & 1 & 0 & 0 & 0 & 0 & 1 & 0 & 0 & 1 & 0 & 1 & 1 & 1 & 0 & 0 \\
X_{2,7} & 1 & 0 & 0 & 0 & 0 & 1 & 0 & 0 & 0 & 0 & 1 & 0 & 0 & 1 & 0 & 0 & 0 & 0 & 1 & 0 \\
X_{3,2} & 0 & 1 & 0 & 1 & 0 & 0 & 1 & 1 & 0 & 0 & 0 & 1 & 0 & 0 & 1 & 0 & 0 & 0 & 0 & 0 \\
X_{3,5} & 1 & 1 & 0 & 1 & 0 & 0 & 0 & 0 & 0 & 0 & 1 & 1 & 0 & 0 & 0 & 1 & 0 & 0 & 0 & 0 \\
X_{4,1} & 0 & 0 & 0 & 1 & 1 & 0 & 0 & 1 & 0 & 1 & 0 & 0 & 0 & 0 & 0 & 0 & 0 & 0 & 1 & 0 \\
X_{4,3} & 0 & 0 & 1 & 0 & 1 & 0 & 0 & 0 & 1 & 1 & 0 & 0 & 1 & 0 & 0 & 0 & 0 & 0 & 1 & 1 \\
X_{6,3} & 0 & 0 & 1 & 0 & 1 & 0 & 0 & 0 & 0 & 0 & 0 & 0 & 1 & 0 & 0 & 0 & 0 & 1 & 0 & 0 \\
X_{7,1} & 0 & 0 & 0 & 1 & 1 & 0 & 0 & 1 & 0 & 1 & 0 & 0 & 0 & 0 & 0 & 1 & 1 & 1 & 0 & 0 \\
 \hline
X_{5,4} & 0 & 0 & 0 & 0 & 0 & 1 & 1 & 1 & 0 & 0 & 0 & 0 & 0 & 1 & 1 & 0 & 1 & 1 & 0 & 0 \\
X_{5,6} & 0 & 0 & 0 & 0 & 0 & 1 & 1 & 1 & 1 & 1 & 0 & 0 & 0 & 1 & 1 & 0 & 1 & 0 & 1 & 1 \\
X_{7,6} & 0 & 0 & 0 & 0 & 0 & 0 & 0 & 0 & 1 & 1 & 0 & 0 & 0 & 0 & 0 & 1 & 1 & 0 & 0 & 1 \\
X_{8,4} & 0 & 0 & 0 & 0 & 0 & 0 & 0 & 0 & 0 & 0 & 1 & 1 & 1 & 1 & 1 & 1 & 1 & 1 & 0 & 1 \\
X_{8,7} & 0 & 0 & 0 & 0 & 0 & 0 & 0 & 0 & 0 & 0 & 1 & 1 & 1 & 1 & 1 & 0 & 0 & 0 & 1 & 1 \\
\end{array}
\right)
\eeq}
where we have grouped the columns associated to perfect matchings that sit on the same point of the matroid polytope. The horizontal line separates internal edges from external legs.\footnote{This organization of rows and columns in $P$ is not obligatory, but it is convenient for simplifying our analysis.} Using \eref{P_ialpha}, we obtain:
{\footnotesize
\beq
\mathscr{P}=\left(
\begin{array}{l|cccccccccc}
& \ \pi_1 \ & \ \pi_2 \ & \ \pi_3 \ & \ \pi_4 \ & \ \pi_5 \ & \ \pi_6 \ & \ \pi_7 \ & \ \pi_8 \ & \ \pi_9 \ & \pi_{10} \\ \hline
X_{1,2} \ \ & 0 & 0 & 0 & 0 & 0 & 0 & 0 & 0 & 0 & 1 \\
\rowcolor{cyan!90!blue!60} X_{18} & 0 & 0 & 0 & 0 & 0 & 0 & 0 & 0 & 0 & 0 \\
X_{2,4} & 0 & 0 & 0 & 0 & 0 & 1 & 1 & 1 & 0 & 0 \\
X_{2,7} & 0 & 0 & 0 & 0 & 0 & 0 & 0 & 0 & 1 & 0 \\
\rowcolor{cyan!90!blue!60} X_{32} & 0 & 0 & 0 & 0 & 0 & 0 & 0 & 0 & 0 & 0 \\
X_{3,5} & 0 & 0 & 0 & 0 & 0 & 1 & 0 & 0 & 0 & 0 \\
X_{4,1} & 0 & 0 & 0 & 0 & 0 & 0 & 0 & 0 & 1 & 0 \\
X_{4,3} & 0 & 0 & 1 & 0 & 0 & 0 & 0 & 0 & 1 & 1 \\
X_{6,3} & 0 & 0 & 0 & 0 & 0 & 0 & 0 & 1 & 0 & 0 \\
X_{7,1} & 0 & 0 & 0 & 0 & 0 & 1 & 1 & 1 & 0 & 0 \\ \hline
X_{5,4} & 0 & 1 & 0 & 0 & 1 & 0 & 1 & 1 & 0 & 0 \\
X_{5,6} & 0 & 1 & 1 & 0 & 1 & 0 & 1 & 0 & 1 & 1 \\
X_{7,6} & 0 & 0 & 1 & 0 & 0 & 1 & 1 & 0 & 0 & 1 \\
X_{8,4} & 0 & 0 & 0 & 1 & 1 & 1 & 1 & 1 & 0 & 1 \\
X_{8,7} & 0 & 0 & 0 & 1 & 1 & 0 & 0 & 0 & 1 & 1 \\
\end{array}
\right)
\eeq}
\noindent This matrix contains rows of zeroes, so we conclude the graph is reducible. $X_{1,8}$ or $X_{3,2}$ can be removed without eliminating points from the matroid polytope.

Finally, we remark that $\mathscr{P}$ is also useful for finding those edges which, in the language of \cite{ArkaniHamed:2012nw}, are {\it removable edges}. Removable edges are defined as those which, starting from a reduced graph, yield a reduced graph after being removed.\footnote{It is important not to confuse these edges with the ones discussed in previous paragraphs, which are edges that can be removed from a reducible graph to produce an equivalent one.} In order to identify removable edges, we first generate a new perfect matching matrix $P'$ from $P$, by removing the corresponding row $k$ and every column $\mu$ for which $P_{k \mu}=1$. Next, we construct the corresponding $\mathscr{P}'$ matrix.  
Removable edges are those 
whose $\mathscr{P}'$ does not display reducibility. This procedure applies to general, not necessarily planar, graphs.

\bigskip

\section{Stratification: New Regions and New Methods}
\label{Strat_NewRegionNewMethod}

We have already had a glimpse that the connection between the Grassmannian and bipartite graphs provides interesting avenues for decomposing the former using the latter. In \sref{singularities} we discussed a decomposition of planar bipartite graphs which is of physical interest due to its connections to the singularity structure of scattering amplitudes. It can be summarized as follows:

\smallskip
\begin{itemize} 
\item[{\bf 1)}] Start from a reduced graph. 
\item[{\bf 2)}] Sequentially delete removable edges.
\end{itemize}
\smallskip

\noindent From a mathematical viewpoint such decomposition is interesting because, for planar graphs, it corresponds to the positroid stratification of the totally non-negative Grassmannian. Recall that the positroid stratification can also be regarded as the intersection between the matroid stratification and the totally non-negative Grassmannian. More generically, as we discuss in \sref{section_partial_matroid_stratification}, for arbitrary graphs the decomposition considered in this section can be regarded as a {\it partial matroid decomposition}, which we shall call the \textit{combinatorial decomposition}.

It is reasonable to only focus on reduced graphs, since it avoids the redundancies in the description of the Grassmannian associated to reducible graphs. It is natural to extend the decomposition defined by the two steps above to arbitrary reduced bipartite graphs and to investigate its implications. This will allows us to go beyond the positive regions of the Grassmannian, which are specific to the planar case. In analogy with the reasoning of \sref{singularities}, it is reasonable to expect that this decomposition is a natural candidate for capturing the singularity structure of on-shell diagrams beyond the planar limit.

\begin{figure}[h]
\begin{center}
\includegraphics[width=15cm]{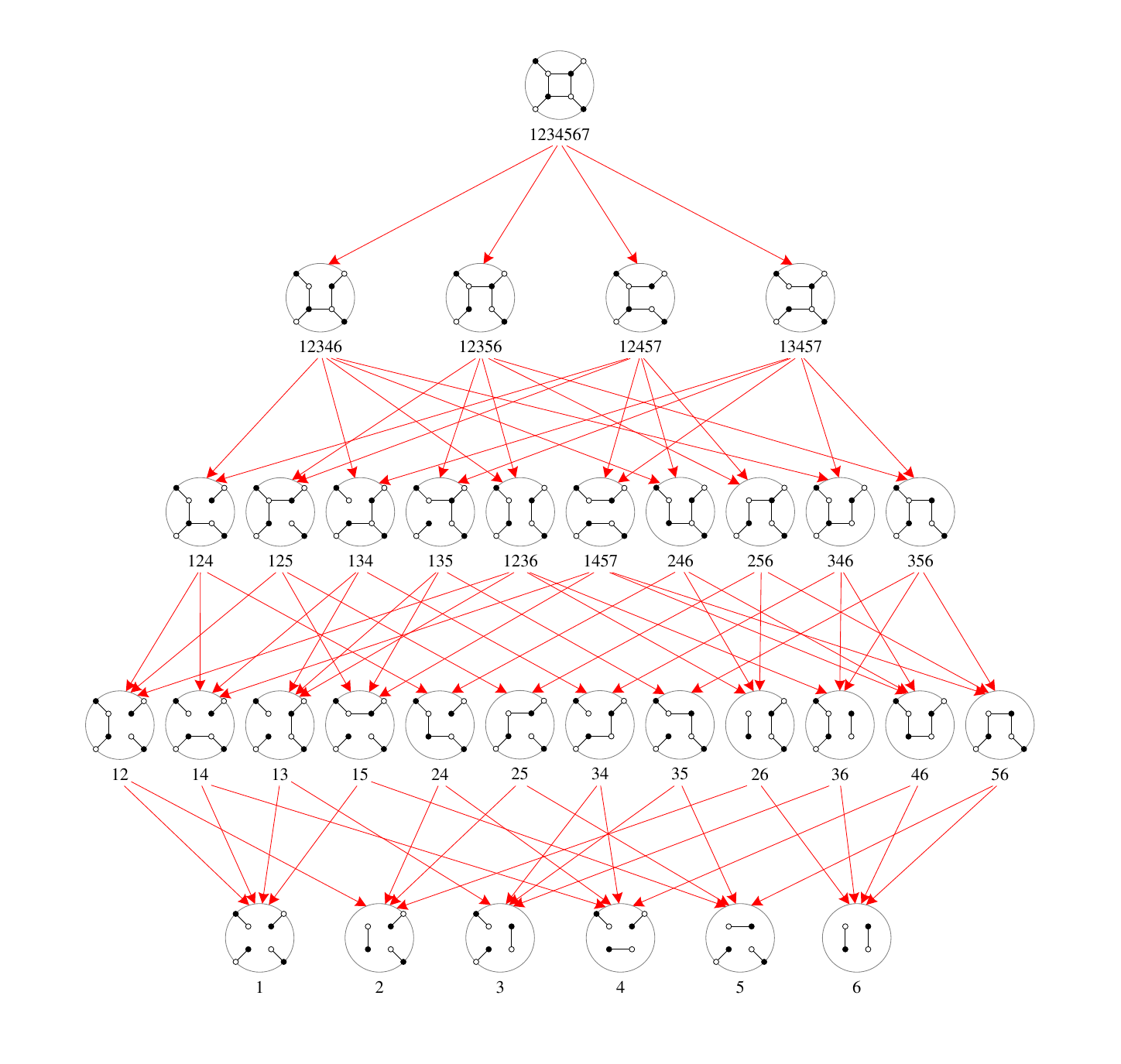}
\caption{Positroid decomposition of $Gr_{2,4}$. Each site corresponds to a positroid stratum, and we indicate the associated graph and surviving perfect matchings.}
\label{G24posi}
\end{center}
\end{figure}

The combinatorial decomposition can be nicely visualized in terms of a poset, in which every node corresponds to a reduced graph and arrows indicate the deletion of a removable edge. For planar graphs, every site in the poset corresponds to a positroid stratum, represented by a specific matroid stratum. \fref{G24posi} presents the simple example of the positroid decomposition of the top-cell of $Gr_{2,4}$, obtained by this procedure.\footnote{In the physics literature, this example has appeared in \cite{ArkaniHamed:2012nw}.} 

In the following, we will first apply our ideas to planar graphs, which are well-known to experts. In coming sections we will also consider the non-planar case, which deserves a detailed study of its own, since it remains relatively unexplored.
In practice, it is useful to exploit the algorithm in \sref{Algoredu} for identifying removable edges.

\bigskip

\subsection{Combinatorial Decomposition Via Polytopes}

\label{section_decomposition_polytopes}

In this section we introduce an alternative implementation of the combinatorial decomposition. It exploits the matroid and matching polytopes, making the connection to the Grassmannian more transparent. In addition, it does not rely on reducibility or removability.

\bigskip

\subsubsection{Step 1: Edge Removal}

The first step of the process corresponds to removing every possible edge of the graph, one at a time. The process terminates when the surviving graph coincides with a perfect matching of the original one, i.e.\ to a vertex in the matching polytope. Notice that {\it any} edge can be removed, i.e.\ there is no restriction to removable edges. The graphs generated by this procedure and their relations can be organized into an Eulerian poset, which is different from the poset discussed in the previous section.

Interestingly, for planar graphs, removing edges is equivalent to constructing the {\it face lattice}\footnote{In the face lattice we do not include the empty set.} of the matching polytope \cite{2007arXiv0706.2501P}. In the next sections we argue and provide evidence that this is also valid for non-planar bipartite graphs. Let us explain in more detail the structure of the poset for the matching polytope. Consider a matching polytope of dimension $d_{{\rm matching}}$. Its boundary has dimension equal to $(d_{{\rm matching}}-1)$ and is a union of facets. Each facet is defined as the intersection of the boundary with a $(d_{{\rm matching}}-1)$-dimensional hyperplane. In turn, each of these facets has a $(d_{{\rm matching}}-2)$-dimensional boundary, which can also be decomposed into faces, and so on. The face lattice of the matching polytope is generated by iterating the boundary operator until reaching 0-dimensional faces. 

In this approach, faces are directly determined from the positions of points in the matching polytope. Computer applications constructing the set of faces for arbitrary polytopes are publicly available, see e.g.\ Polymake \cite{polymake}. Contrary to the method based on removing edges, a single bipartite graph is only used at the initial step, for determining the matching polytope.

Let us consider the planar graph associated to the top-cell of $Gr_{2,4}$, which is shown in \fref{fig:sqbpo}. The matching polytope has seven different points corresponding to its perfect matchings and is given by the following perfect matching matrix

\be
P=
\left(
\begin{array}{c|ccccccc}
 & \ p_1 \ & \ p_2 \ & \ p_3 \ & \ p_4 \ & \ p_5 \ & \ p_6 \ & \ p_7 \ \\
\hline
 X_{1,2} & 0 & 1 & 0 & 0 & 0 & 1 & 0 \\
 X_{1,4} & 0 & 0 & 1 & 0 & 0 & 1 & 0 \\
 X_{3,1} & 0 & 0 & 0 & 0 & 1 & 0 & 1 \\
 X_{5,1} & 0 & 0 & 0 & 1 & 0 & 0 & 1 \\
 X_{2,3} & 1 & 0 & 1 & 1 & 0 & 0 & 0 \\
 X_{2,5} & 1 & 0 & 1 & 0 & 1 & 0 & 0 \\
 X_{4,5} & 1 & 1 & 0 & 0 & 1 & 0 & 0 \\
 X_{4,3} & 1 & 1 & 0 & 1 & 0 & 0 & 0 \\
\end{array}
\right).
\label{P_matrix_G24}
\ee
This matrix defines a 4d polytope. This becomes clearer by row-reducing it, after which we obtain

\beq
G_{\text{matching}} = \left(
\begin{array}{ccccccc}
\ p_1 \ & \ p_2 \ & \ p_3 \ & \ p_4 \ & \ p_5 \ & \ p_6 \ & \ p_7 \ \\
\hline
 1 & 0 & 0 & 0 & 0 & -1 & -1 \\
 0 & 1 & 0 & 0 & 0 & 1 & 0 \\
 0 & 0 & 1 & 0 & 0 & 1 & 0 \\
 0 & 0 & 0 & 1 & 0 & 0 & 1 \\
 0 & 0 & 0 & 0 & 1 & 0 & 1 \\
\end{array}
\right) .
\label{G_matching_G24}
\eeq

\smallskip

Let us briefly discuss the relation between edge removal and lower dimensional faces of the matching polytope. Recall that removing an edge $X_i$ results in eliminating the perfect matchings $p_\mu$ for which the corresponding entry $P_{i\mu}$ is equal to 1. In this example, we obtain eight different subgraphs at the first level, corresponding to eight 3d faces. We then continue removing additional edges, successively obtaining lower dimensional faces until reaching the vertices of the matching polytope, which correspond to the 7 perfect matchings. The resulting face lattice is shown in \fref{FullLatticeSpaced2}. The previous discussion was phrased in terms of edge deletions. As we explained, the face lattice can be determined directly, without referring to edge removals. 

\bigskip

\subsubsection{Step 2: Identification}

The final step in the combinatorial decomposition involves identifying perfect matchings associated to the same point in the matroid polytope, equivalently to the same \pl coordinate. This results in the identification, or more precisely merging, of nodes in the poset for the face lattice of the matching polytope we constructed in the previous section. 

The identification of perfect matchings can give rise to two qualitatively different types of identifications. We refer to them as {\it horizontal} and {\it vertical} identifications, following their effect on points on the poset. They are defined as follows:

\medskip

\begin{itemize}
\item{\bf Horizontal identifications:} they merge nodes in the poset that sit at the same level. Their effect on the matching polytope is to identify different faces without affecting their dimensionalities.

\item{\bf Vertical identifications:} from the viewpoint of the poset, they merge nodes at different levels. They identify different points in a given face of the matching polytope and result in a lower dimensional one.
\end{itemize}

\medskip

\begin{figure}[h]
\begin{center}
\includegraphics[width=11cm]{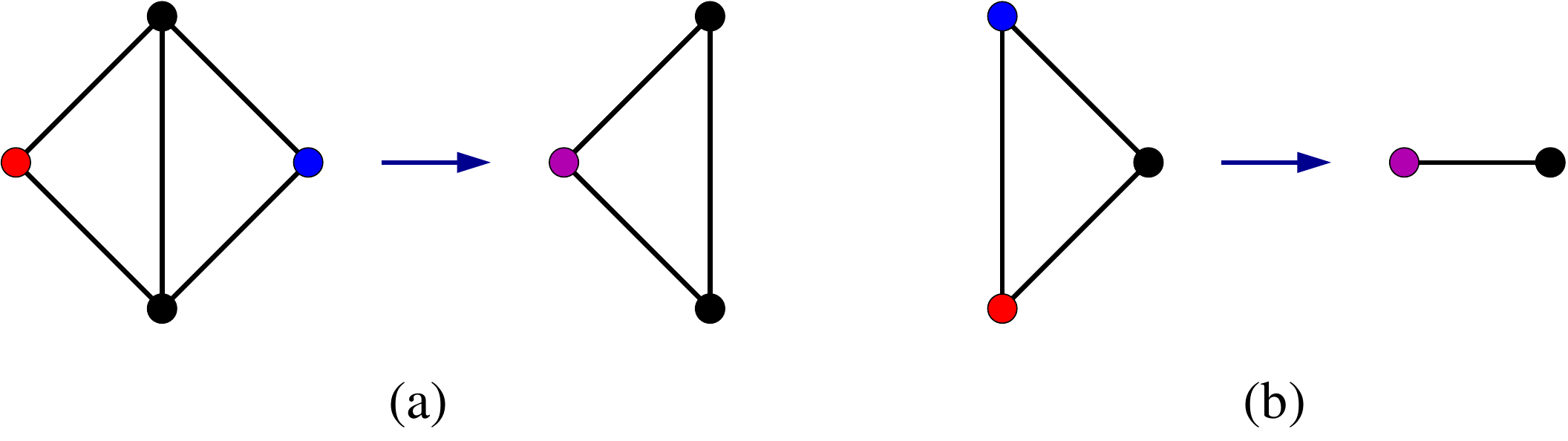}
\caption{Two types of identifications: a) horizontal and b) vertical. Here we show the action on points in the matching polytope. Points, i.e.\ perfect matchings, to be identified are shown in blue and red. Purple dots indicate the resulting points after identification.}
\label{horizontal_vertical_identifications}
\end{center}
\end{figure}

\fref{horizontal_vertical_identifications} shows simple examples of each class of identification at the level of the matching polytope. Generically, more than two perfect matchings can be simultaneously involved in identifications.

This approach to decomposition makes certain general properties of the final poset obtained after identifications rather clear. In particular:

\medskip

\begin{itemize}
\item The number of levels is equal to the dimensions of the matching polytope of a reduced graph in the equivalence class of the starting point plus one. This number is invariant under graph equivalence, and does not depend on the initial graph being reduced. 
\item The number of sites in the lowest level of the poset is equal to the number of points in the matroid polytope.
\end{itemize}

\medskip

\begin{figure}[h]
\hspace{-1.15cm}\includegraphics[width=18cm]{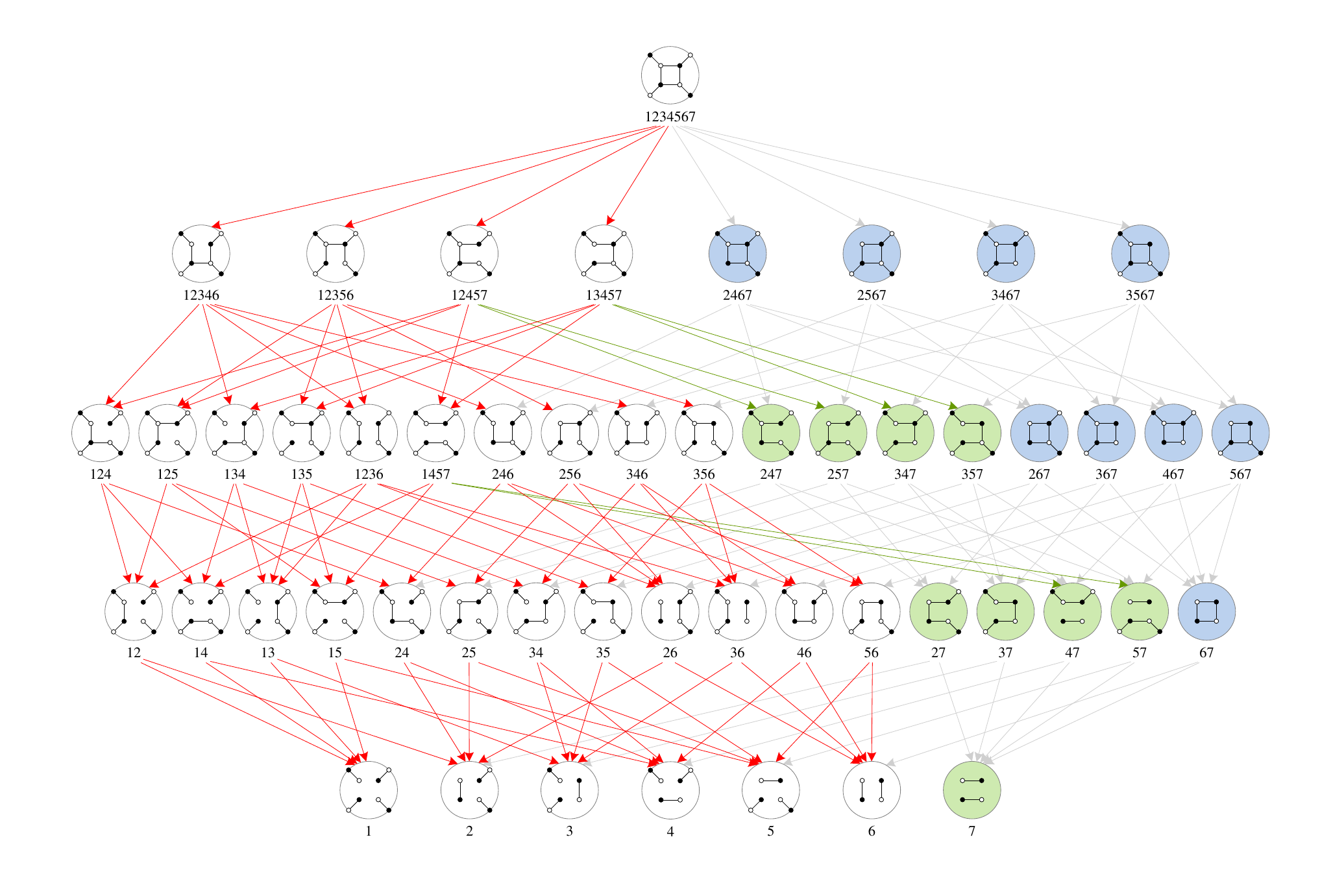}
\begin{center}
\caption{Face lattice of the matching polytope for the graph in \fref{fig:sqbpo}. At each point, we indicate the corresponding graph and the surviving perfect matchings. When $p_6$ and $p_7$ are identified, green and blue nodes in the poset are subject to horizontal and vertical identifications, respectively.}
\label{FullLatticeSpaced2}
\end{center}
\end{figure}

Returning to the $Gr_{2,4}$ example, the matroid polytope in this case is given by:

\beq
G_{\text{matroid}}=
\left(
\begin{array}{c|ccccccc}
 & \ p_1 \ & \ p_2 \ & \ p_3 \ & \ p_4 \ & \ p_5 \ & \ p_6 \ & \ p_7 \ \\
 \hline
 X_{2,3} & 1 & 0 & 1 & 1 & 0 & 0 & 0 \\
 X_{2,5} & 1 & 0 & 1 & 0 & 1 & 0 & 0 \\
 X_{4,5} & 1 & 1 & 0 & 0 & 1 & 0 & 0 \\
 X_{4,3} & 1 & 1 & 0 & 1 & 0 & 0 & 0 \\
\end{array}
\right).
\label{G_matroid_G24}
\eeq
The 7 perfect matchings are mapped to 6 points, with $p_6$ and $p_7$ becoming coincident. \fref{FullLatticeSpaced2} shows the face lattice for the matching polytope. Colored nodes need to be merged with some of the white ones, following the identification of $p_6$ and $p_7$: green and blue nodes are subject to horizontal and vertical identifications, respectively. White nodes correspond to the nodes in \fref{G24posi}. It is straightforward to verify that the entire structure of \fref{G24posi}, i.e.\ including its arrows, is recovered by the identifications.

\bigskip

\subsubsection{Reducible Starting Points}

It is important to stress that the combinatorial decomposition does not require irreducibility at any step. Not only restricting to removable edges, i.e.\ to reduced graphs at intermediate steps, is not necessary, but the starting point does not need to be a reduced graph. As we explained in \sref{sec:QuantGraphRed}, 
the redundancy in reducible graphs is accounted for by the identification of perfect matchings according to the matroid polytope.

\begin{figure}[h]
\begin{center}
\includegraphics[width=5cm]{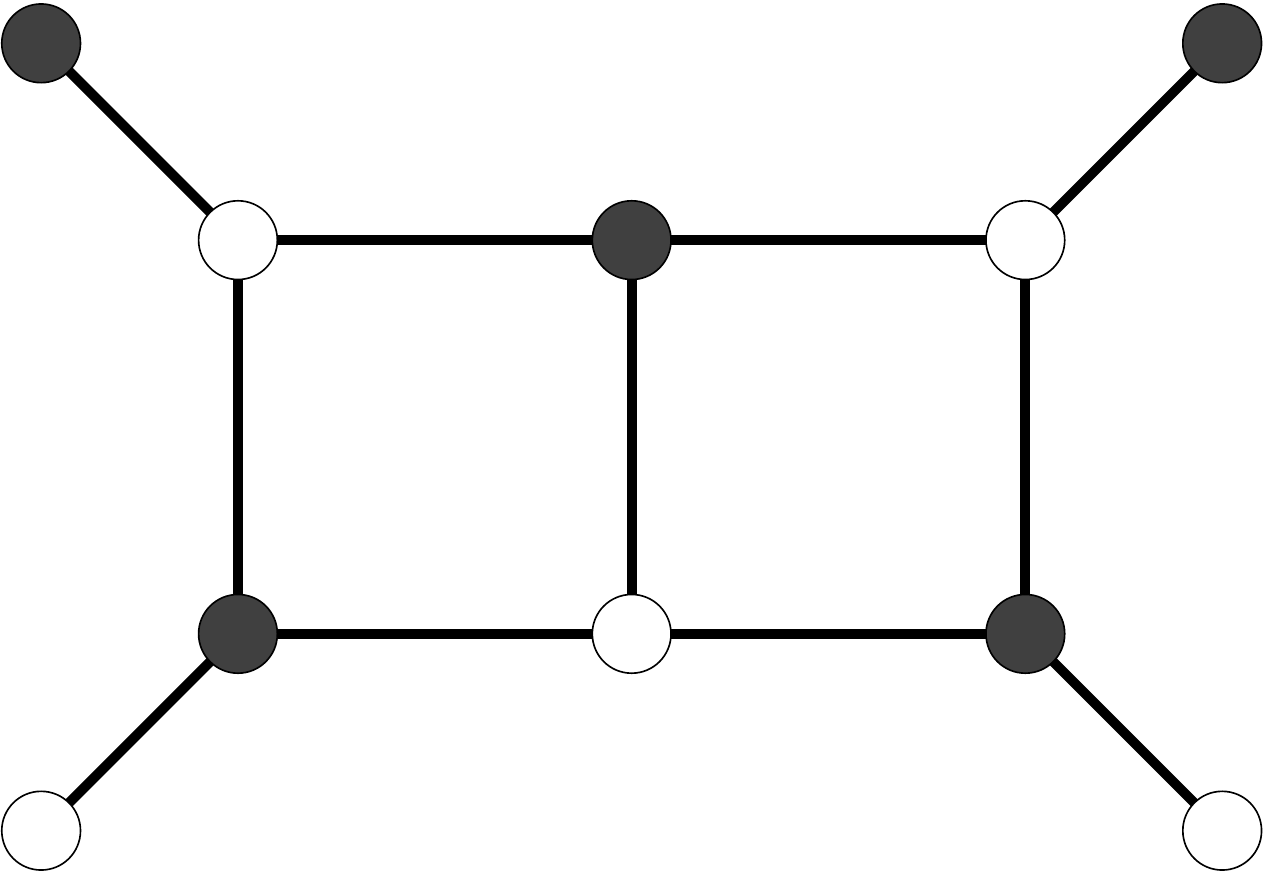}
\caption{A reducible graph for the top-cell of $Gr_{2,4}$.}
\label{G24_2_boxes}
\end{center}
\end{figure}

To see how things work in an explicit example, let us consider the reducible graph in \fref{G24_2_boxes} which is equivalent to the single square box graph studied in the previous sections, and corresponds to the top-dimensional cell of $Gr_{2,4}$. 

This graph has 10 perfect matchings, a relatively small increase with respect to the 7 perfect matchings of the single box graph. However, there is an explosion in the number of possibilities for removing edges. The corresponding poset is shown in \fref{DoubleSqbLatticeOrderedNicev6} of Appendix \ref{section_appendix_poset_double_box}.

The matching polytope is 5d. The difference in dimensions with respect to an equivalent reduced graph, which has a 4d matching polytope as in \eref{G_matching_G24}, is equal to the number of additional faces. This implies that, before identifications, the face lattice has an additional level.

The matroid polytope coincides with the one for the reduced graph given by \eref{G_matroid_G24}, but with larger multiplicities. Perfect matchings are identified as follows:

\beq
\begin{array}{ccc}
\{p_1,p_7,p_9\} & \ \ \ \ \ & \{p_4\} \\
\{p_2,p_8\} & & \{p_5\} \\
\{p_3,p_{10}\} & & \{p_6\}
\end{array}
\label{identifications_G24_2_boxes}
\eeq
from which we determine the horizontal and vertical identifications shown in \fref{DoubleSqbLatticeOrderedNicev6}. These identifications lead to a vast reduction of the poset. The result contains only the white sites in \fref{DoubleSqbLatticeOrderedNicev6} and agrees, once again, with \fref{G24posi}.

\bigskip

\subsubsection{Relation to the Matroid Stratification}

\label{section_partial_matroid_stratification}

In the previous section we introduced the combinatorial decomposition of a bipartite diagram and discuss different implementations.

Here we consider another natural decomposition we can relate to a bipartite graph, which is the matroid stratification of the associated Grassmannian element, and comment on their relations. The boundary measurement provides the necessary map between a graph and the Grassmannian. For planar graphs, we obtain the non-negative Grassmannian from non-negative edge weights. Explicit details of its generalization to non-planar graphs are given in \sref{section_boundary_measurement_non-planar}. In both cases perfect matchings can be mapped to \pl coordinates by referring to the source set specified by them, as already reviewed in \sref{section_graphs_polytopes_geometry}. Multiple perfect matchings can correspond to the same \pl coordinate, which is associated to a point of the matroid polytope. This prescription results in a map
\be
\label{slickmap}
\Delta_{I}  \leftrightarrow \{ p^I_i \} ,
\ee
where $i$ runs over the multiplicity of the corresponding vertex in the matroid polytope. The map identifies the non-vanishing \pl coordinates of the element of the Grassmannian associated to a bipartite graph. Next, we can follow \sref{matroidstra} and construct the matroid stratification of this element of the Grassmannian.

For instance, let us return to the square box diagram in \fref{fig:sqbpo} and \fref{fig:sqbpms} for the top-cell of $Gr_{2,4}$. With the methods in \sref{section_graphs_polytopes_geometry}, we can easily obtain:
\be
\label{PluckerPM24}
\begin{array}{ccccccccccc}
\Delta_{24} & \leftrightarrow & \{p_6, p_7 \} & \ \ \ & 
\Delta_{34} & \leftrightarrow & \{p_2\} & \ \ \ &
\Delta_{12} & \leftrightarrow & \{p_3\} \\
\Delta_{14} & \leftrightarrow & \{p_4\} & \ \ \ &
\Delta_{23} & \leftrightarrow & \{p_5\} & \ \ \ &
\Delta_{13} & \leftrightarrow & \{p_1\} 
\end{array}
\ee
It is now possible to produce the matroid stratification, which is given by:
 \be
 \label{matroidstr}
 \begin{array}{|c|c|}
 \hline
 \ d=4 \ \ &\{12,13,14,23,24,34\}\\
 \hline
 \ d=3 \ \ & \{12,13,14,23,24\},
 {\color{red}{\{12,13,14,23,34\}}},\{12,13,14,24,34\}\\
 &\{12,13,23,24,34\}, {\color{red}{\{12,14,23,24,34\}}},\{13,14,23,24,34\}\\
\hline
 \ d=2 \ \
&
 \{12,13,14\},
 \{12,13,23\},
 \{12,14,24\},
 \{12,23,24\},
 \{13,14,34\},
 \{13,23,34\} \\
 &
 \{14,24,34\},
 \{23,24,34\}, 
 \{12,13,24,34\},
  {\color{red}{\{12,14,23,34\}}},
 \{13,14,23,24\} \\
  \hline
\ d=1 \ \ &
   \{12,13\},\{12,14\},\{12,23\},\{12,24\},
 \{13,14\}, 
 \{13,23\}, 
 \{13,34\}, 
 \{14,24\} \\
 &
 \{14,34\},
 \{23,24\},
 \{23,34\}, 
 \{24,34\} \\
 \hline
 \ d=0 \ \ &
  \{12\},\{13\},\{14\},\{23\},\{24\},\{34\} \\
 \hline
  \end{array}
 \ee
Note that we have used the \pl relation
\be
\label{PMPLagain}
\Delta_{12} \, \Delta_{34} +\Delta_{23} \, \Delta_{14}=\Delta_{13} \, \Delta_{24}
\ee
in order to recognize the dimension of each matroid stratum and to arrange it at the correct level.

We are now in a position to discuss the relation between the combinatorial decomposition and the matroid stratification. Components in the combinatorial decomposition are matroid strata, i.e.\ they are defined by specifying sets of non-vanishing \pl coordinates. However, generically not all matroid strata can be generated by removing edges from a fixed starting graph. The combinatorial decomposition can thus be regarded as a {\it partial matroid decomposition}. In \sref{section_matroid_graphs}, we speculate on possible ways to achieve the complete matroid stratification in terms of bipartite graphs. 

In practical terms, the combinatorial decomposition is given by the intersection between the matroid stratification and the lattice generated by all possible edge removals. For planar graphs, this reduction can be alternatively obtained by intersecting the matroid stratification with the totally non-negative Grassmannian, as explained in \sref{section_positroid_cells}.

The matroid interpretation of the polytope implementation in \sref{section_decomposition_polytopes} for the combinatorial decomposition is clear. The first step restricts the space of strata to those which are reachable by removing edges. The second step eliminates the redundancy in the description of these strata arising from equivalent graphs.

\begin{figure}[h]
\begin{center}
\includegraphics[width=15cm]{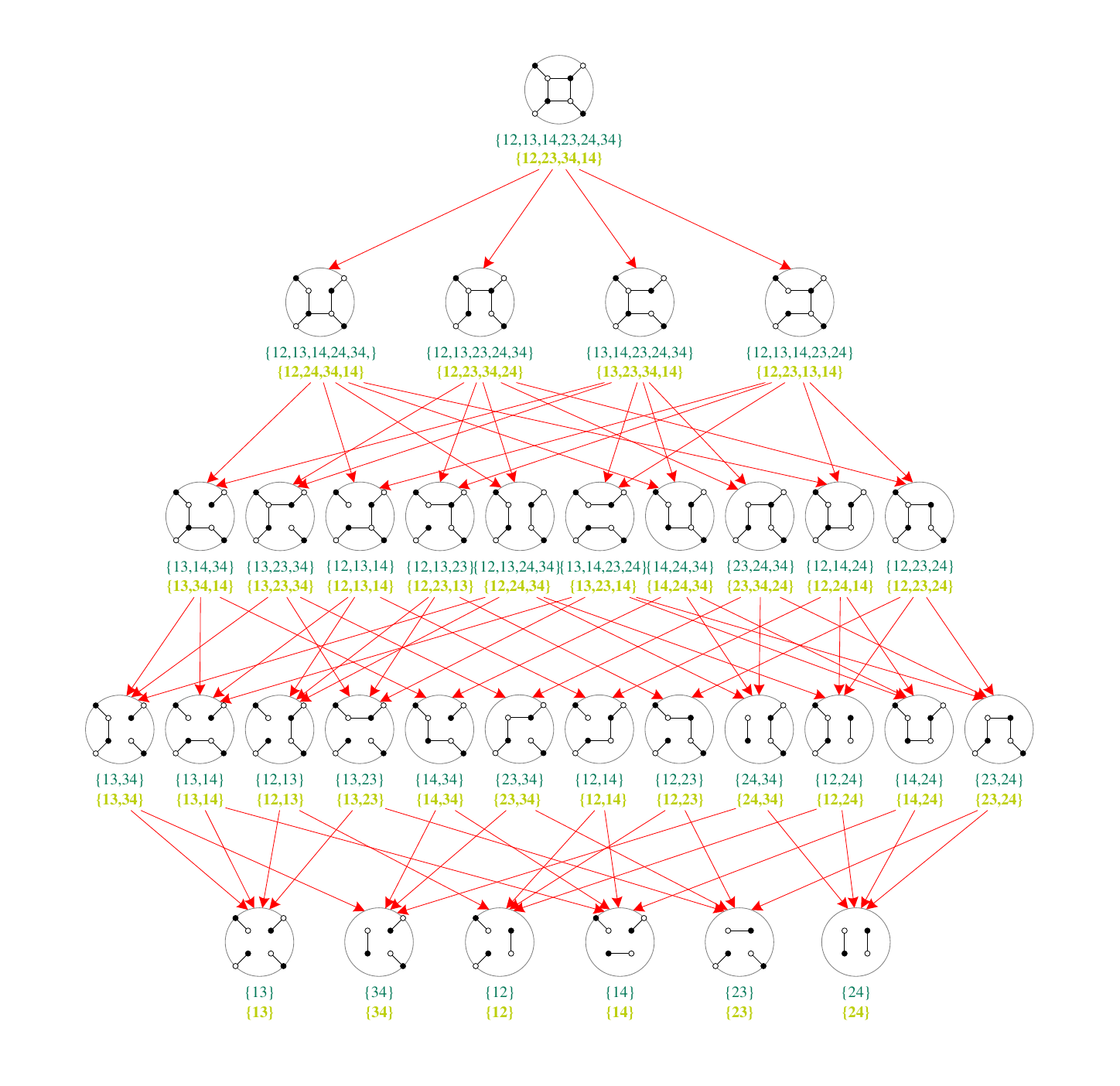}
\caption{Positroid stratification of $Gr_{2,4}$. Each graph maps to a matroid stratum whose matroid is indicated in dark green. The positroid stratum containing the matroid stratum is shown in light green. We see that all positroid strata are present, and no two graphs are in the same positroid stratum.}
\label{G24posiexplicit}
\end{center}
\end{figure}

Returning to the example, let us take \eref{matroidstr} and keep only objects appearing in \fref{FullLatticeSpaced2}. In order to do so, we use the map between perfect matchings and \pl coordinates given by \eref{PluckerPM24}. The strata indicated in red in \eref{matroidstr} disappear, and we are left with:
\be
 \begin{array}{|c|c|}
 \hline
 \ d=4 \ \ &\{12,13,14,23,24,34\}\\
 \hline
 \ d=3 \ \ & \{12,13,14,23,24\},
 \{12,13,14,24,34\},\\
 &\{12,13,23,24,34\},
 \{13,14,23,24,34\}\\
\hline
 \ d=2 \ \ 
 &
 \{12,13,14\},
 \{12,13,23\},
 \{12,14,24\},
 \{12,23,24\},
 \{13,14,34\},
 \{13,23,34\},\\
 &
 \{14,24,34\},
 \{23,24,34\}, 
 \{12,13,24,34\},
 \{13,14,23,24\},\\
  \hline
 \ d=1 \ \ &
   \{12,13\},\{12,14\},\{12,23\},\{12,24\},
 \{13,14\}, 
 \{13,23\}, 
 \{13,34\}, 
 \{14,24\}, \\
 &
 \{14,34\},
 \{23,24\},
 \{23,34\}, 
 \{24,34\},\\
 \hline
 \ d=0 \ \ &
  \{12\},\{13\},\{14\},\{23\},\{24\},\{34\},\\
 \hline
  \end{array}
 \ee
This is indeed the positroid stratification depicted in \fref{G24posiexplicit}, which is identical to \fref{G24posi}. For each graph we show its matroid labels (dark green) and its positroid labels (light green).

\bigskip

\section{Boundary Measurement for Non-Planar Graphs}

\label{section_boundary_measurement_non-planar}

In this section we extend the definition of the {\it boundary measurement} beyond the planar case. This is a crucial element necessary for extending the map between {\it general} bipartite graphs and the Grassmannian. The boundary measurement has been already defined for planar graphs \cite{2006math09764P} and the annulus \cite{2009arXiv0901.0020G}. Here we generalize it to the case of graphs on the plane with an arbitrary number of boundaries.

\begin{figure}[htb!]
\centering
\includegraphics[scale=0.65]{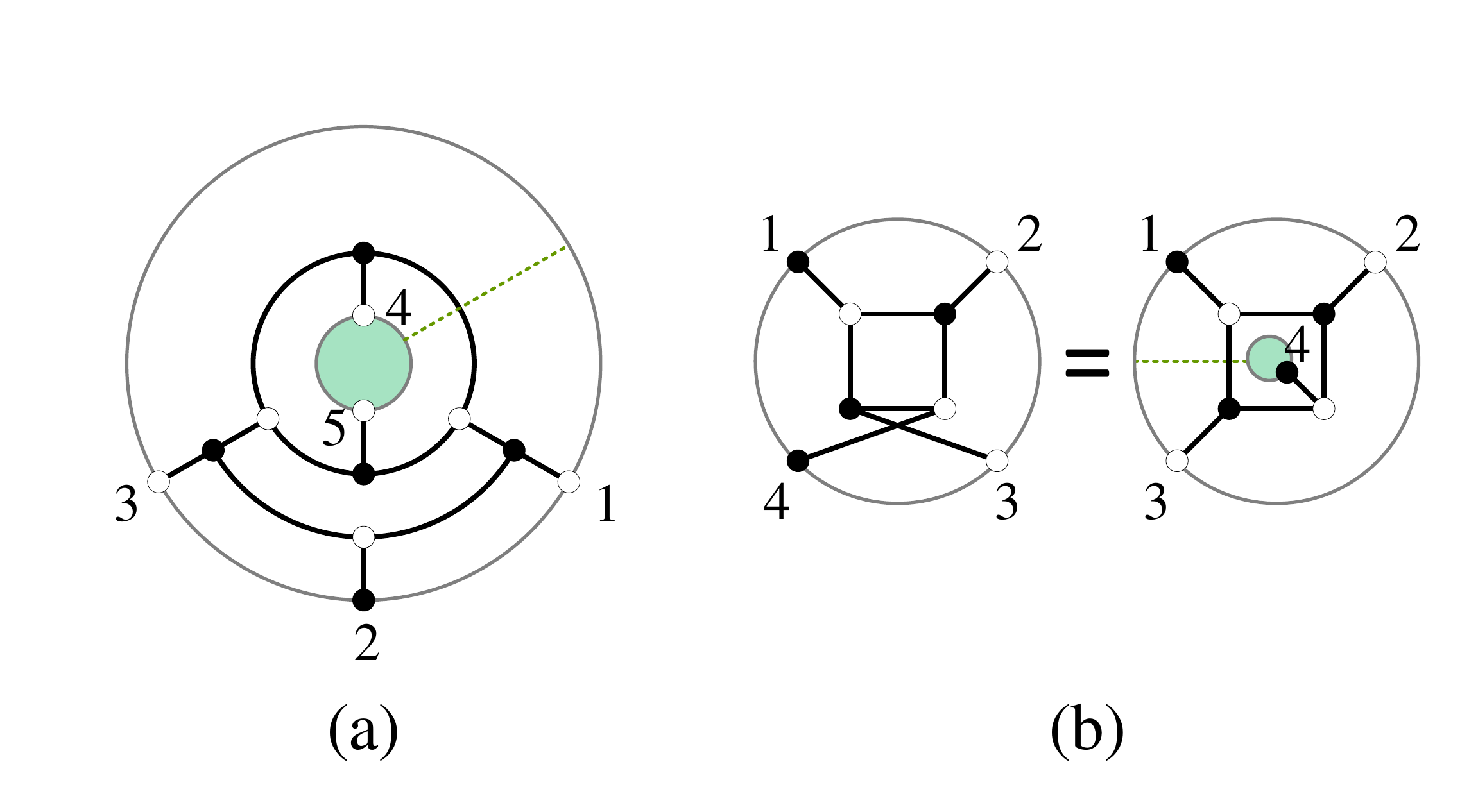}
\caption{(a) A graph with two boundaries. (b) Crossing external legs can be eliminated by introducing a new boundary.}
\label{fig:nonplanars}
\end{figure}

\fref{fig:nonplanars}.a shows an example with two boundaries. \fref{fig:nonplanars}.b illustrates how crossing external legs can be traded by additional boundaries.

A desirable property of a well-behaved boundary measurement is that the matroid polytope derived from the graph should coincide with the one for the corresponding Grassmannian element. This in particular implies that the boundary measurement should realize the map between \pl coordinates and perfect matchings already mentioned in \sref{section_Plucker_coordinates}. 

As we show in the next subsections, our generalization of the boundary measurement to multiple boundaries obeys this property and, moreover, nicely contains as subcases the boundary measurement for graphs on the disk and the annulus.

It is important to note that in the non-planar case the \pl coordinates are no longer positive definite,
given positive edge weights, as will be shown explicitly in the following examples. Thus, the image of the map is no longer restricted to the positive part of the Grassmannian.

\bigskip

\subsection{Boundary Measurement for the Annulus}

Initiating our discussion of multiple boundaries, in this subsection we shall review a method by Gekhtman, Shapiro and Vainshtein \cite{2009arXiv0901.0020G} that maps graphs on the annulus to elements of the Grassmannian. Every perfect matching gives rise to a perfect orientation. As in the planar case, we construct a matrix $C$ whose rows correspond to sources of the perfect orientation and columns correspond to all external nodes. In analogy with what discussed in \sref{sec:planarboundmeas}, $C$ is constructed by selecting certain entries of the more general path matrix.
Each matrix entry in $C$ may be composed of several terms, reflecting the fact that there may be multiple ways to flow along the perfect orientation from a given source to a given sink. For non-planar graphs the boundary measurement needs to deal with two subtle points: 

\bigskip

\begin{itemize}
\item The {\it ordering} of external edges follows a specific prescription when there are multiple boundaries.
\item The {\it signs} assigned to the matrix entries require a careful treatment. 
\end{itemize}
For the annulus, tackling these issues demands the introduction of a \textit{cut} connecting the two boundaries, shown as a green dotted line in \fref{fig:nonplanars}.

\bigskip

Regarding the first point, the canonical ordering on the annulus is to start from the cut and go clockwise around the outer boundary, followed by counterclockwise counting from the cut around the inner boundary.\footnote{Note that this convention is opposite to the one presented in \cite{2009arXiv0901.0020G} and was chosen in order to be consistent with the case of the disk.} In the next subsection we will introduce a generalization for graphs with an arbitrary number of boundaries.

To address the second point, signs in the matrix $C$ have two distinct origins. The first type of signs is the same as that present in the planar case; these are overall signs which all terms in a given matrix entry $C_{ij}$ are subject to. As in the planar case, the overall sign of each entry is $(-1)^{s(i,j)}$, where $s(i,j)$ is the number of sources strictly between $i$ and $j$, neglecting periodicity.

The second type of sign comes from the rotation number of the actual path connecting a source and a sink. In order to find the sign for each path it is necessary to first complete the path to form a closed loop. The prescription for closing the loop is as follows:

\bigskip
\begin{itemize}
\item If the source and the sink are both on the same boundary, the path is closed by adding a segment from the sink to the source which runs clockwise along the boundary.
\item If the source and the sink are on different boundaries, the path is closed by adding a segment that runs clockwise from the sink to the cut, traverses along the cut to the other boundary, and runs clockwise along this boundary until reaching the source. 
\end{itemize}
\bigskip

The sign of a path $P$ is given by $(-1)^{r(P)+1}$, where $r(P)$ is the rotation number of the closed path \cite{2009arXiv0901.0020G}, which can be easily calculated by splitting it at each self-intersection. This gives a number of closed loops that have clockwise or counterclockwise orientation. The rotation number is given by the difference of the number of clockwise loops with the number of counterclockwise loops. Note that this sign automatically accounts for the sign $(-1)$ introduced for a path which runs over a loop in a perfect orientation, reviewed in \sref{sec:planarboundmeas}.

The cut essentially measures the non-planarity of a path, by counting how many times it goes around the non-trivial direction of the annulus. For this reason, it is heuristically clear that the results cannot depend on the choice of cut. This is shown to be the case in \cite{2009arXiv0901.0020G}.

\bigskip

\newpage

\begin{figure}[htb!]
\centering
\includegraphics[scale=0.8]{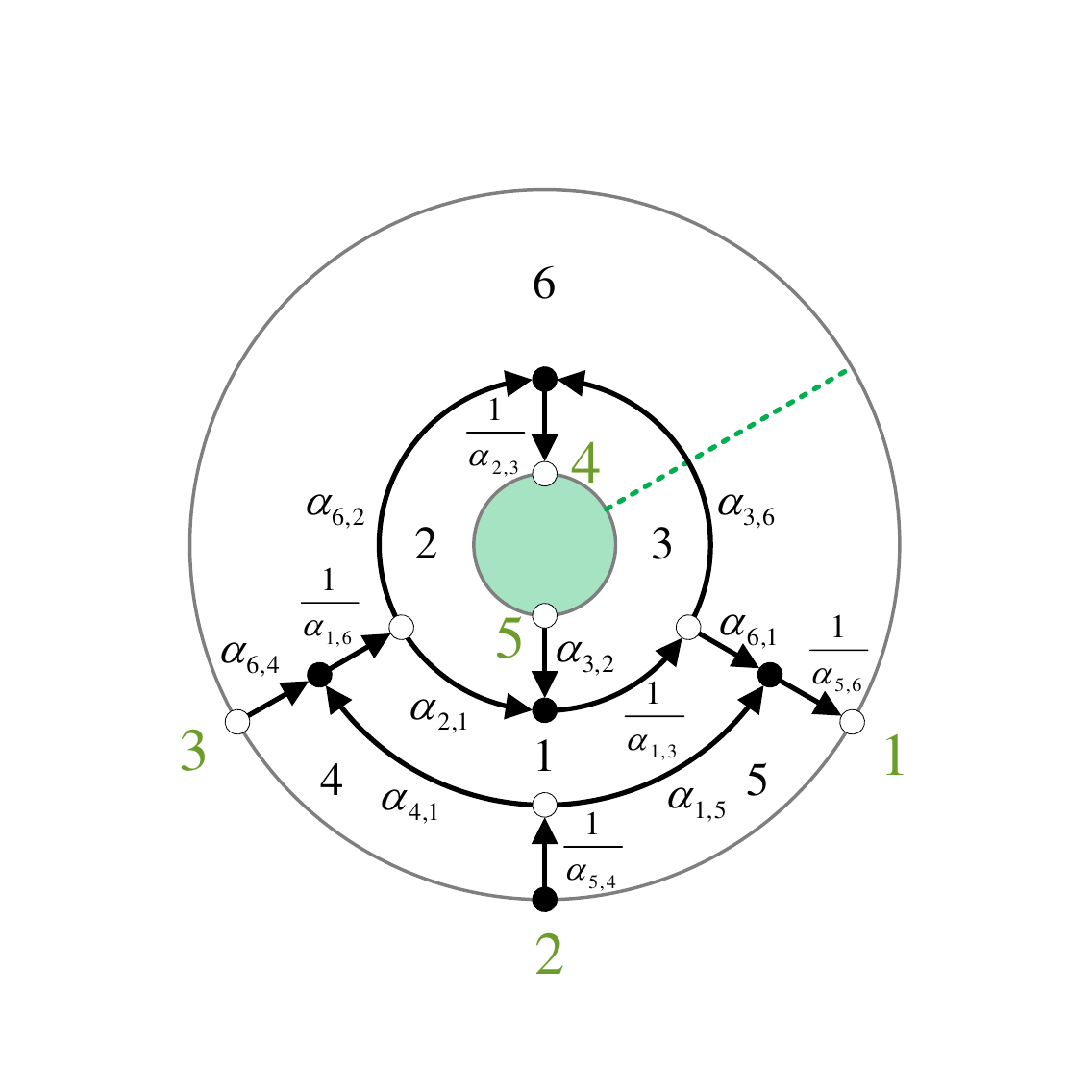}
\caption{A non-planar graph for a top-dimensional region of $Gr_{3,5}$. The cut is indicated by a green dotted line. Arrows show the perfect orientation associated to the perfect matching $p_1$, which contains edges $X_{1,3}$, $X_{1,6}$, $X_{2,3}$ $X_{5,4}$ and $X_{5,6}$.}
\label{fig:nonplanarfriend}
\end{figure}

\paragraph{Example 1.} We shall illustrate the method using the example in \fref{fig:nonplanars}.a, which is shown in more detail in \fref{fig:nonplanarfriend}.

The perfect matchings for this case are given by:

\bigskip

{\footnotesize
\begin{equation}
P = 
\left(
\begin{array}{c|ccccccccccccccc}
  & \ p_1 \ & \ p_2 \ & \ p_3 \ & \ p_4 \ & \ p_5 \ & \ p_6 \ & \ p_7 \ & \ p_8 \ & \ p_9 \ & p_{10} & p_{11} & p_{12} & p_{13} & p_{14} & p_{15} \\
 \hline
 X_{1,3} \ & 1 & 1 & 1 & 1 & 1 & 0 & 0 & 0 & 0 & 0 & 0 & 0 & 0 & 0 & 0 \\
 X_{1,6} \ & 1 & 1 & 0 & 0 & 0 & 1 & 1 & 1 & 0 & 0 & 0 & 0 & 0 & 0 & 0 \\
 X_{3,6} \ & 0 & 0 & 0 & 0 & 0 & 0 & 1 & 1 & 0 & 0 & 1 & 0 & 1 & 0 & 1 \\
 X_{6,1} \ & 0 & 0 & 0 & 0 & 0 & 1 & 0 & 0 & 1 & 1 & 0 & 1 & 0 & 1 & 0 \\
 X_{1,5} \ & 0 & 1 & 0 & 0 & 1 & 0 & 0 & 1 & 0 & 0 & 0 & 0 & 0 & 0 & 1 \\
 X_{2,1} \ & 0 & 0 & 0 & 0 & 0 & 0 & 0 & 0 & 1 & 1 & 1 & 0 & 1 & 0 & 1 \\
 X_{4,1} \ & 0 & 0 & 0 & 1 & 0 & 0 & 0 & 0 & 0 & 1 & 0 & 0 & 1 & 1 & 0 \\
 X_{6,2} \ & 0 & 0 & 1 & 1 & 1 & 0 & 0 & 0 & 0 & 0 & 0 & 1 & 0 & 1 & 0 \\
 X_{2,3} \ & 1 & 1 & 0 & 0 & 0 & 1 & 0 & 0 & 1 & 1 & 0 & 0 & 0 & 0 & 0 \\
 X_{5,4} \ & 1 & 0 & 1 & 0 & 0 & 1 & 1 & 0 & 1 & 0 & 1 & 1 & 0 & 0 & 0 \\
 X_{5,6} \ & 1 & 0 & 1 & 1 & 0 & 0 & 1 & 0 & 0 & 0 & 1 & 0 & 1 & 0 & 0 \\
 X_{3,2} \ & 0 & 0 & 0 & 0 & 0 & 1 & 1 & 1 & 0 & 0 & 0 & 1 & 0 & 1 & 0 \\
  X_{6,4} \ \ & 0 & 0 & 1 & 0 & 1 & 0 & 0 & 0 & 1 & 0 & 1 & 1 & 0 & 0 & 1 \\
\end{array}
\right) \; . 
\label{annulusPmatrix} 
\end{equation}}

\bigskip

We take as reference perfect matching $p_1$, which leads to the perfect orientation displayed in the figure, and hence the source set $\{ 2,3,5\}$. Thus, $C$ takes the form
\begin{equation}
C =
\left(
\begin{array}{c|ccccc}
  & 1 & \ 2 \ & \ 3 \ & 4 & \ 5 \ \\ \hline
2 \ \ & * & 1 & 0 & -* & 0 \\
3 \ \ & -* & 0 & 1 & * & 0 \\
5 \ \ & * & 0 & 0 & * & 1
\end{array}
\right) ,
\end{equation}
where generically non-zero entries have been marked with an asterisk, and the signs $(-1)^{s(i,j)}$ have been inserted. We now proceed to introduce relative signs for the matrix entries. Computing the path matrix, we see that there are precisely two paths between source 2 and sink 1: $\frac{\alpha_{1,5}}{\alpha_{5,4}\alpha_{5,6}}$ and $\frac{\alpha_{2,1}\alpha_{4,1}\alpha_{6,1}}{\alpha_{1,3}\alpha_{1,6}\alpha_{5,4}\alpha_{5,6}}$. In both cases the closed loop is formed as described above, and since this forms a single circle, there are no additional signs. 

\begin{figure}[htb!]
\centering
\includegraphics[scale=0.65]{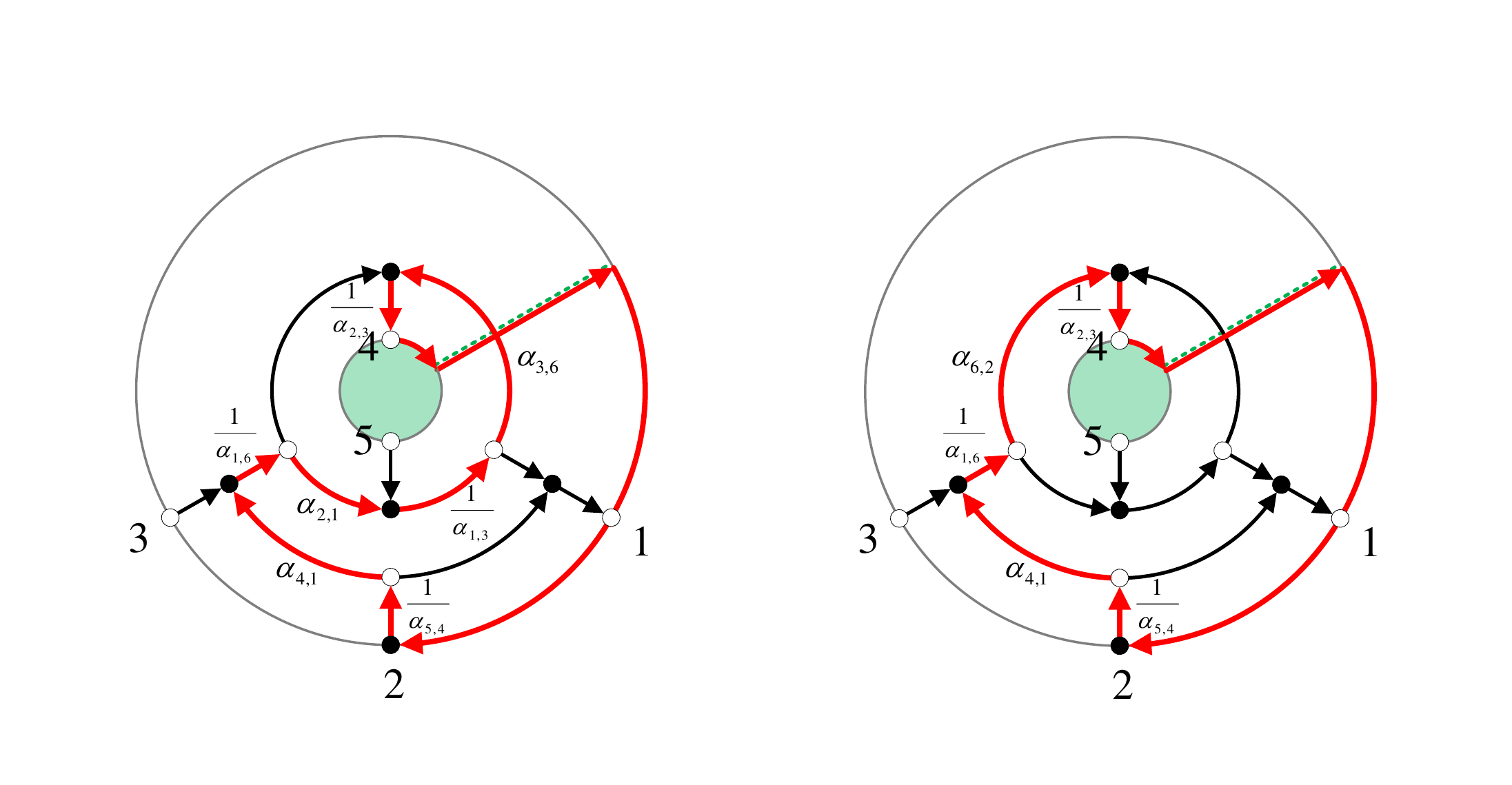}
\caption{Closing the loop for the paths $\frac{\alpha_{2,1} \alpha_{3,6} \alpha_{4,1}}{\alpha_{1,3} \alpha_{1,6} \alpha_{2,3} \alpha_{5,4}}$ (left) and $\frac{\alpha_{4,1} \alpha_{6,2}}{\alpha_{1,6} \alpha_{2,3} \alpha_{5,4}}$ (right).}
\label{fig:nonplanarclosedloop}
\end{figure}

The $C_{14}$ entry is different. Again, there are two paths between source 2 and sink 4: $\frac{\alpha_{2,1} \alpha_{3,6} \alpha_{4,1}}{\alpha_{1,3} \alpha_{1,6} \alpha_{2,3} \alpha_{5,4}}$ and $\frac{\alpha_{4,1} \alpha_{6,2}}{\alpha_{1,6} \alpha_{2,3} \alpha_{5,4}}$. Closing the path following the prescription above, we obtain the loops shown in \fref{fig:nonplanarclosedloop}.

As we see, for the first path there is a clockwise loop and a counterclockwise loop, together forming rotation zero. Hence, we get a sign $(-1)^{0+1}=-1$. For the second path we get a single clockwise loop, which gives $(-1)^{1+1}=1$. Following this procedure for all paths in the path matrix gives
\begin{eqnarray}
C &=
\left(
\begin{array}{ccccc}
 \frac{\alpha_{1,5}}{\alpha_{5,4} \alpha_{5,6}}+\frac{\alpha_{2,1} \alpha_{4,1} \alpha_{6,1}}{\alpha_{1,3} \alpha_{1,6} \alpha_{5,4} \alpha_{5,6}} & 1 & 0 & \frac{\alpha_{2,1} \alpha_{3,6} \alpha_{4,1}}{\alpha_{1,3} \alpha_{1,6} \alpha_{2,3} \alpha_{5,4}}-\frac{\alpha_{4,1} \alpha_{6,2}}{\alpha_{1,6} \alpha_{2,3} \alpha_{5,4}} & 0 \\
 -\frac{\alpha_{2,1} \alpha_{6,1} \alpha_{6,4}}{\alpha_{1,3} \alpha_{1,6} \alpha_{5,6}} & 0 & 1 & \frac{\alpha_{6,2} \alpha_{6,4}}{\alpha_{1,6} \alpha_{2,3}}-\frac{\alpha_{2,1} \alpha_{3,6} \alpha_{6,4}}{\alpha_{1,3} \alpha_{1,6} \alpha_{2,3}} & 0 \\
 \frac{\alpha_{3,2} \alpha_{6,1}}{\alpha_{1,3} \alpha_{5,6}} & 0 & 0 & \frac{\alpha_{3,2} \alpha_{3,6}}{\alpha_{1,3} \alpha_{2,3}} & 1 \\
\end{array}
\right) \nonumber \\ 
&= \left(
\begin{array}{ccccc}
 \mathfrak{p}_2+\mathfrak{p}_{10} & 1 & 0 & \mathfrak{p}_{13}-\mathfrak{p}_4 & 0 \\
 -\mathfrak{p}_9 & 0 & 1 & \mathfrak{p}_3-\mathfrak{p}_{11} & 0 \\
 \mathfrak{p}_6 & 0 & 0 & \mathfrak{p}_7 & 1 \\
\end{array}
\right),
\label{CG35NP}
\end{eqnarray}
where all signs have been included, and the paths have been written as ratios of oriented perfect matchings with the oriented reference matching $\mathfrak{p}_i = \tilde{p}_i / \tilde{p}_1$. This is the element of the Grassmannian associated to this specific graph on an annulus. Note that only $\mathfrak{p_i}$'s associated to single paths are contained in $C$. Those consisting of multiple disjoint components are absent. All perfect matchings will however contribute to the \pl coordinates.

It is a non-trivial fact that the \pl coordinates of \eref{CG35NP} can be written as sums of perfect matchings (or more precisely flows given by the ratio between perfect matchings and the reference matching), whose source set is precisely the set of columns involved in the \pl coordinate in question. For example, $\Delta_{123}$ is given by a sum of contributions from perfect matchings whose flows have source set $\{1,2,3\}$. In fact, it is a \textit{requirement} of a well-behaved boundary measurement that its \pl coordinates have this property.\footnote{As mentioned in \sref{sec:planarboundmeas} for planar graphs, this property is also achieved by not adding any sign to the $\mathcal{M}^C$ matrix. A delicate choice of non-trivial signs is however needed for \pl coordinates to become sums of contributions from perfect matchings while other nice properties are realized.} This is because we identify \pl coordinates $\Delta_I$ with elements $I \in \mathcal{M}$ of a matroid, which in turn are identified with points of the matroid polytope arising from the graph.
These points are formed by the union of perfect matchings which share the same source set $I$. 

The \pl coordinates are:
\beq
\begin{array}{llcrl}
\Delta _{123}=& \mathfrak{p}_6 & \ \ \ \ \ & \Delta _{145}=& \mathfrak{p}_5-\mathfrak{p}_{15} \\
 \Delta _{124}=& \mathfrak{p}_{12} & & \Delta _{234}=& \mathfrak{p}_7 \\
 \Delta _{125}=& \mathfrak{p}_9 & & \Delta _{235}=& \mathfrak{p}_1 \\
 \Delta _{134}=& \mathfrak{p}_8+\mathfrak{p}_{14} & & \Delta _{245}=& \mathfrak{p}_3-\mathfrak{p}_{11} \\
 \Delta _{135}=& \mathfrak{p}_2+\mathfrak{p}_{10} & & \Delta _{345}=& \mathfrak{p}_4-\mathfrak{p}_{13}
\end{array} .
\label{annulusPluckerMap}
\eeq

Multiplying all \pl coordinates by the oriented reference matching $\tilde{p}_1$ we obtain the desired map between \pl coordinates and perfect matchings. It is straightforward to check that all perfect matchings contributing to a \pl coordinate have the correct source set.

As an additional check, we will now show that the removable edges found using the 
technique expounded at the end of \sref{section_efficient_reducibility} are the correct ones, i.e.\ are those that only kill one \pl coordinate each, thus decreasing the dimension by 1. The predicted removable edges are $X_{1,3}$, $X_{3,6}$, $X_{1,5}$, $X_{2,1}$, $X_{4,1}$ and $X_{6,2}$. Removing them results in:

\bigskip
$$
\begin{array}{|c|c|c|}
\hline \text{  Edge  } & \begin{array}{c}\text{Deleted perfect} \\ \text{matchings} \end{array} & \ \begin{array}{c}\text{Vanishing \pl} \\ \text{coordinate} \end{array} \ \\
\hline X_{1,3} & p_1,p_2,p_3,p_4,p_5 & \Delta_{235} \\ 
\hline X_{3,6} & p_7,p_8,p_{11},p_{13},p_{15} & \Delta_{234} \\ 
\hline X_{1,5} & p_2,p_5,p_8,p_{15} & \Delta_{145} \\ 
\hline X_{2,1} & \ p_9,p_{10},p_{11},p_{13},p_{15} \ & \Delta_{125} \\ 
\hline X_{4,1} & p_4,p_{10},p_{13},p_{14} & \Delta_{345} \\ 
\hline X_{6,2} & p_3,p_4,p_5,p_{12},p_{14} & \Delta_{124} \\
\hline
\end{array}
$$
\bigskip

\noindent
It is easy to verify that there are no other edges that only kill a single \pl coordinate. 

\bigskip

\paragraph{Example 2.}
For the example shown in \fref{fig:nonplanars}.b we have the perfect matchings:

\bigskip
{\footnotesize
\begin{equation}
P = 
\left(
\begin{array}{c|ccccccc}
  & \ p_1 \ & \ p_2 \ & \ p_3 \ & \ p_4 \ & \ p_5 \ & \ p_6 \ & \ p_7 \ \\
 \hline
 X_{1,2} \ & 1 & 1 & 0 & 0 & 0 & 0 & 0 \\
 X_{1,4} \ & 1 & 0 & 1 & 0 & 0 & 0 & 0 \\
 X_{3,1} \ & 0 & 0 & 0 & 0 & 0 & 1 & 1 \\
 X_{4,1} \ & 0 & 0 & 0 & 0 & 1 & 0 & 1 \\
 X_{2,3} \ & 0 & 0 & 1 & 1 & 1 & 0 & 0 \\
 X_{2,4} \ & 0 & 0 & 1 & 1 & 0 & 1 & 0 \\
 X_{1,1} \ & 0 & 1 & 0 & 1 & 0 & 1 & 0 \\
 X_{4,3} \ & 0 & 1 & 0 & 1 & 1 & 0 & 0 \\
\end{array}
\right) \quad . 
\end{equation}}

\bigskip

Choosing as reference perfect matching $p_2$, the boundary measurement maps the graph to the Grassmannian element

\begin{equation}
C =
\left(
\begin{array}{cccc}
 \frac{\alpha_{2,3} \alpha_{2,4}}{\alpha_{1,2}} & \frac{\alpha_{2,4} \alpha_{3,1}}{\alpha_{1,2} \alpha_{4,3}} & 1 & 0 \\
 \frac{\alpha_{2,3} \alpha_{4,1}}{\alpha_{1,1} \alpha_{1,2}} & \frac{\alpha_{3,1} \alpha_{4,1}}{\alpha_{1,1} \alpha_{1,2} \alpha_{4,3}}-\frac{\alpha_{1,4}}{\alpha_{1,1} \alpha_{4,3}} & 0 & 1 \\
\end{array}
\right) = \left(
\begin{array}{cccc}
 \mathfrak{p}_4 & \mathfrak{p}_6 & 1 & 0 \\
 \mathfrak{p}_5 & \mathfrak{p}_7-\mathfrak{p}_1 & 0 & 1 \\
\end{array}
\right)
\end{equation}
which gives rise to the following \pl coordinates:

\bigskip

\beq
\begin{array}{llcrl}
 \Delta _{12} =& -\mathfrak{p}_3 & \ \ \ \ \ & \Delta _{23} =& \mathfrak{p}_1-\mathfrak{p}_7 \\
 \Delta _{13} =& -\mathfrak{p}_5 & \ \ \ \ \ & \Delta _{24} =& \mathfrak{p}_6 \\
 \Delta _{14} =& \mathfrak{p}_4 & \ \ \ \ \ & \Delta _{34} =& \mathfrak{p}_2 \\
\end{array} . 
\eeq

\noindent Note that contrary to the planar diagram of $Gr_{2,4}$ studied in \sref{sec:planarboundmeas}, here the \pl coordinates are no longer positive definite for positive edge weights.

\bigskip 

\subsection{Boundary Measurement Beyond the Annulus}

\label{section_bm_beyond_annulus}

In this section we introduce a boundary measurement for graphs on the plane with an arbitrary number of boundaries. The new map reduces to the previously known cases when restricted to the disk or the annulus. As previously mentioned, the map must be insensitive to graphical equivalences and its minors must be identifiable with linear combinations of perfect matchings.\footnote{Whether such a map is unique is an interesting question, beyond the scope of this article.} Additionally, for diagrams on the disk we require that all minors are manifestly non-negative, for positive edge weights.

As we saw in the previous subsection, the success of the boundary measurement is crucially reliant on a delicate assignment of signs to entries in the path matrix. When going from the disk to the annulus, the difficulties of introducing an additional boundary were twofold: first, the ordering of external nodes was sensitively fixed according to the prescription in \cite{2009arXiv0901.0020G}; secondly, it was necessary to complete the path (possibly using the cut) and form a loop in order to count additional loops which are not naturally present in the chosen perfect orientation.

Introducing more boundaries has similar difficulties. The ordering of the external nodes for a generic number of boundaries can be fixed in a way which is reminiscent of going around cuts in complex analysis. The algorithm is as follows:
\begin{itemize}
\item Start at a cut on one of the boundaries. We will preferably choose the outer one.
\item Follow the boundary in a clockwise fashion, until reaching a cut.
\item Follow it to the next boundary, without crossing over it.
\item Follow the next boundary until reaching another cut.
\item Follow the cut to the next boundary, once again without crossing it, and continue in this fashion until reaching the starting point.
\end{itemize}
For the disk and annulus, this procedure fixes a clockwise ordering for the external boundary, followed by a counter-clockwise ordering for the internal boundary, in agreement with the previous section. 

The assignment of signs in the matrix $C$ works similarly to our discussion for the annulus:
there is the usual overall sign $(-1)^{s(i,j)}$ to the entry $C_{ij}$, where $s(i,j)$ counts the number of sources strictly between $i$ and $j$, neglecting periodicity. There is also a sign related to the loops which compose the path. In order to compute it, we close the path by going from the sink clockwise around the corresponding boundary, and then following the necessary cuts and boundaries, 
always going clockwise, until reaching the source. The sign is then $(-1)^{r(P)+1}$, where $r(P)$ is the rotation number of this closed path, obtained by counting the number of clockwise 
loops minus the number of counterclockwise loops, as already explained in the previous section.

For computational convenience, there is a significantly faster way to compute these second type of signs, which does not involve drawing and analyzing the path. Each time a path runs across a cut, it picks up a minus sign iff it is going between two boundaries that can only be reached using this cut. Each entry in $C$ is specified by its source and sink; it is easy then to identify which cuts are going to be actively used in this matrix entry. Thus, each matrix entry activates sign flips for only those edges that run across the relevant cuts. In addition to these signs, it is necessary to add signs to closed loops that are present in the perfect orientation. From a computational standpoint, it is then only necessary to provide information on how nodes are distributed over the different boundaries, which cuts are activated by each pair of boundaries, and which edges are crossed by the respective cuts. 

\bigskip

\begin{figure}[htb!]
\centering
\includegraphics[scale=0.4]{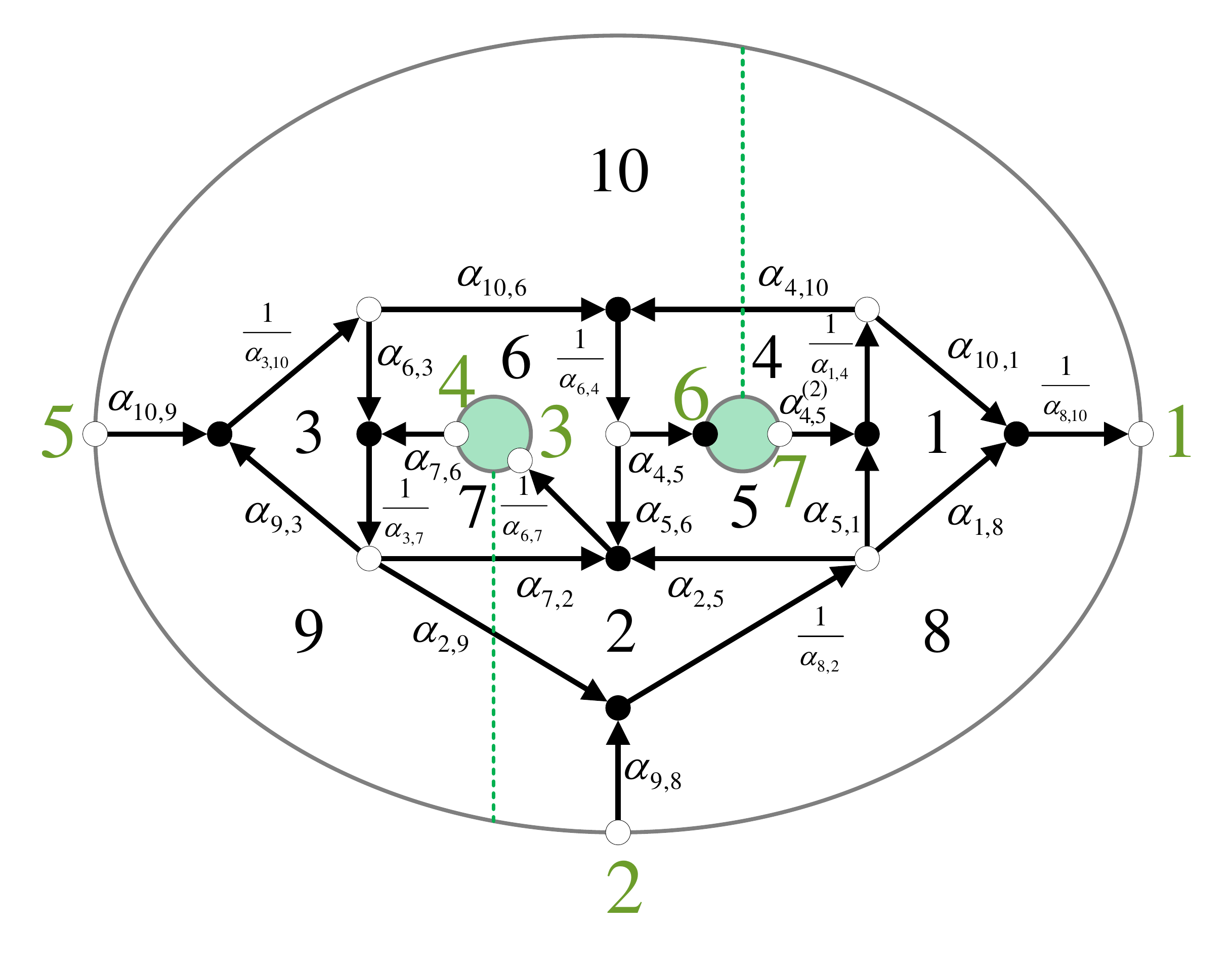}
\caption{A graph with 3 boundaries. The perfect orientation shown corresponds to the oriented perfect matching $p_1$, which contains edges $X_{1,4}$, $X_{3,7}$, $X_{3,10}$, $X_{6,4}$, $X_{6,7}$, $X_{8,2}$ and $X_{8,10}$.}
\label{fig:3Bguy}
\end{figure}

\paragraph{Example: 3 Boundaries.}
To illustrate the method above let us consider the example in \fref{fig:3Bguy}. This is a reduced graph with three boundaries. This is the minimum number of boundaries for this graph, i.e.\ it is impossible to reduce it by flipping external legs. We will later investigate the effect of redistributing external edges over boundaries.

\begin{figure}[htb!]
\centering
\includegraphics[scale=0.32]{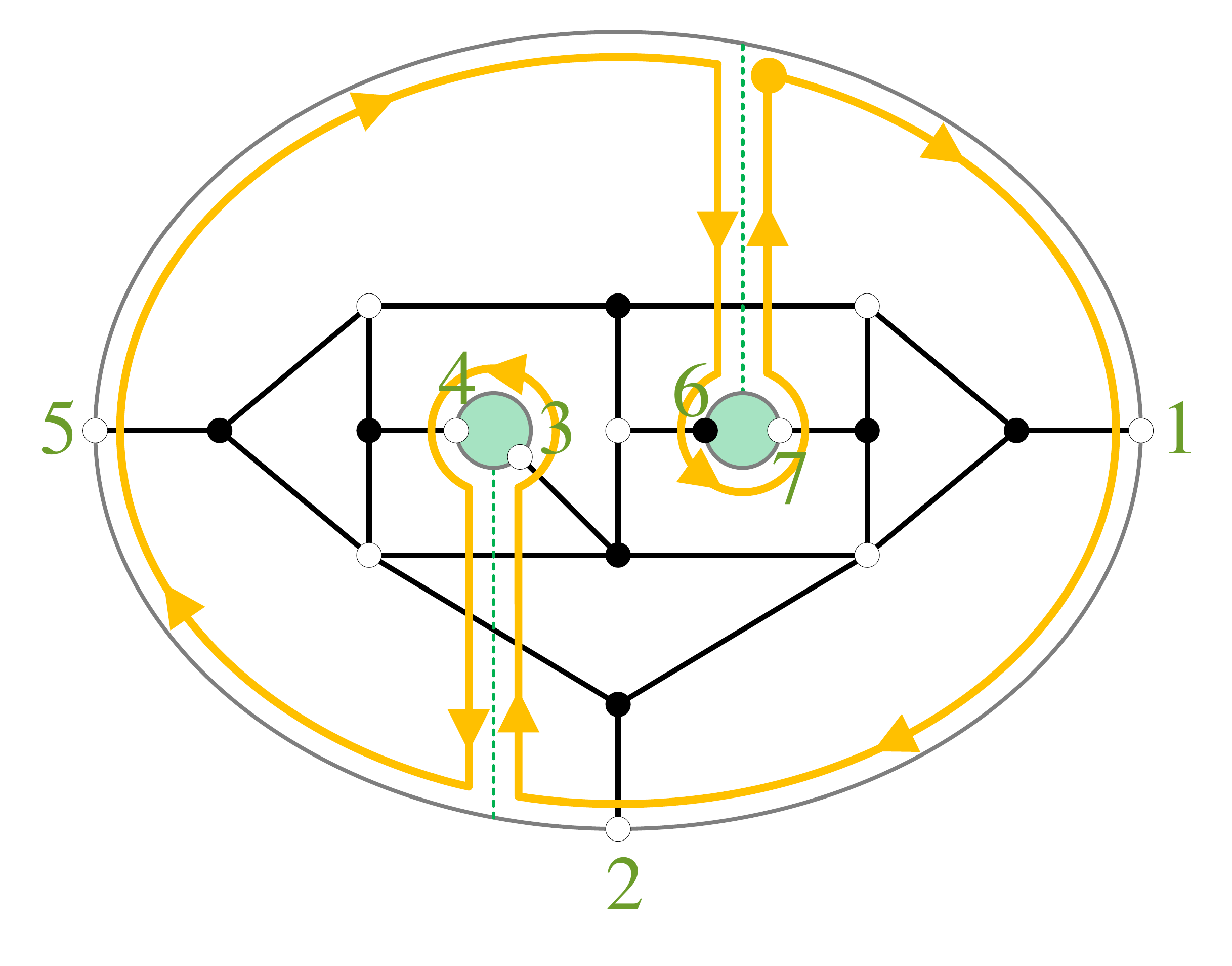}
\caption{Fixing the ordering for three boundaries. The starting point is marked by the large yellow dot.}
\label{fig:3Bcounting}
\end{figure}

The ordering of external nodes is determined by starting at the upper cut on the outer boundary and proceeding according to the algorithm above. This is shown in \fref{fig:3Bcounting}.

This diagram has 88 perfect matchings. For amusement, and to show it is straightforward to explicitly deal with such large graphs using our tools, we provide the perfect matching matrix in Appendix \ref{section_P_matrix_3_boundaries}.

The reference perfect matching was chosen to be $p_1$, which gives rise to the perfect orientation in \fref{fig:3Bguy}. This example exhibits an interesting phenomenon: the perfect orientation contains a closed oriented loop $ \frac{\alpha_{6,3} \alpha_{9,3}}{\alpha_{3,7} \alpha_{3,10}} = \mathfrak{p}_3$.\footnote{This fact is totally unrelated to the multiplicity of boundaries. It did not appear in the previous, simpler examples due to our specific choices of perfect orientations.} When this happens, path connecting two nodes can circle an arbitrary number of times around the internal loop. The sum of contributions to entries in the path matrix thus takes the form of a geometric series, leading to non-trivial denominators containing the loop (see Appendix \ref{PathM}). The piece of the path matrix contributing to the boundary measurement takes the form

\begin{equation}
\left(
\begin{array}{c|ccccccc}
 & 1 & 2 & 3 & 4 & 5 & 6 & 7 \\
\hline 
2 \ \ & \mathfrak{p}_{11}+\mathfrak{p}_{19} & 1 & \mathfrak{p}_{33}+\mathfrak{p}_{47} & 0 & 0 & \mathfrak{p}_6 & 0 \\
4 \ \ &  \frac{\mathfrak{p}_{13}}{1-\mathfrak{p}_3}+\frac{\mathfrak{p}_{23}}{1-\mathfrak{p}_3} & 0 & \frac{\mathfrak{p}_{36}}{1-\mathfrak{p}_3}+\frac{\mathfrak{p}_{37}}{1-\mathfrak{p}_3}+\frac{\mathfrak{p}_{41}}{1-\mathfrak{p}_3}+\frac{\mathfrak{p}_{51}}{1-\mathfrak{p}_3} & 1 & 0 & \frac{\mathfrak{p}_4}{1-\mathfrak{p}_3}+\frac{\mathfrak{p}_7}{1-\mathfrak{p}_3} & 0 \\
5 \ \ &  \frac{\mathfrak{p}_{15}}{1-\mathfrak{p}_3}+\frac{\mathfrak{p}_{27}}{1-\mathfrak{p}_3} & 0 & \frac{\mathfrak{p}_{34}}{1-\mathfrak{p}_3}+\frac{\mathfrak{p}_{38}}{1-\mathfrak{p}_3}+\frac{\mathfrak{p}_{40}}{1-\mathfrak{p}_3}+\frac{\mathfrak{p}_{58}}{1-\mathfrak{p}_3} & 0 & 1 & \frac{\mathfrak{p}_2}{1-\mathfrak{p}_3}+\frac{\mathfrak{p}_{10}}{1-\mathfrak{p}_3} & 0 \\
7 \ \ & \mathfrak{p}_{18} & 0 & \mathfrak{p}_{45} & 0 & 0 & \mathfrak{p}_5 & 1 \\
\end{array}
\right) ,
\end{equation}
where the $(1-\mathfrak{p}_3)^{-1}$ factors arise due to the infinite number of paths involving the closed loop.

Signs are introduced in three steps: first to overall entries according to $(-1)^{s(i,j)}$, then to loops that are present in the perfect orientation, in this case $\mathfrak{p}_3$, and finally to the edges that cross the cuts, in the relevant entries.\footnote{The first step is straightforward, but the second step can in general be subtle; sometimes there are 
flows $\mathfrak{p}_i$ which contain loops, and can be written as a product $\mathfrak{p}_i = \mathfrak{p}_j \mathfrak{p}_{\text{loop}}$. In these cases, when replacing $\mathfrak{p}_{\text{loop}} \to -\mathfrak{p}_{\text{loop}}$, we should also replace $\mathfrak{p}_i \to -\mathfrak{p}_i$. This does not happen in the specific example at hand.}

After introducing the first two types of signs, the matrix becomes 

\begin{equation}
\left(
\begin{array}{c|ccccccc}
& 1 & 2 & 3 & 4 & 5 & 6 & 7 \\ \hline
2 \ \ & \mathfrak{p}_{11}+\mathfrak{p}_{19} & 1 & \mathfrak{p}_{33}+\mathfrak{p}_{47} & 0 & 0 & \mathfrak{p}_6 & 0 \\
4 \ \ & -\frac{\mathfrak{p}_{13}}{1+\mathfrak{p}_3}-\frac{\mathfrak{p}_{23}}{1+\mathfrak{p}_3} & 0 & \frac{\mathfrak{p}_{36}}{1+\mathfrak{p}_3}+\frac{\mathfrak{p}_{37}}{1+\mathfrak{p}_3}+\frac{\mathfrak{p}_{41}}{1+\mathfrak{p}_3}+\frac{\mathfrak{p}_{51}}{1+\mathfrak{p}_3} & 1 & 0 & -\frac{\mathfrak{p}_4}{1+\mathfrak{p}_3}-\frac{\mathfrak{p}_7}{1+\mathfrak{p}_3} & 0 \\
5 \ \ & \frac{\mathfrak{p}_{15}}{1+\mathfrak{p}_3}+\frac{\mathfrak{p}_{27}}{1+\mathfrak{p}_3} & 0 & -\frac{\mathfrak{p}_{34}}{1+\mathfrak{p}_3}-\frac{\mathfrak{p}_{38}}{1+\mathfrak{p}_3}-\frac{\mathfrak{p}_{40}}{1+\mathfrak{p}_3}-\frac{\mathfrak{p}_{58}}{1+\mathfrak{p}_3} & 0 & 1 & \frac{\mathfrak{p}_2}{1+\mathfrak{p}_3}+\frac{\mathfrak{p}_{10}}{1+\mathfrak{p}_3} & 0 \\
7 \ \ &  -\mathfrak{p}_{18} & 0 & \mathfrak{p}_{45} & 0 & 0 & \mathfrak{p}_5 & 1 \\
\end{array}
\right) .
\end{equation}

For the third type of signs there are two cuts, the one to the left $c_L$ which reaches between the outer boundary and the left-most boundary, and the one to the right $c_R$ which reaches between the outer boundary and the right-most boundary. To go from the right-most and the left-most boundary it is necessary to use both cuts. The relevant cuts for each entry are summarized in the following matrix 

\begin{equation}
\left(
\begin{array}{c|ccccccc}
 & \ \ 1 \ \ & \ \ 2 \ \ & 3 & 4 & \ \ 5 \ \ & 6 & 7 \\
 \hline
 2 \ \ & \bullet & \bullet & c_L & c_L & \bullet & c_R & c_R \\
 4 \ \ & c_L & c_L & \bullet & \bullet & c_L & c_L,c_R & c_L,c_R \\
 5 \ \ & \bullet & \bullet & c_L & c_L & \bullet & c_R & c_R \\
 7 \ \ & c_R & c_R & c_L,c_R & c_L,c_R & c_R & \bullet & \bullet
\end{array}
\right) .
\end{equation} 

The action of the cuts are $c_L: \{ \alpha_{7,2}, \alpha_{2,9} \} \to \{ -\alpha_{7,2}, -\alpha_{2,9} \} $ and $c_R: \alpha_{4,10}\to -\alpha_{4,10}$. Applying the action to the path matrix, we finally obtain the desired element of the Grassmannian:

\begin{equation}
C = 
\left(
\begin{array}{c|ccccccc}
& 1 & 2 & 3 & 4 & 5 & 6 & 7 \\
\hline 
2 \ \ & \mathfrak{p}_{11}+\mathfrak{p}_{19} & 1 & \mathfrak{p}_{33}+\mathfrak{p}_{47} & 0 & 0 & -\mathfrak{p}_6 & 0 \\
4 \ \ &  \frac{\mathfrak{p}_{13}}{\mathfrak{p}_3+1}+\frac{\mathfrak{p}_{23}}{\mathfrak{p}_3+1} & 0 & \frac{\mathfrak{p}_{36}}{\mathfrak{p}_3+1}+\frac{\mathfrak{p}_{37}}{\mathfrak{p}_3+1}+\frac{\mathfrak{p}_{41}}{\mathfrak{p}_3+1}+\frac{\mathfrak{p}_{51}}{\mathfrak{p}_3+1} & 1 & 0 & -\frac{\mathfrak{p}_4}{\mathfrak{p}_3+1}-\frac{\mathfrak{p}_7}{\mathfrak{p}_3+1} & 0 \\
5 \ \ &  \frac{\mathfrak{p}_{15}}{\mathfrak{p}_3+1}+\frac{\mathfrak{p}_{27}}{\mathfrak{p}_3+1} & 0 & -\frac{\mathfrak{p}_{34}}{\mathfrak{p}_3+1}+\frac{\mathfrak{p}_{38}}{\mathfrak{p}_3+1}+\frac{\mathfrak{p}_{40}}{\mathfrak{p}_3+1}+\frac{\mathfrak{p}_{58}}{\mathfrak{p}_3+1} & 0 & 1 & \frac{\mathfrak{p}_2}{\mathfrak{p}_3+1}-\frac{\mathfrak{p}_{10}}{\mathfrak{p}_3+1} & 0 \\
7 \ \ &  -\mathfrak{p}_{18} & 0 & -\mathfrak{p}_{45} & 0 & 0 & \mathfrak{p}_5 & 1
\end{array}
\right) .
\end{equation}

We note that $\mathfrak{p}_{58}$, despite containing both $\alpha_{2,9}$ and $ \alpha_{4,10}$, only changes sign once because it is only subject to the action of $c_L$; $\mathfrak{p}_{4}$, on the other hand, is subject to both cuts and does not change sign.

The \pl coordinates become:

{\small
\begin{equation}
\begin{array}{llcrl}
 \Delta _{1234}=&\frac{\mathfrak{p}_{67}}{\mathfrak{p}_3+1}+\frac{\mathfrak{p}_{83}}{\mathfrak{p}_3+1}-\frac{\mathfrak{p}_{75}}{\mathfrak{p}_3+1}-\frac{\mathfrak{p}_{78}}{\mathfrak{p}_3+1} & \ \ \ \ \ & \Delta _{1467}=&\frac{\mathfrak{p}_{12}}{\mathfrak{p}_3+1}+\frac{\mathfrak{p}_{22}}{\mathfrak{p}_3+1} \\
 \Delta _{1235}=&\frac{\mathfrak{p}_{70}}{\mathfrak{p}_3+1}+\frac{\mathfrak{p}_{72}}{\mathfrak{p}_3+1}+\frac{\mathfrak{p}_{79}}{\mathfrak{p}_3+1}-\frac{\mathfrak{p}_{74}}{\mathfrak{p}_3+1} & & \Delta _{1567}=&\frac{\mathfrak{p}_{16}}{\mathfrak{p}_3+1}+\frac{\mathfrak{p}_{31}}{\mathfrak{p}_3+1} \\
 \Delta _{1236}=&\frac{\mathfrak{p}_{80}}{\mathfrak{p}_3+1}+\frac{\mathfrak{p}_{87}}{\mathfrak{p}_3+1} & & \Delta _{2345}=&-\mathfrak{p}_{45} \\
 \Delta _{1237}=&\frac{\mathfrak{p}_{64}}{\mathfrak{p}_3+1}+\frac{\mathfrak{p}_{88}}{\mathfrak{p}_3+1} & & \Delta _{2346}=&-\frac{\mathfrak{p}_{53}}{\mathfrak{p}_3+1}-\frac{\mathfrak{p}_{57}}{\mathfrak{p}_3+1} \\
 \Delta _{1245}=&\mathfrak{p}_{18} & & \Delta _{2347}=&\frac{\mathfrak{p}_{34}}{\mathfrak{p}_3+1}-\frac{\mathfrak{p}_{38}}{\mathfrak{p}_3+1}-\frac{\mathfrak{p}_{40}}{\mathfrak{p}_3+1}-\frac{\mathfrak{p}_{58}}{\mathfrak{p}_3+1} \\
 \Delta _{1246}=&\frac{\mathfrak{p}_{21}}{\mathfrak{p}_3+1}+\frac{\mathfrak{p}_{30}}{\mathfrak{p}_3+1} & & \Delta _{2356}=&\frac{\mathfrak{p}_{48}}{\mathfrak{p}_3+1}+\frac{\mathfrak{p}_{50}}{\mathfrak{p}_3+1} \\
 \Delta _{1247}=&\frac{\mathfrak{p}_{15}}{\mathfrak{p}_3+1}+\frac{\mathfrak{p}_{27}}{\mathfrak{p}_3+1} & & \Delta _{2357}=&\frac{\mathfrak{p}_{36}}{\mathfrak{p}_3+1}+\frac{\mathfrak{p}_{37}}{\mathfrak{p}_3+1}+\frac{\mathfrak{p}_{41}}{\mathfrak{p}_3+1}+\frac{\mathfrak{p}_{51}}{\mathfrak{p}_3+1} \\
 \Delta _{1256}=&\frac{\mathfrak{p}_{28}}{\mathfrak{p}_3+1}-\frac{\mathfrak{p}_{24}}{\mathfrak{p}_3+1} & & \Delta _{2367}=&\frac{\mathfrak{p}_{42}}{\mathfrak{p}_3+1}+\frac{\mathfrak{p}_{44}}{\mathfrak{p}_3+1} \\
 \Delta _{1257}=&-\frac{\mathfrak{p}_{13}}{\mathfrak{p}_3+1}-\frac{\mathfrak{p}_{23}}{\mathfrak{p}_3+1} & & \Delta _{2456}=&\mathfrak{p}_5 \\
 \Delta _{1267}=&-\frac{\mathfrak{p}_{17}}{\mathfrak{p}_3+1}-\frac{\mathfrak{p}_{32}}{\mathfrak{p}_3+1} & & \Delta _{2457}=&1 \\
 \Delta _{1345}=&\mathfrak{p}_{65}-\mathfrak{p}_{66} & & \Delta _{2467}=&\frac{\mathfrak{p}_2}{\mathfrak{p}_3+1}-\frac{\mathfrak{p}_{10}}{\mathfrak{p}_3+1} \\
 \Delta _{1346}=&\frac{\mathfrak{p}_{68}}{\mathfrak{p}_3+1}-\frac{\mathfrak{p}_{82}}{\mathfrak{p}_3+1} & & \Delta _{2567}=&\frac{\mathfrak{p}_4}{\mathfrak{p}_3+1}+\frac{\mathfrak{p}_7}{\mathfrak{p}_3+1} \\
 \Delta _{1347}=&\frac{\mathfrak{p}_{59}}{\mathfrak{p}_3+1}+\frac{\mathfrak{p}_{69}}{\mathfrak{p}_3+1}-\frac{\mathfrak{p}_{61}}{\mathfrak{p}_3+1}-\frac{\mathfrak{p}_{77}}{\mathfrak{p}_3+1} & & \Delta _{3456}=&\mathfrak{p}_{46} \\
 \Delta _{1356}=&\frac{\mathfrak{p}_{73}}{\mathfrak{p}_3+1}+\frac{\mathfrak{p}_{84}}{\mathfrak{p}_3+1} & & \Delta _{3457}=&\mathfrak{p}_{33}+\mathfrak{p}_{47} \\
 \Delta _{1357}=&\frac{\mathfrak{p}_{60}}{\mathfrak{p}_3+1}+\frac{\mathfrak{p}_{62}}{\mathfrak{p}_3+1}+\frac{\mathfrak{p}_{71}}{\mathfrak{p}_3+1}+\frac{\mathfrak{p}_{85}}{\mathfrak{p}_3+1} & & \Delta _{3467}=&\frac{\mathfrak{p}_{35}}{\mathfrak{p}_3+1}+\frac{\mathfrak{p}_{56}}{\mathfrak{p}_3+1} \\
 \Delta _{1367}=&\frac{\mathfrak{p}_{63}}{\mathfrak{p}_3+1}+\frac{\mathfrak{p}_{86}}{\mathfrak{p}_3+1} & & \Delta _{3567}=&\frac{\mathfrak{p}_{43}}{\mathfrak{p}_3+1}-\frac{\mathfrak{p}_{49}}{\mathfrak{p}_3+1} \\
 \Delta _{1456}=&\mathfrak{p}_{20} & & \Delta _{4567}=&-\mathfrak{p}_6 \\
 \Delta _{1457}=&\mathfrak{p}_{11}+\mathfrak{p}_{19} & & \\
\end{array}
\label{Plucker_3_boundaries_1}
\end{equation}}
Modulo the denominators, the \pl coordinates take a remarkably simple form, becoming sums of $\mathfrak{p}_{i}$ contributions from {\it individual} perfect matchings. Recalling that \pl coordinates are given by maximal sub-determinants of the boundary measurement, it is worthwhile to note that the cancellations required to achieve this result are highly non-trivial and are very sensitive to the sign assignment. It is thus, in particular, extremely sensitive to the ordering of external nodes, which indirectly affects the signs $(-1)^{s(i,j)}$.

At first sight, \eref{Plucker_3_boundaries_1} does not include contributions from all perfect matchings. For example, $\mathfrak{p}_{25}$ does not appear anywhere.  This is a result of the fact that the flow $\mathfrak{p}_1$  associated to the reference perfect matching we chose, has the same sources and sinks as the flow $\mathfrak{p}_3$, which corresponds to a different perfect matching. Equivalently, $p_1$ and $p_3$ correspond to the same point in the matroid polytope. In order to accurately obtain the map between \pl coordinates and perfect matchings it is necessary to multiply \eref{Plucker_3_boundaries_1} by $\tilde{p}_1 +\tilde{p}_3$, after which we obtain:

\bigskip

{\small
\begin{equation}
\begin{array}{llcrl}
 \Delta _{1234} \leftrightarrow & p_{67}- p_{75}- p_{78}+ p_{83} & \ \ \ \ \ & \Delta _{1467} \leftrightarrow & p_{12}+ p_{22} \\
 \Delta _{1235} \leftrightarrow & p_{70}+ p_{72}- p_{74}+ p_{79} & & \Delta _{1567} \leftrightarrow & p_{16}+ p_{31} \\
 \Delta _{1236} \leftrightarrow & p_{80}+ p_{87} & & \Delta _{2345} \leftrightarrow &- p_{45}- p_{52} \\
 \Delta _{1237} \leftrightarrow & p_{64}+ p_{88} & & \Delta _{2346} \leftrightarrow &- p_{53}- p_{57} \\
 \Delta _{1245} \leftrightarrow & p_{18}+ p_{25} & & \Delta _{2347} \leftrightarrow & p_{34}- p_{38}- p_{40}- p_{58} \\
 \Delta _{1246} \leftrightarrow & p_{21}+ p_{30} & & \Delta _{2356} \leftrightarrow & p_{48}+ p_{50} \\
 \Delta _{1247} \leftrightarrow & p_{15}+ p_{27} & & \Delta _{2357} \leftrightarrow & p_{36}+ p_{37}+ p_{41}+ p_{51} \\
 \Delta _{1256} \leftrightarrow & p_{28}- p_{24} & & \Delta _{2367} \leftrightarrow & p_{42}+ p_{44} \\
 \Delta _{1257} \leftrightarrow &- p_{13}- p_{23} & & \Delta _{2456} \leftrightarrow & p_5+ p_8 \\
 \Delta _{1267} \leftrightarrow &- p_{17}- p_{32} & & \Delta _{2457} \leftrightarrow & p_1+ p_3 \\
 \Delta _{1345} \leftrightarrow & p_{65}- p_{66}+ p_{76}- p_{81} & & \Delta _{2467} \leftrightarrow & p_2- p_{10} \\
 \Delta _{1346} \leftrightarrow & p_{68}- p_{82} & & \Delta _{2567} \leftrightarrow & p_4+ p_7 \\
 \Delta _{1347} \leftrightarrow & p_{59}- p_{61}+ p_{69}- p_{77} & & \Delta _{3456} \leftrightarrow & p_{46}+ p_{54} \\
 \Delta _{1356} \leftrightarrow & p_{73}+ p_{84} & & \Delta _{3457} \leftrightarrow & p_{33}+ p_{39}+ p_{47}+ p_{55} \\
 \Delta _{1357} \leftrightarrow & p_{60}+ p_{62}+ p_{71}+ p_{85} & & \Delta _{3467} \leftrightarrow & p_{35}+ p_{56} \\
 \Delta _{1367} \leftrightarrow & p_{63}+ p_{86} & & \Delta _{3567} \leftrightarrow & p_{43}- p_{49} \\
 \Delta _{1456} \leftrightarrow & p_{20}+ p_{29} & & \Delta _{4567} \leftrightarrow &- p_6- p_9 \\
 \Delta _{1457} \leftrightarrow & p_{11}+ p_{14}+ p_{19}+ p_{26} & & \\
\end{array}
\end{equation}}

\bigskip

\noindent All perfect matchings nicely appear now. It is straightforward to verify that all perfect matchings indeed have the source sets associated to the corresponding \pl coordinate.

\bigskip

\paragraph{Example: 4 Boundaries.}

\begin{figure}[htb!]
\centering
\includegraphics[scale=0.33]{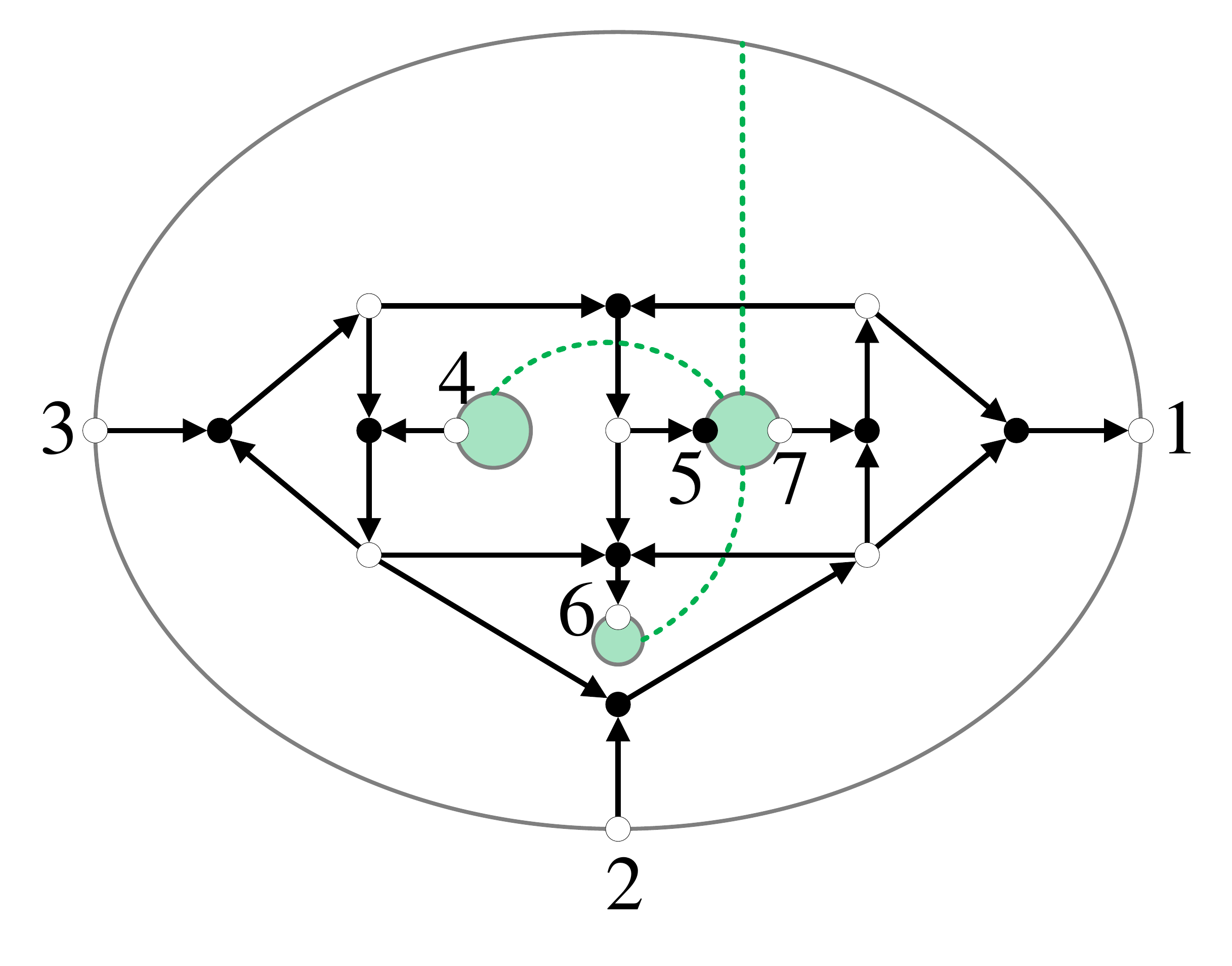}
\caption{A bipartite graph with 4 boundaries. It is related to the one in \fref{fig:3Bguy} by redistributing external nodes over boundaries. As before, we pick a perfect orientation given by the perfect matching $p_1$, which contains the edges $X_{1,4}$, $X_{3,7}$, $X_{3,10}$, $X_{6,4}$, $X_{6,7}$, $X_{8,2}$ and $X_{8,10}$.}
\label{fig:4Bguy}
\end{figure}

To illustrate our methods, let us consider the example with 4 boundaries shown in \fref{fig:4Bguy}. This is basically a formal exercise, mainly intended to see once again the general techniques at work, since, as the alert reader might easily realize, the new example only differs from \fref{fig:3Bguy} by changing the distribution of external nodes over boundaries. Such reorganization can be regarded as an elaborate generalization of external leg crossing. Having noticed the relation to the previous example, our discussion will be briefer.

Perfect matchings are the same as for the previous example. Even choosing the same reference perfect matching, the sign assignment is completely changed due to the new cuts and ordering of external nodes. The new signs however conspire to generate simple expressions for the \pl coordinates in terms of perfect matchings. The boundary measurement is given by:

\begin{equation}
C = 
\left(
\begin{array}{c|ccccccc}
 & 1 & 2 & 3 & 4 & 5 & 6 & 7 \\
 \hline
2 \ \ & \mathfrak{p}_{11}+\mathfrak{p}_{19} & 1 & 0 & 0 & -\mathfrak{p}_6 & -\mathfrak{p}_{33}-\mathfrak{p}_{47} & 0 \\
3 \ \ & -\frac{\mathfrak{p}_{15}}{\mathfrak{p}_3+1}-\frac{\mathfrak{p}_{27}}{\mathfrak{p}_3+1} & 0 & 1 & 0 & \frac{\mathfrak{p}_{10}}{\mathfrak{p}_3+1}-\frac{\mathfrak{p}_2}{\mathfrak{p}_3+1} & -\frac{\mathfrak{p}_{34}}{\mathfrak{p}_3+1}+\frac{\mathfrak{p}_{40}}{\mathfrak{p}_3+1}+\frac{\mathfrak{p}_{58}}{\mathfrak{p}_3+1}-\frac{\mathfrak{p}_{38}}{\mathfrak{p}_3+1} & 0 \\
4 \ \ &  \frac{\mathfrak{p}_{13}}{\mathfrak{p}_3+1}+\frac{\mathfrak{p}_{23}}{\mathfrak{p}_3+1} & 0 & 0 & 1 & -\frac{\mathfrak{p}_4}{\mathfrak{p}_3+1}-\frac{\mathfrak{p}_7}{\mathfrak{p}_3+1} & \frac{\mathfrak{p}_{36}}{\mathfrak{p}_3+1}-\frac{\mathfrak{p}_{37}}{\mathfrak{p}_3+1}-\frac{\mathfrak{p}_{41}}{\mathfrak{p}_3+1}-\frac{\mathfrak{p}_{51}}{\mathfrak{p}_3+1} & 0 \\
7 \ \ &  -\mathfrak{p}_{18} & 0 & 0 & 0 & \mathfrak{p}_5 & \mathfrak{p}_{45} & 1 
\end{array}
\right),
\end{equation}
which gives the following map between \pl coordinates and perfect matchings:

{\small
\begin{equation}
\begin{array}{llcrl}
 \Delta _{1234} \leftrightarrow & p _{18}+ p _{25} & \ \ \ \ \ & \Delta _{1467} \leftrightarrow & p _{59}+ p _{61}+ p _{69}+ p _{77} \\
 \Delta _{1235} \leftrightarrow & p _{24}- p _{28} & & \Delta _{1567} \leftrightarrow &- p _{63}- p _{86} \\
 \Delta _{1236} \leftrightarrow & p _{70}- p _{72}+ p _{74}- p _{79} & & \Delta _{2345} \leftrightarrow & p _5+ p _8 \\
 \Delta _{1237} \leftrightarrow & p _{13}+ p _{23} & & \Delta _{2346} \leftrightarrow & p _{45}+ p _{52} \\
 \Delta _{1245} \leftrightarrow & p _{21}+ p _{30} & & \Delta _{2347} \leftrightarrow & p_1 + p_3 \\
 \Delta _{1246} \leftrightarrow & p _{67}+ p _{75}- p _{78}+ p _{83} & & \Delta _{2356} \leftrightarrow & p _{50}- p _{48} \\
 \Delta _{1247} \leftrightarrow & p _{15}+ p _{27} & & \Delta _{2357} \leftrightarrow &- p _4- p _7 \\
 \Delta _{1256} \leftrightarrow & p _{87}- p _{80} & & \Delta _{2367} \leftrightarrow & p _{36}- p _{37}- p _{41}- p _{51} \\
 \Delta _{1257} \leftrightarrow &- p _{17}- p _{32} & & \Delta _{2456} \leftrightarrow & p _{57}- p _{53} \\
 \Delta _{1267} \leftrightarrow &- p _{64}- p _{88} & & \Delta _{2457} \leftrightarrow & p _2- p _{10} \\
 \Delta _{1345} \leftrightarrow & p _{20}+ p _{29} & & \Delta _{2467} \leftrightarrow & p _{34}+ p _{38}- p _{40}- p _{58} \\
 \Delta _{1346} \leftrightarrow &- p _{65}+ p _{66}- p _{76}+ p _{81} & & \Delta _{2567} \leftrightarrow & p _{44}- p _{42} \\
 \Delta _{1347} \leftrightarrow & p _{11}+ p _{14}+ p _{19}+ p _{26} & & \Delta _{3456} \leftrightarrow & p _{46}+ p _{54} \\
 \Delta _{1356} \leftrightarrow & p _{84}- p _{73} & & \Delta _{3457} \leftrightarrow &- p _6- p _9 \\
 \Delta _{1357} \leftrightarrow &- p _{16}- p _{31} & & \Delta _{3467} \leftrightarrow &- p _{33}- p _{39}- p _{47}- p _{55} \\
 \Delta _{1367} \leftrightarrow & p _{60}- p _{62}+ p _{71}- p _{85} & & \Delta _{3567} \leftrightarrow & p _{43}+ p _{49} \\
 \Delta _{1456} \leftrightarrow &- p _{68}- p _{82} & & \Delta _{4567} \leftrightarrow & p _{56}- p _{35} \\
 \Delta _{1457} \leftrightarrow & p _{12}+ p _{22} & & & \\
\end{array}
\end{equation}}

\bigskip
\bigskip

Before closing, let us briefly discuss the effect of two operations that can affect the ordering of external nodes: modification of cuts and redistribution of external nodes over boundaries, including the possibility of creating new ones. Changing cuts has no net effect: once the labels of \pl coordinates have been permuted to the original order, one finds the same map between minors and perfect matchings. Changing the actual distribution of external nodes over boundaries by flipping external legs produces a new map, in which the \textit{relative} signs of the perfect matchings are different.

For planar graphs the latter operation has a simultaneously strong and irrelevant effect. Such a change in general implies the loss of positivity. The decomposition of the flipped diagram will not be the positroid stratification, because each irreducible subgraph will not correspond to a different positroid. However, it would be positroid-like: apart from the matroid labels of each irreducible subgraph, the poset for the non-planar case would be identical to that of the positroid stratification. In fact, permuting the labels of the matroid strata will reproduce the positroid stratification. This is further discussed in \sref{section_matroid_graphs}, where the case of $Gr_{2,4}$ is shown explicitly.

\bigskip

\section{Combinatorial Decomposition of Non-Planar Graphs}
\label{sec:NonplDecomp}

In this section we will apply the techniques introduced in \sref{Strat_NewRegionNewMethod} to non-planar diagrams. We present in detail a few examples and construct their decomposition. As we will show in these examples, the combinatorial decomposition of non-planar on-shell diagrams does not correspond to the positroid stratification of the Grassmannian, 
but is still a subset of the matroid stratification. \sref{section_matroid_graphs} collects some ideas about how the full matroid stratification might be achieved by combining different graphs.

\bigskip

\subsection*{Example 1: Graph on the Annulus}
We begin by illustrating our techniques with the example displayed in \fref{fig:nonplanarfriend}. This example has 15 perfect matchings. The matching polytope is given by \eref{annulusPmatrix} and is 6-dimensional. The matroid polytope is 

\bigskip

{\small
\begin{equation}
G_{\text{matroid}} = 
\left(
\begin{array}{c|ccccccccccccccc}
  & \ p_1 \ & \ p_2 \ & \ p_3 \ & \ p_4 \ & \ p_5 \ & \ p_6 \ & \ p_7 \ & \ p_8 \ & \ p_9 \ & p_{10} & p_{11} & p_{12} & p_{13} & p_{14} & p_{15} \\
 \hline
 X_{2,3} \ & 1 & 1 & 0 & 0 & 0 & 1 & 0 & 0 & 1 & 1 & 0 & 0 & 0 & 0 & 0 \\
 X_{5,4} \ & 1 & 0 & 1 & 0 & 0 & 1 & 1 & 0 & 1 & 0 & 1 & 1 & 0 & 0 & 0 \\
 X_{5,6} \ & 1 & 0 & 1 & 1 & 0 & 0 & 1 & 0 & 0 & 0 & 1 & 0 & 1 & 0 & 0 \\
 X_{3,2} \ & 0 & 0 & 0 & 0 & 0 & 1 & 1 & 1 & 0 & 0 & 0 & 1 & 0 & 1 & 0 \\
 X_{6,4} \ \ & 0 & 0 & 1 & 0 & 1 & 0 & 0 & 0 & 1 & 0 & 1 & 1 & 0 & 0 & 1 \\
\end{array}
\right)
\label{GmatroidAnnulusSec11}
\end{equation}}

\bigskip

\noindent and has dimension 4. This example has 10 non-vanishing \pl coordinates, and the following \pl relations:
\begin{eqnarray}
\Delta _{125} \Delta _{134}-\Delta _{124} \Delta _{135}+\Delta _{123} \Delta _{145}=0 \nonumber \\
\Delta _{125} \Delta _{234}-\Delta _{124} \Delta _{235}+\Delta _{123} \Delta _{245}=0 \nonumber \\
\Delta _{135} \Delta _{234}-\Delta _{134} \Delta _{235}+\Delta _{123} \Delta _{345}=0 \nonumber \\
\Delta _{145} \Delta _{234}-\Delta _{134} \Delta _{245}+\Delta _{124} \Delta _{345}=0 \nonumber \\
\Delta _{145} \Delta _{235}-\Delta _{135} \Delta _{245}+\Delta _{125} \Delta _{345}=0
\label{annulusPluckerRels}
\end{eqnarray}
of which only $3$ are independent.

The face lattice of the matching polytope contains 412 elements of various dimensions; it is therefore very impractical to draw the full poset. Below we present the first level in detail, subsequent levels follow analogously. 

\bigskip

\paragraph{First Level: Dimension 5.} This example has 13 edges. We now proceed by removing them to obtain the first level of the face lattice of the matching polytope, which contains the following faces:

\bigskip

{\small
\begin{equation}
\begin{array}{|c|c|}
\hline
\text{Removed} & \multirow{2}{*}{Face} \\ 
\text{edge}  &  \\
\hline
X_{1,3}  & \; p_6,p_7,p_8,p_9,p_{10},p_{11},p_{12},p_{13},p_{14},p_{15} \\
X_{1,6}  & \; p_3,p_4,p_5,p_9,p_{10},p_{11},p_{12},p_{13},p_{14},p_{15} \\
X_{3,6}  & \; p_1,p_2,p_3,p_4,p_5,p_6,p_9,p_{10},p_{12},p_{14} \\
X_{6,1}  & \; p_1,p_2,p_3,p_4,p_5,p_7,p_8,p_{11},p_{13},p_{15} \\
X_{1,5}  & \; p_1,p_3,p_4,p_6,p_7,p_9,p_{10},p_{11},p_{12},p_{13},p_{14} \\
X_{2,1}  & \; p_1,p_2,p_3,p_4,p_5,p_6,p_7,p_8,p_{12},p_{14} \\
X_{4,1}  & \; p_1,p_2,p_3,p_5,p_6,p_7,p_8,p_9,p_{11},p_{12},p_{15} \\
X_{6,2}  & \; p_1,p_2,p_6,p_7,p_8,p_9,p_{10},p_{11},p_{13},p_{15} \\
X_{2,3}  & \; p_3,p_4,p_5,p_7,p_8,p_{11},p_{12},p_{13},p_{14},p_{15} \\
X_{5,4}  & \; p_2,p_4,p_5,p_8,p_{10},p_{13},p_{14},p_{15} \\
X_{5,6}  & \; p_2,p_5,p_6,p_8,p_9,p_{10},p_{12},p_{14},p_{15} \\
X_{3,2}  & \; p_1,p_2,p_3,p_4,p_5,p_9,p_{10},p_{11},p_{13},p_{15} \\
X_{6,4}  & \; p_1,p_2,p_4,p_6,p_7,p_8,p_{10},p_{13},p_{14} \\
\hline
\end{array} 
\label{annulusmatroidpoly}
\end{equation} 
}
\bigskip

\noindent where the faces in the table show the surviving perfect matchings after removal of the corresponding edge. In order to find the decomposition we are interested in, we proceed by identifying perfect matchings which have the same coordinate in the matroid polytope, as explained in \sref{Strat_NewRegionNewMethod}.
This can be done by looking at \eref{GmatroidAnnulusSec11}, or directly from \eref{annulusPluckerMap}, and is:

\begin{equation}
\begin{array}{ccccccccc}
\{p_1\} & \ \ \ \ & \{p_2,p_{10}\} & \ \ \ \ & \{p_3,p_{11}\} & \ \ \ \ & \{p_4,p_{13}\} & \ \ \ \ & \{p_5,p_{15}\} \\ 
\{p_6\} & & \{p_7\} & & \{p_8,p_{14}\} &  & \{p_9\} & & \{p_{12}\} 
\end{array} .
\label{AnnulusIdentify}
\end{equation}

\noindent The faces then become:

{\small
\begin{equation}
\hspace{-.225cm}\begin{array}{|c|c|c|}
\hline
\text{Removed} & \multirow{2}{*}{Face} & \multirow{2}{*}{\pl Coordinates} \\ 
\text{edge}  &  &   \\
\hline
X_{1,3} & \; p_6,p_7,p_8,p_9,p_{10},p_{11},p_{12},p_{13},p_{15} & \Delta _{123},\Delta _{234},\Delta _{134},\Delta _{125},\Delta _{135},\Delta _{245},\Delta _{124},\Delta _{345},\Delta _{145} \\
X_{3,6} & \; p_1,p_2,p_3,p_4,p_5,p_6,p_9,p_{12},p_{14} & \Delta _{235},\Delta _{135},\Delta _{245},\Delta _{345},\Delta _{145},\Delta _{123},\Delta _{125},\Delta _{124},\Delta _{134} \\
X_{1,5} & \; p_1,p_3,p_4,p_6,p_7,p_9,p_{10},p_{12},p_{14} & \Delta _{235},\Delta _{245},\Delta _{345},\Delta _{123},\Delta _{234},\Delta _{125},\Delta _{135},\Delta _{124},\Delta _{134} \\
X_{2,1} & \; p_1,p_2,p_3,p_4,p_5,p_6,p_7,p_8,p_{12} & \Delta _{235},\Delta _{135},\Delta _{245},\Delta _{345},\Delta _{145},\Delta _{123},\Delta _{234},\Delta _{134},\Delta _{124} \\
X_{4,1} & \; p_1,p_2,p_3,p_5,p_6,p_7,p_8,p_9,p_{12} & \Delta _{235},\Delta _{135},\Delta _{245},\Delta _{145},\Delta _{123},\Delta _{234},\Delta _{134},\Delta _{125},\Delta _{124} \\
X_{6,2} & \; p_1,p_2,p_6,p_7,p_8,p_9,p_{11},p_{13},p_{15} & \Delta _{235},\Delta _{135},\Delta _{123},\Delta _{234},\Delta _{134},\Delta _{125},\Delta _{245},\Delta _{345},\Delta _{145} \\
\hline
X_{1,6} & \; p_3,p_4,p_5,p_9,p_{10},p_{12},p_{14} & \Delta _{245},\Delta _{345},\Delta _{145},\Delta _{125},\Delta _{135},\Delta _{124},\Delta _{134} \\
X_{6,1} & \; p_1,p_2,p_3,p_4,p_5,p_7,p_8 & \Delta _{235},\Delta _{135},\Delta _{245},\Delta _{345},\Delta _{145},\Delta _{234},\Delta _{134} \\
X_{2,3} & \; p_3,p_4,p_5,p_7,p_8,p_{12} & \Delta _{245},\Delta _{345},\Delta _{145},\Delta _{234},\Delta _{134},\Delta _{124} \\
X_{5,4} & \; p_2,p_4,p_5,p_8 & \Delta _{135},\Delta _{345},\Delta _{145},\Delta _{134} \\
X_{5,6} & \; p_2,p_5,p_6,p_8,p_9,p_{12} & \Delta _{135},\Delta _{145},\Delta _{123},\Delta _{134},\Delta _{125},\Delta _{124} \\
X_{3,2} & \; p_1,p_2,p_3,p_4,p_5,p_9 & \Delta _{235},\Delta _{135},\Delta _{245},\Delta _{345},\Delta _{145},\Delta _{125} \\
X_{6,4} & \; p_1,p_2,p_4,p_6,p_7,p_8 & \Delta _{235},\Delta _{135},\Delta _{345},\Delta _{123},\Delta _{234},\Delta _{134} \\
\hline
\end{array} \quad .
\label{nonplBoundLevel1Plucker}
\end{equation} 
}

\noindent In the table above we show the surviving perfect matchings after removing the corresponding edge in the graph, and after the identifications in \eref{AnnulusIdentify}. We also show the non-vanishing \pl coordinates for each subgraph.

As a consequence of the identifications, the faces in the lower half of the table are of dimension lower than 5 and get identified with other lower-dimensional ones, i.e.\ they are subject to vertical identifications. This can be deduced by counting the surviving \pl coordinates and relevant \pl relations \eref{annulusPluckerRels}. Hence $X_{1,6}$, $X_{6,1}$, $X_{2,3}$, $X_{5,4}$, $X_{5,6}$, $X_{3,2}$ and $X_{6,4}$ are not removable edges. For the remaining 6 boundaries there is no horizontal identification at this level, so the 6 removable edges are $X_{1,3}$, $X_{3,6}$, $X_{1,5}$, $X_{2,1}$, $X_{4,1}$ and $X_{6,2}$. The removal of any of these edges yields a 5-dimensional element of the Grassmannian. Each of these corresponds to a differential form which is a singularity in the sense explained in \sref{singularities}. Moreover, each of the boundaries also corresponds to a matroid stratum with 9 elements each, where the elements are given by the indices of the \pl coordinates in \eref{nonplBoundLevel1Plucker}. 

\bigskip

\paragraph{Full Combinatorial Decomposition.} To represent the boundaries of the entire poset, we group the elements in each level of the poset by how many perfect matchings they have, thus presenting the information of each level by pairs of numbers, where the first specifies the number of faces of a certain type and the second specifies the type. For example, $14[6]$ means there are 14 faces, each containing 6 perfect matchings. This information is presented in \tref{table:facelattice2Bguy}.

\begin{table}[htt!!]
\begin{center}
\begin{tabular}{|c|c|}
\hline
$d$ & {\bf Faces of matching polytope}  \\ \hline \hline
{\bf 5} & 1[8], 2[9], 8[10], 2[11] \\ \hline
{\bf 4} & 11[5] 14[6], 23[7], 12[8] \\ \hline
{\bf 3} & 67[4], 46[5], 13[6] \\ \hline
{\bf 2} & 112[3], 19[4] \\ \hline
{\bf 1} & 67[2] \\ \hline
{\bf 0} & 15[1] \\ \hline
\end{tabular}
\caption{Faces of the matching polytope. At each level of dimension $d$, a pair of numbers $m[n]$ indicates that there are $m$ boundaries consisting of $n$ perfect matchings. \label{table:facelattice2Bguy}
}
\end{center}
\end{table}

After the identification \eref{AnnulusIdentify}, 272 of the faces get identified with other boundaries, to yield a poset with 140 elements, described by \tref{table:GrassmannianPoset2Bguy}.
\begin{table}[htt!!]
\begin{center}
\begin{tabular}{|c|c|}
\hline
$d$ & {\bf Matroids}  \\ \hline \hline
{\bf 5} & 6[9] \\ \hline
{\bf 4} & 5[6], 6[7], 6[8] \\ \hline
{\bf 3} & \ \ 5[4], 5[24], 6[6] \ \ \\ \hline
{\bf 2} & 30[3], 12[4] \\ \hline
{\bf 1} & 30[2] \\ \hline
{\bf 0} & 10[1] \\ \hline
\end{tabular}
\caption{Matroids in the decomposition of the diagram shown in \fref{fig:nonplanarfriend}. At each level, a pair of numbers $m[n]$ indicates that there are $m$ matroids consisting of $n$ bases. \label{table:GrassmannianPoset2Bguy}
}
\end{center} 
\end{table}
It is straightforward to verify that these tables agree with the detailed analysis of the first level presented before.

As a further check, using the methods introduced in \sref{sec:QuantGraphRed} and applying the identification \eref{AnnulusIdentify} it is straightforward to check that \tref{table:GrassmannianPoset2Bguy} is consistent with the poset obtained by deleting only removable edges.

\bigskip

\subsection*{Example 2: Graph with 3 Boundaries}
To further illustrate the computational power of these techniques, we treat the example presented in \fref{fig:3Bguy}, which has 88 perfect matchings. The matching polytope has in total $74670$ faces, which after identification reduces to $8585 $ faces. The face lattice information, before and after identification, is presented in the table below.

\bigskip
\begin{center}
{\small
$
\begin{array}{|c|c|c|}
\hline
d & \text{\bf Faces of matching polytope} & \text{\bf Matroids} \\ \hline \hline
\multirow{2}{*} {\bf 10} & 2[48], 4[52], 1[56], 2[58], 4[59], & 1[30], 2[31], 5[33], 6[34] \\
   & 1[60], 2[64], 4[65], 1[68], 2[70]  & \\ \hline
\multirow{5}{*} {\bf 9}  & 4[24], 6[28], 4[29], 8[30], 15[32], & 6[20], 5[25], 20[28], 6[29], 22[30], \\
   & 2[34], 4[35], 9[36], 16[37], 8[38], & 4[31], 20[32], 4[33] \\
   & 16[39], 17[40], 8[41], 16[42], 8[43], & \\
   & 30[44], 12[45], 4[46], 12[47], 6[48], & \\
   & 8[49], 6[50], 9[52], 2[54] &  \\ \hline
\multirow{6}{*} {\bf 8} & 4[10], 4[12], 16[14], 12[16], 16[17], & 7[15], 60[19], 5[20], 12[22], 42[23], \\
  & 10[18], 8[19], 51[20], 44[21], 54[22], & 12[24], 8[25], 38[26], 80[27], 16[28], \\
  & 52[23], 98[24], 40[25], 92[26], 112[27], & 32[29], 8[31] \\
  & 83[28], 60[29], 122[30], 52[31], 98[32], &  \\
  & 60[33], 100[34], 16[35], 66[36], 8[37], &  \\
  & 18[38], 20[39], 3[40] &  \\ \hline
\multirow{5}{*} {\bf 7} & 40[8], 48[9], 36[10], 24[11], 204[12], & 77[14], 114[16], 154[18], 74[19], 5[20], \\
  & 48[13], 182[14], 216[15], 251[16], 488[17], & 106[21], 62[22], 58[23], 48[24], 68[25], \\
  & 518[18], 264[19], 602[20], 284[21], 432[22], & 20[26] \\
  & 292[23], 265[24], 140[25], 246[26], 72[27], &  \\
  & 84[28], 36[29], 8[30] &  \\ \hline
\multirow{4}{*} {\bf 6} & 424[7], 292[8], 216[9], 988[10], 724[11], & 63[10], 100[12], 163[13], 292[14], 274[15], \\
  & 1079[12], 1720[13], 1742[14], 1296[15], 849[16], & 24[16], 146[17], 140[18], 100[19], 22[20], \\
  & 656[17], 728[18], 236[19], 226[20], 192[21], & 70[21] \\
  & 32[22] &  \\ \hline
\multirow{3}{*} {\bf 5} & 1880[6], 892[7], 2636[8], 2656[9], 4618[10], & 611[9], 90[10], 230[11], 352[12], 396[13], \\
  & 2012[11], 1686[12], 952[13], 410[14], 228[15], & 66[14], 68[15], 68[16] \\
  & 177[16] &  \\ \hline
\multirow{2}{*} {\bf 4} & 4452[5], 3170[6], 5876[7], 3859[8], 788[9], & 21[5], 105[6], 534[7], 731[8], 140[9], \\
  & 908[10], 116[12] & 322[10], 41[12] \\ \hline
{\bf 3} & 6242[4], 4044[5], 2622[6], 135[8] & 140[4], 586[5], 534[6], 61[8] \\ \hline
{\bf 2} & 4260[3], 1077[4] & 350[3], 293[4] \\ \hline
{\bf 1} & 1134[2] & 210[2] \\ \hline
{\bf 0} & 88[1] & 35[1] \\ \hline
\end{array}
$
}
\end{center}
\bigskip

\subsection{Non-Eulerian Posets}
The face lattice of a convex polytope is a graded poset. Moreover this poset is Eulerian, which means that the number of elements of even dimension is one more than the number of elements of odd dimension, i.e.\
\begin{equation}
\sum_{i=0}^d (-1)^i N_B^{(i)} = 1 ,
\end{equation}
where $d$ is the dimension of the polytope and $N_B^{(i)}$ is the number of faces of the polytope of dimension $i$.\footnote{If we were to include the empty set in our face lattice, the number of boundaries would sum to 0 rather than 1.} 

As a check that the face lattice of the matching polytope for non-planar graphs can be obtained through successive edge removal, we evaluate the Eulerian number in the two previous examples:
\begin{eqnarray}
\text{\textbf{Example 1:}} \quad \sum_{i=0}^6 (-1)^i N_B^{(i)} &= 15-67+131-\ldots +1= 1 \\
\text{\textbf{Example 2:}} \quad \sum_{i=0}^{11} (-1)^i N_B^{(i)} &= 88 - 1134 + \ldots -1 = 1 .
\end{eqnarray}

While the positroid stratification was shown to be Eulerian \cite{2005arXiv09129W}, for non-planar cases the combinatorial decomposition is in general not Eulerian. This can be seen for example by computing the Eulerian number for the two examples above:
\begin{eqnarray}
\text{\textbf{Example 1:}} \quad \sum_{i=0}^6 (-1)^i N_B^{(i)} &= 10-30+42-\ldots +1= -1 , \\
\text{\textbf{Example 2:}} \quad \sum_{i=0}^{11} (-1)^i N_B^{(i)} &= 35 - 210 + \ldots -1 = 14 .
\end{eqnarray}

The appearance of non-Eulerian posets should not be surprising. Due to the identifications involved in the combinatorial decomposition, the resulting poset might not describe the face lattice of a geometric polytope.

\bigskip

\section{Matroid Stratification from Multiple Graphs}
\label{section_matroid_graphs}

As already explained in \sref{section_partial_matroid_stratification}, the combinatorial decomposition yields a subset of the matroid stratification: only certain strata appear in the decomposition. It is then natural to ask whether it is possible to extend it such that it produces the full matroid stratification. This leads us to the following reasonable conjecture:

\bigskip
\begin{itemize}
\item{\bf Conjecture:} The full matroid stratification can be obtained by {\it simultaneously} considering the combinatorial decomposition of {\it multiple} bipartite graphs associated to Grassmannian elements with a maximal number of degrees of freedom. Some of these graphs are non-planar. The matroid stratification is given by the union of the resulting strata.
\end{itemize}
\bigskip

This proposal follows from the definition of the matroid stratification in \sref{matroidstra}. Analogously to the positroid stratification, where we take the common refinement of $n$ \textit{cyclically permuted} Schubert cells, hence $n$ cyclic permutations, the matroid stratification is in general the refinement over \textit{all} $n!$  permutations. 
Here we remind that every permutation specifies a lexicographic order that characterizes the Schubert cell, analogously to \sref{Schubert_dec}. The distribution of external nodes over boundaries gives rise, following the discussion in \sref{section_boundary_measurement_non-planar}, to different orderings, which we map to these permutations.

In essence, 
to access all the permutations and hence all the matroids, we have to consider 
permutations which cannot be obtained by cyclic rotations of $1,2,\ldots,n$, which are the only ones
that can be realized on planar graphs. The other permutations can be obtained only by introducing new boundaries, thus making the graphs non-planar.

\begin{figure}[htb!]
\centering
\includegraphics[scale=1]{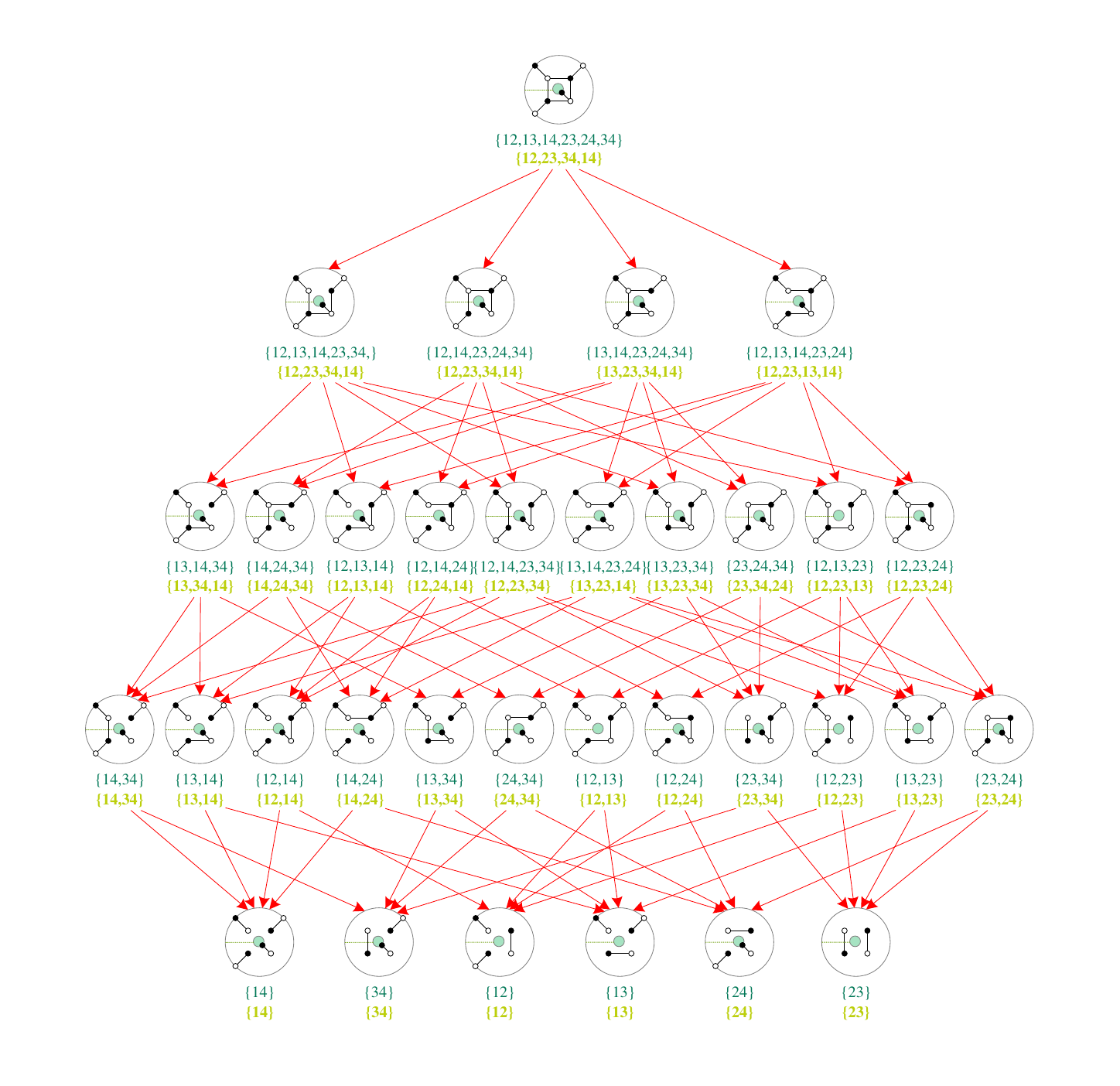}
\caption{Decomposition of the square box with flipped legs and two boundaries. It corresponds to the permutation $1243$. The dark green label indicates the matroid stratum corresponding to the graph, the light green label indicates the positroid stratum.}
\label{fig:SqbPermutedStrat}
\end{figure}

To illustrate this idea, let us consider the decomposition of the diagram in \fref{fig:nonplanars}.b which, after introducing an additional boundary and the corresponding cut, is the same as the square box but with ordering $1243$. The decomposition is obtained through the procedure explained in \sref{Strat_NewRegionNewMethod} and is shown in \fref{fig:SqbPermutedStrat}, where the matroid label is given in dark green and the positroid label is in light green. 
The matroid labels are identical to those of \fref{G24posi}, 
but with 3 and 4 interchanged,
as mentioned at the end of \sref{section_bm_beyond_annulus}. The fact that we no longer have the positroid stratification is confirmed by the fact that the positroid stratum $\{ C \in Gr_{2,4} \mid \Delta_{12} \neq 0 , \Delta_{23} \neq 0 , \Delta_{34} \neq 0 , \Delta_{14} \neq 0\}$ has multiple representatives, and some positroid strata are missing, e.g.\ $\{ C \in Gr_{2,4} \mid \Delta_{12} \neq 0 , \Delta_{24} \neq 0 , \Delta_{34} \neq 0 , \Delta_{14} \neq 0\}$. However, we note that the decomposition just obtained is precisely the same as that of \sref{sec:PositroidStrat} but where each component is the simultaneous refinement of 4 cyclically permuted Schubert cells with respect to the lexicographic order specified by the permutation $1243$.

\begin{figure}[htb!]
\centering
\includegraphics[scale=0.55]{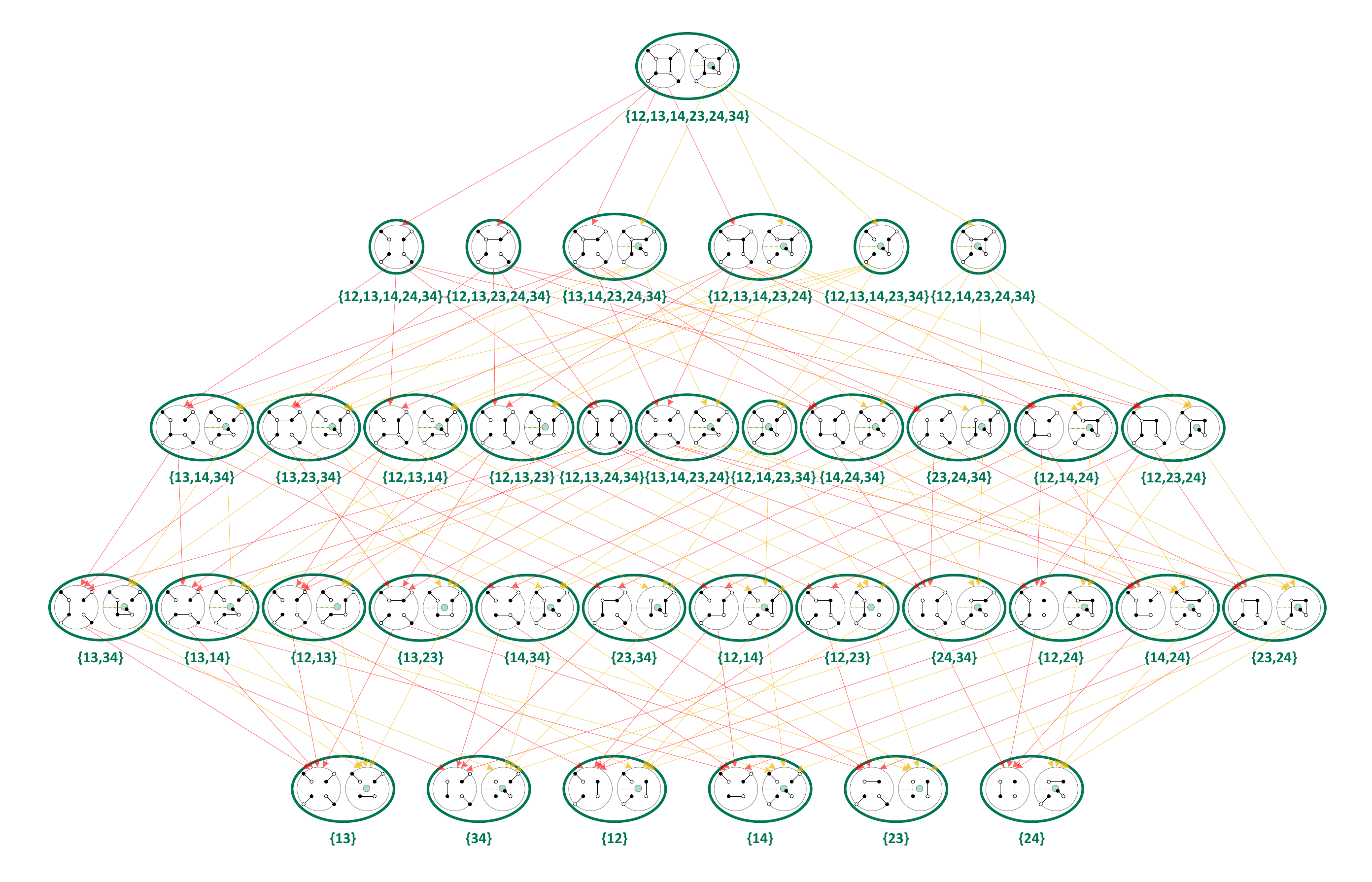}
\caption{Matroid stratification of $Gr_{2,4}$ via a pair of graphs, both planar and non-planar. Matroid strata are indicated by green circles. Red and yellow arrows belong to the combinatorial decompositions of the planar and non-planar graphs, respectively.}
\label{fig:Gr24MatroidStrat}
\end{figure}

In the decomposition of the non-planar graph, the matroid strata that were missing from the decomposition of the planar case with ordering $1234$, marked in red in \eref{matroidstr}, are now present. Hence we conclude that the union of the matroid strata of the decomposition in \fref{G24posi} and \fref{fig:SqbPermutedStrat} gives the entire matroid stratification,
at least at the combinatorial level.
We provide in \fref{fig:Gr24MatroidStrat} a depiction of how the two decompositions together form the entire matroid stratification. The matroid strata are marked by a green circle, where the matroid labels have been included underneath.  

Generally, including all $n!$ permutations of external edges modulo cyclicity will include all matroid strata, but in practice it can be sufficient to consider fewer permutations. 

Let us explain why this is the case and show how to determine the diagrams required for the matroid stratification in the case of $Gr_{2,4}$, whose matroid contains the 6 bases $12$, $13$, $14$, $23$, $24$ and $34$. We begin by only discussing the problem in terms of permutations and lexicographic orders, 
and explain how the graphs fit into this picture at a secondary stage.

Each permutation of $1,2,3,4$ specifies a lexicographic order, to which we can associate a
Schubert cell analogously to the definition in \sref{Schubert_dec}. The positroid stratification uses $n$ permutations, related to each other by cyclic shifts, and the corresponding
Schubert cells, and is then specified by $n$ entries.
To put a label in each entry, 
we select the lexicographically minimal non-zero element with respect to the permutation in question. 
For example, the permutation 2413 will select the matroid element $(24)$, if present, otherwise select $(21)$, if present, etc. 

The matroid stratification generically uses $n !$ permutations. However, in order
to find all the strata, it is sufficient to refine over the set of Schubert cells such that for each base there exists a Schubert cell whose lexicographic order has that base as minimal element. Thus, to specify all matroids in the example at hand, we will need 6 permutations, each permutation having a different lexicographically minimal order of the form:
\begin{equation}
\label{permu_matr}
\begin{array}{cccccc}
12XX, & 13XX, & 14XX, & 23XX, & 24XX, & 34XX ,  
\end{array}
\end{equation}
where $XX$ may be any order of the remaining two digits, e.g.\ it does not matter whether we choose 1342 or 1324. For example, 
the first lexicographic order will always find the matroid base $12$, regardless of the presence or absence of other bases; 
the second one will always find $13$ regardless of the other matroid bases, and so on. 
Strictly speaking the order of the first two digits is also irrelevant, since either order specifies the same matroid element. 
In this way, each matroid base, if present, will appear in one of the six entries associated to the different lexicographic orders. 
A set of $6$ permutations as in (\ref{permu_matr}) are sufficient 
for labeling all matroids with the correct matroid labels.

Graphs fit into this picture as follows. Each graph specifies an ordering, dictated by the arrangement of the external edges. Because of cyclicity of the starting point, 
the graph actually specifies $n$ orderings, related to each other by cyclic shifts.
 In this example, the planar graph has the ordering 1234, which specifies the permutations
\begin{equation}
\begin{array}{cccc}
1234, & 2341, & 3412, & 4123 ,
\end{array} 
\label{ordering1}
\end{equation}
which simply differ in which edge of the graph we call ``1''. We see that such a graph contains $4$ of the required lexicographic orders.\footnote{We remind once more that it does not matter whether it is 4123 or 1423: either way the lexicographically minimal element will be the one corresponding to the \pl coordinate $\Delta_{14}$.} We are however still missing a permutation of the form $13XX$ and one of the form $24XX$. If we introduce a second graph with the ordering 1243, we obtain the permutations
\begin{equation}
\begin{array}{cccc}
1243, & 2431, & 4312, & 3124 ,
\end{array}
\label{ordering2}
\end{equation}
which contain the lexicographic orders given by $3124$ and $2431$ as desired, and two more which were already covered by the previous graph. Thus, we see that the two graphs with ordering 1234 and 1243 are sufficient to cover all lexicographic orders and corresponding Schubert cells which are required to specify the matroids. 
We then argue that their decomposition will cover the combinatoric structure of the entire matroid stratification.

As a check at the first level, we indeed see that the decomposition of the two diagrams does indeed overlap in the matroids obtained by removing $12$ or removing $34$, which are precisely the lexicographically minimal sets of those permutations which in the arguments above were covered by both orders 1234 and 1243,
and by 3412 and 4312, respectively. Likewise, at the first level the decompositions do not overlap precisely on the matroid labels which are lexicographically minimal to those permutations which do not overlap for the two orderings. This is also true at the second level, where $\{12,13,24,34\}$ is missing $14$ and $23$, which are precisely those which are not lexicographically minimal of any permutation in equation \eref{ordering2}. Also, $\{12,14,23,34\}$ is missing $13$ and $24$, which are precisely those which are not lexicographically minimal of any permutation in equation \eref{ordering1}.

It is reasonable to expect that it might be possible to find which graphs are necessary to cover the entire matroid stratification by simply listing the set of all possible matroid elements, 
a set of permutations for which these elements are the lexicographically minimal subsets, and finding graphs whose ordering can achieve these permutations. We leave a detailed study of this interesting possibility for future investigation.

\bigskip

\section{Conclusions}

\label{section_conclusions}

We presented a detailed investigation of the geometric and combinatorial structures, such as the Grassmannian and toric Calabi-Yaus, which are ingrained in quantum field theory at a fundamental level. Such objects become manifest when formulating gauge theories in terms of on-shell diagrams, equivalently bipartite graphs. We extended these correspondences along various directions, most notably by the inclusion of non-planarity. In our opinion, the new structures we uncovered are natural candidates to arise in scattering amplitudes beyond the planar limit. This is certainly one of the most interesting questions in this area worth pursuing in the future.

As part of our investigation, we introduced a new combinatorial decomposition of the Grassmannian, which reduces to its positroid stratification for planar graphs. We explained how this decomposition can be directly obtained from the matching and matroid polytopes. We also extended the boundary measurement, which maps bipartite graphs to the Grassmannian, to graphs with an arbitrary number of boundaries. We discussed a quantitative measurement of graph reducibility and introduced several efficient algorithms for computing the boundary measurement, and for constructing the matroid and matching polytopes.

Our work suggests that general bipartite graphs, i.e.\ including non-planar ones, can lead to a more refined description of the Grassmannian.  It would be extremely interesting to continue investigating, along the lines of \sref{section_matroid_graphs}, how they can be exploited for the matroid stratification of the Grassmannian.

Finally, it would also be interesting to determine whether our ideas are relevant for the most recent geometric understanding of scattering amplitudes based on the amplituhedron.

\bigskip

\section*{Acknowledgements}

We would like to thank J. Bourjaily for useful discussions. The work of S.F and D. G. is supported by the U.K. Science and Technology Facilities Council (STFC). A.M. acknowledges funding by the Durham International Junior Research Fellowship.


\appendix

\newpage

\section{The Path Matrix}

\label{PathM}

In this appendix we describe an efficient algorithm to extract the paths for a given perfect orientation of a bipartite diagram, planar or non-planar. This is an important step of the boundary measurement which maps bipartite graphs to elements of the Grassmannian. The path matrix $\mathcal{M}$ is an $n_v \times n_v$ matrix, where $n_v$ is the number of vertices in the diagram. Given a perfect orientation, each entry $\mathcal{M}_{ab}$ contains the sum of edge weights for all oriented paths connecting vertices $a$ and $b$. We shall now show how this matrix can be obtained using the Kasteleyn matrix.

The perfect orientation is determined in terms of a reference perfect matching $p_{\text{ref}}$ as explained in \sref{RelationFlowPerfOrientMatch}. We now construct two matrices as follows: we define $K^{r}$ as the Kasteleyn matrix where we have set to zero the edge weights $X_{i,j} \in p_{\text{ref}}$ and replaced all other $X_{i,j} \to \alpha_{i,j}$; we define $\tilde K^{r}$ as the Kasteleyn matrix where we have set to zero all the edge weights not belonging to $p_{\text{ref}}$, and sent $X_{i,j} \to 1/\alpha_{i,j}$ for the edge weights $X_{i,j} \in p_{\text{ref}}$. We then arrange the following $n_v \times n_v$ matrix:

\begin{equation}
C=\left(
\begin{array} {cc}
\mathbb{I}_{n_w \times n_w} & -K^{r} \\ 
-(\tilde K^{r})^T & \mathbb{I}_{n_b \times n_b}
\end{array}
\right) \; ,
\end{equation}

\noindent where $n_w$ and $n_b$ is the number of white and black nodes, respectively. The path matrix is $\mathcal{M} =C^{-1}$. 

The entries $\mathcal{M}_{ab}$ are generally sums of ratios of edge weights $\alpha_{i,j}$, where the denominator contains those $\alpha_{i,j}$ in $\tilde{p}_{\text{ref}}$ which are relevant to the path. We remind the reader that an edge in the numerator signifies that the direction of that edge is from the white node to the black node, an edge in the denominator signifies the opposite direction. 

Sometimes a path from a vertex $a$ to a vertex $b$ contains a loop. This results in an infinite number of paths from $a$ to $b$, which differ in the number of times the path runs over the loop. The entry $\mathcal{M}_{ab}$ will thus contain the infinite sum of paths: $(1-\text{loop})^{-1} = 1+ \text{loop}+(\text{loop})^2+ \dots$.

Let us consider the non-planar bipartite graph associated with $Gr_{3,5}$, displayed in \fref{fig:nonplanarfriend}. The Kasteleyn matrix is
{\small
\be
K=
\left(
\begin{array}{ccccc}
 \ X_{6,2} \ & \ X_{2,1} \ & \ X_{1,6} \ & 0 & 0 \\
 X_{3,6} & X_{1,3} & 0 & \ X_{6,1} \ & 0 \\
 0 & 0 & X_{4,1} & X_{1,5} & \ X_{5,4} \ \\
 X_{2,3} & 0 & 0 & 0 & 0 \\
 0 & X_{3,2} & 0 & 0 & 0 \\
 0 & 0 & X_{6,4} & 0 & 0 \\
 0 & 0 & 0 & X_{5,6} & 0
\end{array}
\right) \; .
\ee}
Let us consider the perfect orientation in \fref{fig:nonplanarfriend}. The two auxiliary matrices become
{\small
\be
K^{r}=
\left(
\begin{array}{ccccc}
 \ \alpha_{6,2} \ & \ \alpha_{2,1} \ & 0 & 0 & 0 \\
 \alpha_{3,6} & 0 & 0 & \alpha_{6,1} & 0 \\
 0 & 0 & \ \alpha_{4,1} \ & \ \alpha_{1,5} \ & \ \ 0 \ \ \\
 0 & 0 & 0 & 0 & 0 \\
 0 & \alpha_{3,2} & 0 & 0 & 0 \\
 0 & 0 & \alpha_{6,4} & 0 & 0 \\
 0 & 0 & 0 & 0 & 0
\end{array}
\right)
\qquad 
\tilde K^{r}=
\left(
\begin{array}{ccccc}
 0 & 0 & \ \frac{1}{\alpha_{1,6}} \ & 0 & 0 \\
 0 & \ \frac{1}{\alpha_{1,3}} \ & 0 & 0 & 0 \\
 0 & 0 & 0 & 0 & \ \frac{1}{\alpha_{5,4}} \ \\
 \ \frac{1}{\alpha_{2,3}} \ & 0 & 0 & 0 & 0 \\
 0 & 0 & 0 & 0 & 0 \\
 0 & 0 & 0 & 0 & 0 \\
 0 & 0 & 0 & \ \frac{1}{\alpha_{5,6}} \ & 0
\end{array}
\right) \nonumber \; . 
\ee}
The path matrix is
{
\begin{tiny}
\be
\mathcal{M}=C^{-1}=
\left(
\begin{array}{cccccc}
 1 & \frac{\alpha_{2,1}}{\alpha_{1,3}} & 0 & \frac{\alpha_{2,1} \alpha_{3,6}+\alpha_{1,3} \alpha_{6,2}}{\alpha_{1,3} \alpha_{2,3}} & 0 & 0 \\
 0 & 1 & 0 & \frac{\alpha_{3,6}}{\alpha_{2,3}} & 0 & 0 \\
 \frac{\alpha_{4,1}}{\alpha_{1,6}} & \frac{\alpha_{2,1} \alpha_{4,1}}{\alpha_{1,3} \alpha_{1,6}} & 1 & \frac{\alpha_{4,1} \left(\alpha_{2,1} \alpha_{3,6}+\alpha_{1,3} \alpha_{6,2}\right)}{\alpha_{1,3} \alpha_{1,6} \alpha_{2,3}} & 0 & 0 \\
 0 & 0 & 0 & 1 & 0 & 0 \\
 0 & \frac{\alpha_{3,2}}{\alpha_{1,3}} & 0 & \frac{\alpha_{3,2} \alpha_{3,6}}{\alpha_{1,3} \alpha_{2,3}} & 1 & 0 \\
 \frac{\alpha_{6,4}}{\alpha_{1,6}} & \frac{\alpha_{2,1} \alpha_{6,4}}{\alpha_{1,3} \alpha_{1,6}} & 0 & \frac{\left(\alpha_{2,1} \alpha_{3,6}+\alpha_{1,3} \alpha_{6,2}\right) \alpha_{6,4}}{\alpha_{1,3} \alpha_{1,6} \alpha_{2,3}} & 0 & 1 \\
 0 & 0 & 0 & 0 & 0 & 0 \\
 0 & 0 & 0 & \frac{1}{\alpha_{2,3}} & 0 & 0 \\
 0 & \frac{1}{\alpha_{1,3}} & 0 & \frac{\alpha_{3,6}}{\alpha_{1,3} \alpha_{2,3}} & 0 & 0 \\
 \frac{1}{\alpha_{1,6}} & \frac{\alpha_{2,1}}{\alpha_{1,3} \alpha_{1,6}} & 0 & \frac{\alpha_{2,1} \alpha_{3,6}+\alpha_{1,3} \alpha_{6,2}}{\alpha_{1,3} \alpha_{1,6} \alpha_{2,3}} & 0 & 0 \\
 0 & 0 & 0 & 0 & 0 & 0 \\
 \frac{\alpha_{4,1}}{\alpha_{1,6} \alpha_{5,4}} & \frac{\alpha_{2,1} \alpha_{4,1}}{\alpha_{1,3} \alpha_{1,6} \alpha_{5,4}} & \frac{1}{\alpha_{5,4}} & \frac{\alpha_{4,1} \left(\alpha_{2,1} \alpha_{3,6}+\alpha_{1,3} \alpha_{6,2}\right)}{\alpha_{1,3} \alpha_{1,6} \alpha_{2,3} \alpha_{5,4}} & 0 & 0
\end{array}
\right.
\qquad
\dots
\nonumber
\ee
\be
\dots
\left.
\begin{array}{cccccc}
 \frac{\alpha_{2,1} \alpha_{6,1}}{\alpha_{1,3} \alpha_{5,6}} & \frac{\alpha_{2,1} \alpha_{3,6}}{\alpha_{1,3}}+\alpha_{6,2} & \alpha_{2,1} & 0 & \frac{\alpha_{2,1} \alpha_{6,1}}{\alpha_{1,3}} & 0 \\
 \frac{\alpha_{6,1}}{\alpha_{5,6}} & \alpha_{3,6} & 0 & 0 & \alpha_{6,1} & 0 \\
 \frac{\alpha_{1,5}+\frac{\alpha_{2,1} \alpha_{4,1} \alpha_{6,1}}{\alpha_{1,3} \alpha_{1,6}}}{\alpha_{5,6}} & \frac{\alpha_{4,1} \left(\alpha_{2,1} \alpha_{3,6}+\alpha_{1,3} \alpha_{6,2}\right)}{\alpha_{1,3} \alpha_{1,6}} & \frac{\alpha_{2,1} \alpha_{4,1}}{\alpha_{1,6}} & \alpha_{4,1} & \alpha_{1,5}+\frac{\alpha_{2,1} \alpha_{4,1} \alpha_{6,1}}{\alpha_{1,3} \alpha_{1,6}} & 0 \\
 0 & 0 & 0 & 0 & 0 & 0 \\
 \frac{\alpha_{3,2} \alpha_{6,1}}{\alpha_{1,3} \alpha_{5,6}} & \frac{\alpha_{3,2} \alpha_{3,6}}{\alpha_{1,3}} & \alpha_{3,2} & 0 & \frac{\alpha_{3,2} \alpha_{6,1}}{\alpha_{1,3}} & 0 \\
 \frac{\alpha_{2,1} \alpha_{6,1} \alpha_{6,4}}{\alpha_{1,3} \alpha_{1,6} \alpha_{5,6}} & \frac{\left(\alpha_{2,1} \alpha_{3,6}+\alpha_{1,3} \alpha_{6,2}\right) \alpha_{6,4}}{\alpha_{1,3} \alpha_{1,6}} & \frac{\alpha_{2,1} \alpha_{6,4}}{\alpha_{1,6}} & \alpha_{6,4} & \frac{\alpha_{2,1} \alpha_{6,1} \alpha_{6,4}}{\alpha_{1,3} \alpha_{1,6}} & 0 \\
 1 & 0 & 0 & 0 & 0 & 0 \\
 0 & 1 & 0 & 0 & 0 & 0 \\
 \frac{\alpha_{6,1}}{\alpha_{1,3} \alpha_{5,6}} & \frac{\alpha_{3,6}}{\alpha_{1,3}} & 1 & 0 & \frac{\alpha_{6,1}}{\alpha_{1,3}} & 0 \\
 \frac{\alpha_{2,1} \alpha_{6,1}}{\alpha_{1,3} \alpha_{1,6} \alpha_{5,6}} & \frac{\alpha_{2,1} \alpha_{3,6}+\alpha_{1,3} \alpha_{6,2}}{\alpha_{1,3} \alpha_{1,6}} & \frac{\alpha_{2,1}}{\alpha_{1,6}} & 1 & \frac{\alpha_{2,1} \alpha_{6,1}}{\alpha_{1,3} \alpha_{1,6}} & 0 \\
 \frac{1}{\alpha_{5,6}} & 0 & 0 & 0 & 1 & 0 \\
 \frac{\alpha_{1,3} \alpha_{1,5} \alpha_{1,6}+\alpha_{2,1} \alpha_{4,1} \alpha_{6,1}}{\alpha_{1,3} \alpha_{1,6} \alpha_{5,4} \alpha_{5,6}} & \frac{\alpha_{4,1} \left(\alpha_{2,1} \alpha_{3,6}+\alpha_{1,3} \alpha_{6,2}\right)}{\alpha_{1,3} \alpha_{1,6} \alpha_{5,4}} & \frac{\alpha_{2,1} \alpha_{4,1}}{\alpha_{1,6} \alpha_{5,4}} & \frac{\alpha_{4,1}}{\alpha_{5,4}} & \frac{\alpha_{1,5}+\frac{\alpha_{2,1} \alpha_{4,1} \alpha_{6,1}}{\alpha_{1,3} \alpha_{1,6}}}{\alpha_{5,4}} & 1
\end{array}
\right)
\nonumber \; . 
\ee
\end{tiny}}

\smallskip

\section{Combinatorial Reduction for a Reducible Graph}

\label{section_appendix_poset_double_box}

\begin{figure}[H]
\begin{center}
\includegraphics[width=19cm,angle=90]{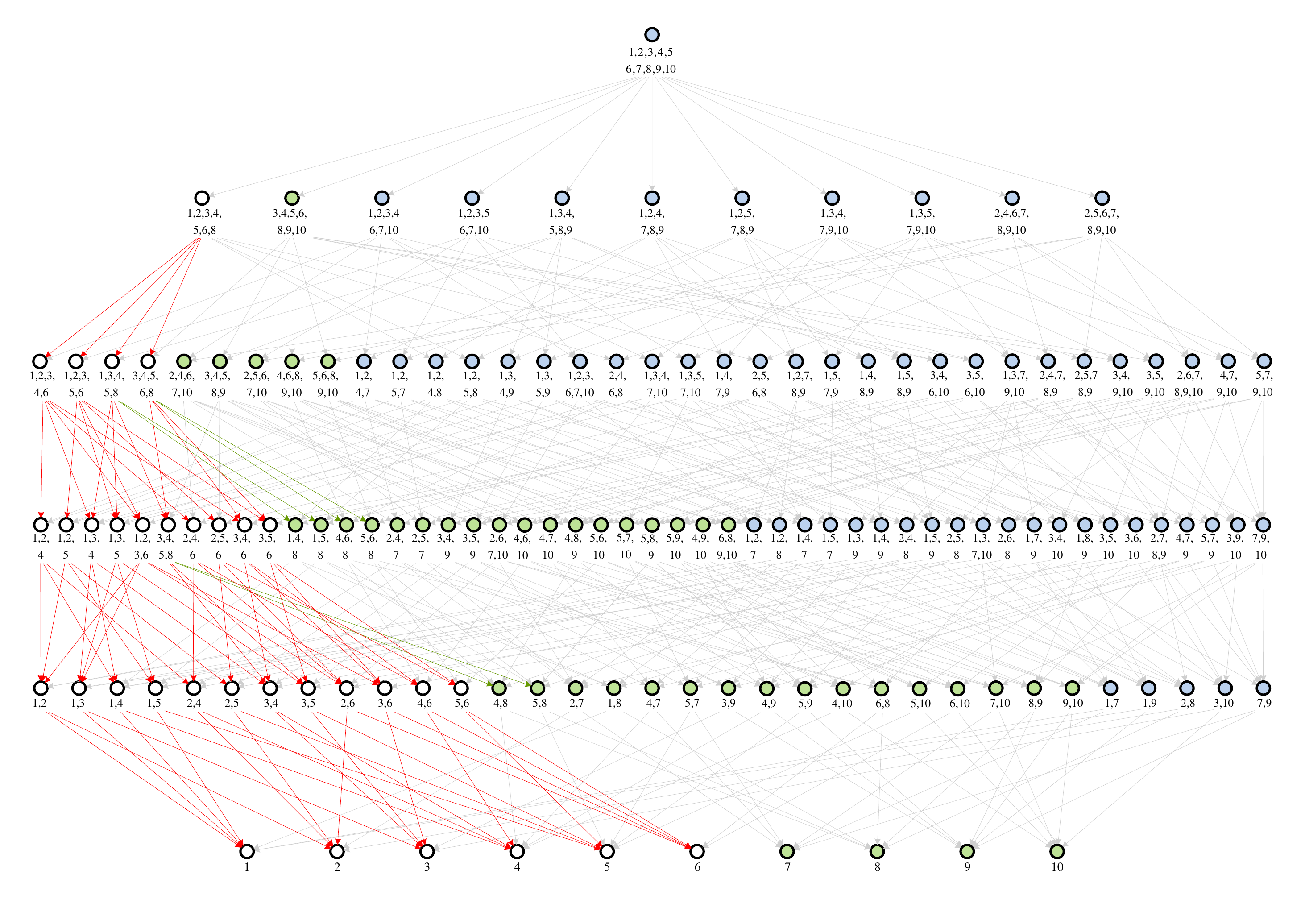}
\caption{Face lattice of the matching polytope for \fref{G24_2_boxes}. At each point, we indicate the surviving perfect matchings. Following the identifications in \eref{identifications_G24_2_boxes}, green and blue nodes in the poset are subject to horizontal and vertical identifications, respectively.}
\label{DoubleSqbLatticeOrderedNicev6}
\end{center}
\end{figure}

\fref{DoubleSqbLatticeOrderedNicev6} shows the face lattice of the matching polytope for the reducible graph in \fref{G24_2_boxes}. We list the surviving perfect matchings for every point in the poset. Due to space limitations, we do not provide the corresponding bipartite graphs. Green and blue dots are merged with white ones under horizontal and vertical identifications, respectively. The identifications are determined by \eref{identifications_G24_2_boxes}. 

\bigskip

\section{Perfect Matching Matrix for an Example With 3 Boundaries}

\label{section_P_matrix_3_boundaries}

For those readers interested in following the details of our calculations, here we provide the perfect matching matrix for the graph in \fref{fig:3Bguy}, which has 88 perfect matchings.

\bigskip

\noindent\makebox[\textwidth]{%
\tiny
$
P =
\left(
\begin{array}{c|ccccccccccccccccc}
 & p_1 & p_2 & p_3 & p_4 & p_5 & p_6 & p_7 & p_8 & p_9 & p_{10} & p_{11} & p_{12} & p_{13} & p_{14} & p_{15} & p_{16} & p_{17} \\
 \hline
 X_{1,4} & 1 & 1 & 1 & 1 & 0 & 0 & 0 & 0 & 0 & 0 & 1 & 1 & 1 & 1 & 1 & 1 & 1 \\
 X_{3,7} & 1 & 1 & 0 & 0 & 1 & 1 & 0 & 0 & 0 & 0 & 1 & 1 & 0 & 0 & 0 & 0 & 0 \\
 X_{3,10} & 1 & 0 & 0 & 0 & 1 & 1 & 1 & 0 & 0 & 0 & 1 & 0 & 1 & 0 & 0 & 0 & 0 \\
 X_{6,4} & 1 & 0 & 1 & 0 & 0 & 0 & 0 & 0 & 0 & 0 & 1 & 0 & 1 & 1 & 1 & 0 & 0 \\
 X_{8,2} & 1 & 1 & 1 & 1 & 1 & 0 & 0 & 1 & 0 & 0 & 0 & 0 & 0 & 0 & 0 & 0 & 0 \\
 X_{6,3} & 0 & 0 & 1 & 0 & 0 & 0 & 0 & 1 & 1 & 1 & 0 & 0 & 0 & 1 & 1 & 0 & 0 \\
 X_{9,3} & 0 & 0 & 1 & 1 & 0 & 0 & 0 & 1 & 1 & 0 & 0 & 0 & 0 & 1 & 0 & 1 & 0 \\
 X_{4,10} & 0 & 0 & 0 & 0 & 1 & 1 & 1 & 1 & 1 & 1 & 0 & 0 & 0 & 0 & 0 & 0 & 0 \\
 X_{5,1} & 0 & 0 & 0 & 0 & 0 & 1 & 1 & 0 & 1 & 1 & 0 & 0 & 0 & 0 & 0 & 0 & 0 \\
 X_{5,6} & 0 & 0 & 0 & 0 & 0 & 0 & 0 & 0 & 0 & 0 & 0 & 0 & 0 & 0 & 0 & 0 & 0 \\
 X_{2,5} & 0 & 0 & 0 & 0 & 0 & 0 & 0 & 0 & 0 & 0 & 0 & 0 & 0 & 0 & 0 & 0 & 0 \\
 X_{1,8} & 0 & 0 & 0 & 0 & 0 & 0 & 0 & 0 & 0 & 0 & 1 & 1 & 1 & 1 & 1 & 1 & 1 \\
 X_{2,9} & 0 & 0 & 0 & 0 & 0 & 0 & 1 & 0 & 0 & 1 & 0 & 0 & 1 & 0 & 1 & 0 & 1 \\
 X_{10,1} & 0 & 0 & 0 & 0 & 0 & 0 & 0 & 0 & 0 & 0 & 0 & 0 & 0 & 0 & 0 & 0 & 0 \\
 X_{7,2} & 0 & 0 & 0 & 0 & 0 & 0 & 0 & 0 & 0 & 0 & 0 & 0 & 0 & 0 & 0 & 0 & 0 \\
 X_{10,6} & 0 & 1 & 0 & 1 & 0 & 0 & 0 & 0 & 0 & 0 & 0 & 1 & 0 & 0 & 0 & 1 & 1 \\
 X_{6,7} & 1 & 1 & 1 & 1 & 1 & 1 & 1 & 1 & 1 & 1 & 1 & 1 & 1 & 1 & 1 & 1 & 1 \\
 X_{8,10} & 1 & 1 & 1 & 1 & 1 & 1 & 1 & 1 & 1 & 1 & 0 & 0 & 0 & 0 & 0 & 0 & 0 \\
 X_{4,5} & 0 & 1 & 0 & 1 & 1 & 1 & 1 & 1 & 1 & 1 & 0 & 1 & 0 & 0 & 0 & 1 & 1 \\
 X_{9,8} & 0 & 0 & 0 & 0 & 0 & 1 & 0 & 0 & 1 & 0 & 1 & 1 & 0 & 1 & 0 & 1 & 0 \\
 X_{7,6} & 0 & 0 & 0 & 1 & 0 & 0 & 1 & 0 & 0 & 0 & 0 & 0 & 1 & 0 & 0 & 1 & 1 \\
 X_{10,9} & 0 & 1 & 0 & 0 & 0 & 0 & 0 & 0 & 0 & 1 & 0 & 1 & 0 & 0 & 1 & 0 & 1 \\
 Y_{4,5} & 0 & 0 & 0 & 0 & 1 & 0 & 0 & 1 & 0 & 0 & 0 & 0 & 0 & 0 & 0 & 0 & 0 \\
\end{array}
\right. \dots
$
}
\noindent\makebox[\textwidth]{%
\tiny
$
\hspace{1cm}
\dots
\left.
\begin{array}{cccccccccccccccccc}
 p_{18} & p_{19} & p_{20} & p_{21} & p_{22} & p_{23} & p_{24} & p_{25} & p_{26} & p_{27} & p_{28} & p_{29} & p_{30} & p_{31} & p_{32} & p_{33} & p_{34} & p_{35} \\
 \hline
 0 & 0 & 0 & 0 & 0 & 0 & 0 & 0 & 0 & 0 & 0 & 0 & 0 & 0 & 0 & 1 & 1 & 1 \\
 1 & 1 & 1 & 1 & 1 & 0 & 0 & 0 & 0 & 0 & 0 & 0 & 0 & 0 & 0 & 1 & 1 & 1 \\
 1 & 1 & 1 & 0 & 0 & 1 & 1 & 0 & 0 & 0 & 0 & 0 & 0 & 0 & 0 & 1 & 0 & 0 \\
 1 & 1 & 0 & 0 & 0 & 1 & 0 & 1 & 1 & 1 & 0 & 0 & 0 & 0 & 0 & 1 & 0 & 0 \\
 1 & 0 & 0 & 1 & 0 & 0 & 0 & 1 & 0 & 0 & 1 & 0 & 0 & 0 & 0 & 0 & 1 & 0 \\
 0 & 0 & 0 & 0 & 0 & 0 & 0 & 1 & 1 & 1 & 0 & 1 & 1 & 0 & 0 & 0 & 0 & 0 \\
 0 & 0 & 0 & 0 & 0 & 0 & 0 & 1 & 1 & 0 & 1 & 1 & 0 & 1 & 0 & 0 & 0 & 0 \\
 0 & 0 & 1 & 0 & 0 & 0 & 1 & 0 & 0 & 0 & 0 & 1 & 1 & 0 & 0 & 0 & 0 & 0 \\
 0 & 1 & 0 & 0 & 1 & 1 & 0 & 0 & 1 & 1 & 0 & 0 & 0 & 1 & 1 & 0 & 0 & 0 \\
 0 & 0 & 0 & 0 & 0 & 0 & 0 & 0 & 0 & 0 & 0 & 0 & 0 & 0 & 0 & 0 & 1 & 0 \\
 0 & 0 & 0 & 0 & 0 & 0 & 0 & 0 & 0 & 0 & 0 & 0 & 0 & 0 & 0 & 1 & 0 & 1 \\
 0 & 0 & 1 & 0 & 0 & 0 & 1 & 0 & 0 & 0 & 0 & 1 & 1 & 0 & 0 & 0 & 0 & 0 \\
 0 & 0 & 0 & 0 & 0 & 1 & 1 & 0 & 0 & 1 & 0 & 0 & 1 & 0 & 1 & 0 & 0 & 0 \\
 1 & 1 & 0 & 1 & 1 & 1 & 0 & 1 & 1 & 1 & 1 & 0 & 0 & 1 & 1 & 0 & 0 & 0 \\
 0 & 0 & 0 & 0 & 0 & 0 & 0 & 0 & 0 & 0 & 0 & 0 & 0 & 0 & 0 & 0 & 0 & 0 \\
 0 & 0 & 0 & 1 & 1 & 0 & 0 & 0 & 0 & 0 & 1 & 0 & 0 & 1 & 1 & 0 & 1 & 1 \\
 1 & 1 & 1 & 1 & 1 & 1 & 1 & 1 & 1 & 1 & 1 & 1 & 1 & 1 & 1 & 0 & 0 & 0 \\
 0 & 0 & 0 & 0 & 0 & 0 & 0 & 0 & 0 & 0 & 0 & 0 & 0 & 0 & 0 & 1 & 1 & 1 \\
 0 & 0 & 1 & 1 & 1 & 0 & 1 & 0 & 0 & 0 & 1 & 1 & 1 & 1 & 1 & 0 & 0 & 1 \\
 0 & 1 & 1 & 0 & 1 & 0 & 0 & 0 & 1 & 0 & 0 & 1 & 0 & 1 & 0 & 1 & 0 & 1 \\
 0 & 0 & 0 & 0 & 0 & 1 & 1 & 0 & 0 & 0 & 1 & 0 & 0 & 1 & 1 & 0 & 0 & 0 \\
 0 & 0 & 0 & 1 & 1 & 0 & 0 & 0 & 0 & 1 & 0 & 0 & 1 & 0 & 1 & 0 & 1 & 1 \\
 1 & 0 & 1 & 1 & 0 & 0 & 1 & 1 & 0 & 0 & 1 & 1 & 1 & 0 & 0 & 0 & 0 & 0 \\
\end{array}
\right. \dots
$
}
\noindent\makebox[\textwidth]{%
\tiny
$
\hspace{1cm}
\dots
\left.
\begin{array}{cccccccccccccccccc}
 p_{36} & p_{37} & p_{38} & p_{39} & p_{40} & p_{41} & p_{42} & p_{43} & p_{44} & p_{45} & p_{46} & p_{47} & p_{48} & p_{49} & p_{50} & p_{51} & p_{52} & p_{53} \\
 \hline
 1 & 1 & 1 & 1 & 1 & 1 & 1 & 1 & 1 & 0 & 0 & 0 & 0 & 0 & 0 & 0 & 0 & 0 \\
 0 & 0 & 0 & 0 & 0 & 0 & 0 & 0 & 0 & 1 & 1 & 1 & 0 & 0 & 0 & 0 & 0 & 0 \\
 1 & 1 & 0 & 0 & 0 & 0 & 0 & 0 & 0 & 1 & 1 & 1 & 1 & 1 & 1 & 1 & 0 & 0 \\
 1 & 1 & 1 & 1 & 1 & 0 & 0 & 0 & 0 & 0 & 0 & 0 & 0 & 0 & 0 & 0 & 0 & 0 \\
 1 & 0 & 1 & 0 & 0 & 1 & 1 & 0 & 0 & 1 & 0 & 0 & 1 & 0 & 0 & 0 & 1 & 1 \\
 0 & 0 & 1 & 1 & 1 & 0 & 0 & 0 & 0 & 0 & 0 & 0 & 0 & 0 & 0 & 0 & 1 & 1 \\
 0 & 0 & 0 & 1 & 0 & 1 & 0 & 1 & 0 & 0 & 0 & 0 & 0 & 0 & 0 & 0 & 1 & 0 \\
 0 & 0 & 0 & 0 & 0 & 0 & 0 & 0 & 0 & 1 & 1 & 1 & 1 & 1 & 1 & 1 & 1 & 1 \\
 0 & 0 & 0 & 0 & 0 & 0 & 0 & 0 & 0 & 0 & 0 & 1 & 0 & 1 & 0 & 1 & 0 & 0 \\
 0 & 0 & 0 & 0 & 0 & 1 & 0 & 0 & 0 & 1 & 0 & 1 & 0 & 0 & 0 & 1 & 1 & 0 \\
 0 & 1 & 0 & 1 & 1 & 0 & 0 & 1 & 1 & 0 & 1 & 0 & 0 & 0 & 1 & 0 & 0 & 0 \\
 0 & 0 & 0 & 0 & 0 & 0 & 0 & 0 & 0 & 0 & 0 & 0 & 0 & 0 & 0 & 0 & 0 & 0 \\
 0 & 1 & 0 & 0 & 1 & 0 & 0 & 0 & 1 & 0 & 0 & 0 & 0 & 0 & 1 & 1 & 0 & 0 \\
 0 & 0 & 0 & 0 & 0 & 0 & 0 & 0 & 0 & 0 & 0 & 0 & 0 & 0 & 0 & 0 & 0 & 0 \\
 1 & 0 & 1 & 0 & 0 & 0 & 1 & 0 & 0 & 0 & 0 & 0 & 1 & 1 & 0 & 0 & 0 & 1 \\
 0 & 0 & 0 & 0 & 0 & 1 & 1 & 1 & 1 & 0 & 0 & 0 & 0 & 0 & 0 & 0 & 0 & 0 \\
 0 & 0 & 0 & 0 & 0 & 0 & 0 & 0 & 0 & 0 & 0 & 0 & 0 & 0 & 0 & 0 & 0 & 0 \\
 1 & 1 & 1 & 1 & 1 & 1 & 1 & 1 & 1 & 1 & 1 & 1 & 1 & 1 & 1 & 1 & 1 & 1 \\
 0 & 0 & 0 & 0 & 0 & 0 & 1 & 1 & 1 & 0 & 1 & 0 & 1 & 1 & 1 & 0 & 0 & 1 \\
 0 & 0 & 0 & 1 & 0 & 0 & 0 & 1 & 0 & 0 & 1 & 1 & 0 & 1 & 0 & 0 & 0 & 0 \\
 1 & 1 & 0 & 0 & 0 & 1 & 1 & 1 & 1 & 0 & 0 & 0 & 1 & 1 & 1 & 1 & 0 & 0 \\
 0 & 0 & 1 & 0 & 1 & 0 & 1 & 0 & 1 & 0 & 0 & 0 & 0 & 0 & 0 & 0 & 0 & 1 \\
 0 & 0 & 0 & 0 & 0 & 0 & 0 & 0 & 0 & 1 & 1 & 0 & 1 & 0 & 1 & 0 & 1 & 1 \\
\end{array}
\right. \dots
$
}
\noindent\makebox[\textwidth]{%
\tiny
$
\hspace{1cm}
\dots
\left.
\begin{array}{cccccccccccccccccc}
 p_{54} & p_{55} & p_{56} & p_{57} & p_{58} & p_{59} & p_{60} & p_{61} & p_{62} & p_{63} & p_{64} & p_{65} & p_{66} & p_{67} & p_{68} & p_{69} & p_{70} & p_{71} \\
 \hline
 0 & 0 & 0 & 0 & 0 & 1 & 1 & 1 & 1 & 1 & 1 & 0 & 0 & 0 & 0 & 0 & 0 & 0 \\
 0 & 0 & 0 & 0 & 0 & 1 & 0 & 0 & 0 & 0 & 0 & 1 & 1 & 1 & 1 & 1 & 0 & 0 \\
 0 & 0 & 0 & 0 & 0 & 0 & 1 & 0 & 0 & 0 & 0 & 1 & 1 & 0 & 0 & 0 & 1 & 1 \\
 0 & 0 & 0 & 0 & 0 & 0 & 1 & 1 & 0 & 0 & 0 & 1 & 0 & 0 & 0 & 0 & 1 & 1 \\
 0 & 0 & 0 & 0 & 0 & 0 & 0 & 0 & 0 & 0 & 0 & 0 & 0 & 1 & 0 & 0 & 1 & 0 \\
 1 & 1 & 1 & 1 & 1 & 0 & 0 & 1 & 0 & 0 & 0 & 0 & 0 & 0 & 0 & 0 & 0 & 0 \\
 1 & 1 & 0 & 0 & 0 & 0 & 0 & 0 & 1 & 0 & 0 & 0 & 0 & 0 & 0 & 0 & 0 & 0 \\
 1 & 1 & 1 & 1 & 1 & 0 & 0 & 0 & 0 & 0 & 0 & 0 & 1 & 0 & 0 & 0 & 0 & 0 \\
 0 & 1 & 1 & 0 & 1 & 0 & 0 & 0 & 0 & 0 & 0 & 0 & 0 & 0 & 0 & 1 & 0 & 1 \\
 0 & 1 & 0 & 0 & 1 & 1 & 0 & 0 & 1 & 0 & 1 & 0 & 1 & 1 & 0 & 1 & 0 & 0 \\
 1 & 0 & 0 & 1 & 0 & 0 & 0 & 0 & 0 & 0 & 0 & 1 & 0 & 0 & 1 & 0 & 0 & 0 \\
 0 & 0 & 0 & 0 & 0 & 1 & 1 & 1 & 1 & 1 & 1 & 0 & 1 & 0 & 0 & 0 & 0 & 0 \\
 0 & 0 & 0 & 1 & 1 & 0 & 0 & 0 & 0 & 0 & 1 & 0 & 0 & 0 & 0 & 0 & 0 & 0 \\
 0 & 0 & 0 & 0 & 0 & 0 & 0 & 0 & 0 & 0 & 0 & 1 & 0 & 1 & 1 & 1 & 1 & 1 \\
 0 & 0 & 1 & 0 & 0 & 0 & 1 & 1 & 0 & 1 & 0 & 0 & 0 & 0 & 0 & 0 & 1 & 1 \\
 0 & 0 & 0 & 0 & 0 & 1 & 0 & 0 & 1 & 1 & 1 & 0 & 0 & 1 & 1 & 1 & 0 & 0 \\
 0 & 0 & 0 & 0 & 0 & 0 & 0 & 0 & 0 & 0 & 0 & 0 & 0 & 0 & 0 & 0 & 0 & 0 \\
 1 & 1 & 1 & 1 & 1 & 0 & 0 & 0 & 0 & 0 & 0 & 0 & 0 & 0 & 0 & 0 & 0 & 0 \\
 1 & 0 & 1 & 1 & 0 & 0 & 0 & 0 & 0 & 1 & 0 & 0 & 0 & 0 & 1 & 0 & 0 & 0 \\
 1 & 1 & 1 & 0 & 0 & 1 & 1 & 1 & 1 & 1 & 0 & 1 & 1 & 0 & 1 & 1 & 0 & 1 \\
 0 & 0 & 0 & 0 & 0 & 0 & 1 & 0 & 1 & 1 & 1 & 0 & 0 & 0 & 0 & 0 & 1 & 1 \\
 0 & 0 & 1 & 1 & 1 & 1 & 0 & 1 & 0 & 1 & 1 & 0 & 0 & 1 & 1 & 1 & 0 & 0 \\
 1 & 0 & 0 & 1 & 0 & 0 & 0 & 0 & 0 & 0 & 0 & 1 & 1 & 1 & 1 & 0 & 1 & 0 \\
\end{array}
\right. \dots
$
}
\noindent\makebox[\textwidth]{%
\tiny
$
\hspace{1cm}
\dots
\left.
\begin{array}{ccccccccccccccccc}
 p_{72} & p_{73} & p_{74} & p_{75} & p_{76} & p_{77} & p_{78} & p_{79} & p_{80} & p_{81} & p_{82} & p_{83} & p_{84} & p_{85} & p_{86} & p_{87} & p_{88} \\
 \hline
 0 & 0 & 0 & 0 & 0 & 0 & 0 & 0 & 0 & 0 & 0 & 0 & 0 & 0 & 0 & 0 & 0 \\
 0 & 0 & 0 & 0 & 0 & 0 & 0 & 0 & 0 & 0 & 0 & 0 & 0 & 0 & 0 & 0 & 0 \\
 1 & 1 & 1 & 0 & 0 & 0 & 0 & 0 & 0 & 0 & 0 & 0 & 0 & 0 & 0 & 0 & 0 \\
 1 & 0 & 0 & 1 & 1 & 1 & 1 & 0 & 0 & 0 & 0 & 0 & 0 & 0 & 0 & 0 & 0 \\
 0 & 0 & 0 & 1 & 0 & 0 & 0 & 1 & 1 & 0 & 0 & 0 & 0 & 0 & 0 & 0 & 0 \\
 0 & 0 & 0 & 1 & 1 & 1 & 1 & 0 & 0 & 1 & 1 & 1 & 0 & 0 & 0 & 0 & 0 \\
 0 & 0 & 0 & 0 & 1 & 0 & 0 & 1 & 0 & 1 & 0 & 0 & 1 & 1 & 0 & 0 & 0 \\
 0 & 1 & 1 & 0 & 0 & 0 & 0 & 0 & 0 & 1 & 1 & 1 & 0 & 0 & 0 & 0 & 0 \\
 0 & 0 & 0 & 0 & 0 & 1 & 0 & 0 & 0 & 0 & 0 & 0 & 0 & 1 & 1 & 0 & 1 \\
 0 & 0 & 1 & 0 & 0 & 0 & 0 & 1 & 0 & 1 & 0 & 1 & 0 & 1 & 0 & 0 & 1 \\
 1 & 0 & 0 & 0 & 1 & 0 & 1 & 0 & 0 & 0 & 0 & 0 & 1 & 0 & 0 & 1 & 0 \\
 0 & 1 & 1 & 0 & 0 & 0 & 0 & 0 & 0 & 1 & 1 & 1 & 0 & 0 & 0 & 0 & 0 \\
 1 & 0 & 1 & 0 & 0 & 0 & 1 & 0 & 0 & 0 & 0 & 1 & 0 & 0 & 0 & 1 & 1 \\
 1 & 0 & 0 & 1 & 1 & 1 & 1 & 1 & 1 & 0 & 0 & 0 & 1 & 1 & 1 & 1 & 1 \\
 0 & 1 & 0 & 1 & 0 & 1 & 0 & 0 & 1 & 0 & 1 & 0 & 0 & 0 & 1 & 0 & 0 \\
 0 & 0 & 0 & 0 & 0 & 0 & 0 & 1 & 1 & 0 & 0 & 0 & 1 & 1 & 1 & 1 & 1 \\
 0 & 0 & 0 & 0 & 0 & 0 & 0 & 0 & 0 & 0 & 0 & 0 & 0 & 0 & 0 & 0 & 0 \\
 0 & 0 & 0 & 0 & 0 & 0 & 0 & 0 & 0 & 0 & 0 & 0 & 0 & 0 & 0 & 0 & 0 \\
 0 & 1 & 0 & 0 & 0 & 0 & 0 & 0 & 1 & 0 & 1 & 0 & 1 & 0 & 1 & 1 & 0 \\
 0 & 1 & 0 & 0 & 1 & 1 & 0 & 0 & 0 & 1 & 1 & 0 & 1 & 1 & 1 & 0 & 0 \\
 1 & 1 & 1 & 0 & 0 & 0 & 0 & 1 & 1 & 0 & 0 & 0 & 1 & 1 & 1 & 1 & 1 \\
 0 & 0 & 0 & 1 & 0 & 1 & 1 & 0 & 1 & 0 & 1 & 1 & 0 & 0 & 1 & 1 & 1 \\
 1 & 1 & 1 & 1 & 1 & 0 & 1 & 1 & 1 & 1 & 1 & 1 & 1 & 0 & 0 & 1 & 0 \\
\end{array}
\right)~~.
$
}

\bibliographystyle{JHEP}
\bibliography{matroref}

\providecommand{\href}[2]{#2}\begingroup\raggedright\begin{thebibliography}{10}

\bibitem{ArkaniHamed:2009dn}
N.~Arkani-Hamed, F.~Cachazo, C.~Cheung, and J.~Kaplan, {\it {A Duality For The
  S Matrix}},  {\em JHEP} {\bf 1003} (2010) 020,
  [\href{http://xxx.lanl.gov/abs/0907.5418}{{\tt arXiv:0907.5418}}].

\bibitem{Mason:2009qx}
L.~Mason and D.~Skinner, {\it {Dual Superconformal Invariance, Momentum
  Twistors and Grassmannians}},  {\em JHEP} {\bf 0911} (2009) 045,
  [\href{http://xxx.lanl.gov/abs/0909.0250}{{\tt arXiv:0909.0250}}].

\bibitem{ArkaniHamed:2009vw}
N.~Arkani-Hamed, F.~Cachazo, and C.~Cheung, {\it {The Grassmannian Origin Of
  Dual Superconformal Invariance}},  {\em JHEP} {\bf 1003} (2010) 036,
  [\href{http://xxx.lanl.gov/abs/0909.0483}{{\tt arXiv:0909.0483}}].

\bibitem{ArkaniHamed:2010kv}
N.~Arkani-Hamed, J.~L. Bourjaily, F.~Cachazo, S.~Caron-Huot, and J.~Trnka, {\it
  {The All-Loop Integrand For Scattering Amplitudes in Planar N=4 SYM}},  {\em
  JHEP} {\bf 1101} (2011) 041, [\href{http://xxx.lanl.gov/abs/1008.2958}{{\tt
  arXiv:1008.2958}}].

\bibitem{Bourjaily:2010kw}
J.~L. Bourjaily, J.~Trnka, A.~Volovich, and C.~Wen, {\it {The Grassmannian and
  the Twistor String: Connecting All Trees in N=4 SYM}},  {\em JHEP} {\bf 1101}
  (2011) 038, [\href{http://xxx.lanl.gov/abs/1006.1899}{{\tt
  arXiv:1006.1899}}].

\bibitem{ArkaniHamed:2012nw}
N.~Arkani-Hamed, J.~L. Bourjaily, F.~Cachazo, A.~B. Goncharov, A.~Postnikov,
  and J.~Trnka, {\it {Scattering Amplitudes and the Positive Grassmannian}},
  \href{http://xxx.lanl.gov/abs/1212.5605}{{\tt arXiv:1212.5605}}.

\bibitem{ref_amplituhedron}
N.~Arkani-Hamed and J.~Trnka, ``The amplituhedron.''
  \url{http://susy2013.ictp.it/} ,
  \url{http://www.staff.science.uu.nl/~tonge105/igst13/Trnka.pdf}, 2013.

\bibitem{Franco:2012mm}
S.~Franco, {\it {Bipartite Field Theories: from D-Brane Probes to Scattering
  Amplitudes}},  {\em JHEP} {\bf 1211} (2012) 141,
  [\href{http://xxx.lanl.gov/abs/1207.0807}{{\tt arXiv:1207.0807}}].

\bibitem{Franco:2012wv}
S.~Franco, D.~Galloni, and R.-K. Seong, {\it {New Directions in Bipartite Field
  Theories}},  {\em JHEP} {\bf 1306} (2013) 032,
  [\href{http://xxx.lanl.gov/abs/1211.5139}{{\tt arXiv:1211.5139}}].

\bibitem{Xie:2012mr}
D.~Xie and M.~Yamazaki, {\it {Network and Seiberg Duality}},  {\em JHEP} {\bf
  1209} (2012) 036, [\href{http://xxx.lanl.gov/abs/1207.0811}{{\tt
  arXiv:1207.0811}}].

\bibitem{Heckman:2012jh}
J.~J. Heckman, C.~Vafa, D.~Xie, and M.~Yamazaki, {\it {String Theory Origin of
  Bipartite SCFTs}},  {\em JHEP} {\bf 1305} (2013) 148,
  [\href{http://xxx.lanl.gov/abs/1211.4587}{{\tt arXiv:1211.4587}}].

\bibitem{Franco:2013pg}
S.~Franco, {\it {Cluster Transformations from Bipartite Field Theories}},
  \href{http://xxx.lanl.gov/abs/1301.0316}{{\tt arXiv:1301.0316}}.

\bibitem{Baur:2013hwa}
K.~Baur, A.~King, and R.~J. Marsh, {\it {Dimer models and cluster categories of
  Grassmannians}},  \href{http://xxx.lanl.gov/abs/1309.6524}{{\tt
  arXiv:1309.6524}}.

\bibitem{Hanany:2005ve}
A.~Hanany and K.~D. Kennaway, {\it {Dimer models and toric diagrams}},
  \href{http://xxx.lanl.gov/abs/hep-th/0503149}{{\tt hep-th/0503149}}.

\bibitem{Franco:2005rj}
S.~Franco, A.~Hanany, K.~D. Kennaway, D.~Vegh, and B.~Wecht, {\it {Brane dimers
  and quiver gauge theories}},  {\em JHEP} {\bf 0601} (2006) 096,
  [\href{http://xxx.lanl.gov/abs/hep-th/0504110}{{\tt hep-th/0504110}}].

\bibitem{Franco:2005sm}
S.~Franco, A.~Hanany, D.~Martelli, J.~Sparks, D.~Vegh, et~al., {\it {Gauge
  theories from toric geometry and brane tilings}},  {\em JHEP} {\bf 0601}
  (2006) 128, [\href{http://xxx.lanl.gov/abs/hep-th/0505211}{{\tt
  hep-th/0505211}}].

\bibitem{Butti:2005sw}
A.~Butti, D.~Forcella, and A.~Zaffaroni, {\it {The Dual superconformal theory
  for L**pqr manifolds}},  {\em JHEP} {\bf 0509} (2005) 018,
  [\href{http://xxx.lanl.gov/abs/hep-th/0505220}{{\tt hep-th/0505220}}].

\bibitem{Franco:2013ana}
S.~Franco and A.~Uranga, {\it {Bipartite Field Theories from D-Branes}},
  \href{http://xxx.lanl.gov/abs/1306.6331}{{\tt arXiv:1306.6331}}.

\bibitem{2006math09764P}
A.~{Postnikov}, {\it {Total positivity, Grassmannians, and networks}},  {\em
  ArXiv Mathematics e-prints} (2006)
  [\href{http://xxx.lanl.gov/abs/math/0609764}{{\tt math/0609764}}].

\bibitem{Postnikovlectures}
A.~Postnikov, ``{Positive Grassmannian}.''
  \url{http://www-math.mit.edu/~ahmorales/18.318lecs/lectures.pdf}, 2013.

\bibitem{2012arXiv1210.5433T}
K.~{Talaska} and L.~{Williams}, {\it {Network Parameterizations for the
  Grassmannian}},  {\em ArXiv e-prints} (2012)
  [\href{http://xxx.lanl.gov/abs/1210.5433}{{\tt arXiv:1210.5433}}].

\bibitem{oxley2006matroid}
J.~Oxley, {\em Matroid Theory}.
\newblock Oxford graduate texts in mathematics. Oxford University Press, 2006.

\bibitem{fink2010MatroidPolytopeSubdivisions}
A.~R. Fink, ``Matroid polytope subdivisions and valuations.''
  \url{http://www.maths.qmul.ac.uk/~fink/thesis.pdf}, 2010.

\bibitem{2012arXiv1204.6446K}
Y.~{Kodama} and L.~{Williams}, {\it {The Deodhar decomposition of the
  Grassmannian and the regularity of KP solitons}},  {\em Advances in
  Mathematics} {\bf 244} (2013) 979 -- 1032,
  [\href{http://xxx.lanl.gov/abs/1204.6446}{{\tt arXiv:1204.6446}}].

\bibitem{2012arXiv1202.3128T}
K.~{Talaska}, {\it {Determinants of weighted path matrices}},  {\em ArXiv
  e-prints} (2012) [\href{http://xxx.lanl.gov/abs/1202.3128}{{\tt
  arXiv:1202.3128}}].

\bibitem{Britto:2004ap}
R.~Britto, F.~Cachazo, and B.~Feng, {\it {New recursion relations for tree
  amplitudes of gluons}},  {\em Nucl.Phys.} {\bf B715} (2005) 499--522,
  [\href{http://xxx.lanl.gov/abs/hep-th/0412308}{{\tt hep-th/0412308}}].

\bibitem{Britto:2005fq}
R.~Britto, F.~Cachazo, B.~Feng, and E.~Witten, {\it {Direct proof of tree-level
  recursion relation in Yang-Mills theory}},  {\em Phys.Rev.Lett.} {\bf 94}
  (2005) 181602, [\href{http://xxx.lanl.gov/abs/hep-th/0501052}{{\tt
  hep-th/0501052}}].

\bibitem{2007arXiv0706.2501P}
A.~{Postnikov}, D.~{Speyer}, and L.~{Williams}, {\it {Matching polytopes, toric
  geometry, and the non-negative part of the Grassmannian}},  {\em Journal of
  Algebraic Combinatorics} {\bf 30} (2009) 173--191,
  [\href{http://xxx.lanl.gov/abs/0706.2501}{{\tt arXiv:0706.2501}}].

\bibitem{2008arXiv0801.4822T}
K.~Talaska, {\it {A formula for Pl\"ucker coordinates associated with a planar
  network}},  {\em International Mathematics Research Notices} {\bf 2008}
  (2008) [\href{http://xxx.lanl.gov/abs/0801.4822}{{\tt arXiv:0801.4822}}].

\bibitem{Amariti:2013ija}
A.~Amariti and D.~Forcella, {\it {Scattering Amplitudes and Toric Geometry}},
  \href{http://xxx.lanl.gov/abs/1305.5252}{{\tt arXiv:1305.5252}}.

\bibitem{Forcella:2008bb}
D.~Forcella, A.~Hanany, Y.-H. He, and A.~Zaffaroni, {\it {The Master Space of
  N=1 Gauge Theories}},  {\em JHEP} {\bf 0808} (2008) 012,
  [\href{http://xxx.lanl.gov/abs/0801.1585}{{\tt arXiv:0801.1585}}].

\bibitem{Seiberg:1994pq}
N.~Seiberg, {\it {Electric - magnetic duality in supersymmetric nonAbelian
  gauge theories}},  {\em Nucl.Phys.} {\bf B435} (1995) 129--146,
  [\href{http://xxx.lanl.gov/abs/hep-th/9411149}{{\tt hep-th/9411149}}].

\bibitem{Feng:2000mi}
B.~Feng, A.~Hanany, and Y.-H. He, {\it {D-brane gauge theories from toric
  singularities and toric duality}},  {\em Nucl.Phys.} {\bf B595} (2001)
  165--200, [\href{http://xxx.lanl.gov/abs/hep-th/0003085}{{\tt
  hep-th/0003085}}].

\bibitem{Feng:2001xr}
B.~Feng, A.~Hanany, and Y.-H. He, {\it {Phase structure of D-brane gauge
  theories and toric duality}},  {\em JHEP} {\bf 0108} (2001) 040,
  [\href{http://xxx.lanl.gov/abs/hep-th/0104259}{{\tt hep-th/0104259}}].

\bibitem{Feng:2001bn}
B.~Feng, A.~Hanany, Y.-H. He, and A.~M. Uranga, {\it {Toric duality as Seiberg
  duality and brane diamonds}},  {\em JHEP} {\bf 0112} (2001) 035,
  [\href{http://xxx.lanl.gov/abs/hep-th/0109063}{{\tt hep-th/0109063}}].

\bibitem{Beasley:2001zp}
C.~E. Beasley and M.~R. Plesser, {\it {Toric duality is Seiberg duality}},
  {\em JHEP} {\bf 0112} (2001) 001,
  [\href{http://xxx.lanl.gov/abs/hep-th/0109053}{{\tt hep-th/0109053}}].

\bibitem{Feng:2002zw}
B.~Feng, S.~Franco, A.~Hanany, and Y.-H. He, {\it {Symmetries of toric
  duality}},  {\em JHEP} {\bf 0212} (2002) 076,
  [\href{http://xxx.lanl.gov/abs/hep-th/0205144}{{\tt hep-th/0205144}}].

\bibitem{polymake}
E.~Gawrilow and M.~Joswig, {\it polymake: a framework for analyzing convex
  polytopes},  in {\em Polytopes Combinatorics and Computation} (G.~Kalai and
  G.~Ziegler, eds.), vol.~29 of {\em DMV Seminar}, pp.~43--73.
\newblock Birkhuser Basel, 2000.

\bibitem{2009arXiv0901.0020G}
M.~{Gekhtman}, M.~{Shapiro}, and A.~{Vainshtein}, {\it {Poisson Geometry of
  Directed Networks in an Annulus}},  {\em Journal of the European Mathematical
  Society} {\bf 14} (2012) 541--570,
  [\href{http://xxx.lanl.gov/abs/0901.0020}{{\tt arXiv:0901.0020}}].

\bibitem{2005arXiv09129W}
L.~K. {Williams}, {\it {Shelling totally nonnegative flag varieties}},  {\em
  J.~Reine~Angew.~Math.} {\bf 609} (2007) 001,
  [\href{http://xxx.lanl.gov/abs/0509129}{{\tt 0509129}}].

\end{thebibliography}\endgroup
\end{document}